\documentclass{emulateapj}
\usepackage{graphicx}
\usepackage{epsfig}
\newcommand{\km}{${\rm km\,s}^{-1}$}
\newcommand{\fuse}{{\em FUSE}}
\newcommand{\hi}{H$\;${\small\rm I}\relax}

\newcommand{\neviii}{Ne$\;${\small\rm VIII}\relax}
\newcommand{\ci}{C$\;${\small\rm I}\relax}
\newcommand{\cii}{C$\;${\small\rm II}\relax}
\newcommand{\ciii}{C$\;${\small\rm III}\relax}

\newcommand{\ovi}{O$\;${\small\rm VI}\relax}
\newcommand{\ovii}{O$\;${\small\rm VII}\relax}
\newcommand{\oviii}{O$\;${\small\rm VIII}\relax}

\newcommand{\siii}{Si$\;${\small\rm II}\relax}

\def\dex#1{10$^{#1}$}

\slugcomment{ApJ: received 2006 November 17; accepted 2006 December 10}
\shortauthors{Lehner et al.}
\shorttitle{Ly$\alpha$ Forest at $z \la 0.4$}
\begin{document}

\title{Physical Properties, Baryon Content, and Evolution of the Ly$\alpha$
Forest: New Insights from High Resolution Observations at $z \la 0.4 $\altaffilmark{1} }
\author{N.\ Lehner\altaffilmark{2},
	B.\ D.\ Savage\altaffilmark{3},
	P.\ Richter\altaffilmark{4},
	K.\ R. Sembach\altaffilmark{5},
	T.\ M. Tripp\altaffilmark{6}, 
	B.\ P.\ Wakker\altaffilmark{3}
	}
   
\altaffiltext{1}{Based on observations made with the NASA-CNES-CSA 
Far Ultraviolet Spectroscopic Explorer. FUSE is operated for NASA by the Johns 
Hopkins University under NASA contract NAS5-32985. Based on observations made with the NASA/ESA Hubble Space Telescope,
obtained at the Space Telescope Science Institute, which is operated by the
Association of Universities for Research in Astronomy, Inc. under NASA
contract No. NAS5-26555.}
\altaffiltext{2}{Department of Physics, University of Notre Dame, 225 Nieuwland Science Hall, Notre Dame, IN 46556}
\altaffiltext{3}{Department of Astronomy, University of Wisconsin, 475 North Charter Street, Madison, WI 53706}
\altaffiltext{4}{Argelander-Institut f\"ur Astronomie, Universit\"at Bonn, Auf dem H\"ugel 71, 53121 Bonn, Germany}
\altaffiltext{5}{Space Telescope Science Institute, 3700 San Martin Drive, Baltimore, MD 21218.}
\altaffiltext{6}{Department of Astronomy, University of Massachusetts, Amherst, MA 01003.}

\begin{abstract}
We present a study of the Ly$\alpha$ forest at $z\la 0.4$ from which we conclude that 
at least 20\% of the total baryons in the universe are located in the
highly-ionized gas traced by broad Ly$\alpha$ absorbers. The cool
photoionized low-$z$ intergalactic medium (IGM) probed by 
narrow Ly$\alpha$ absorbers contains about 30\% of the baryons. 
We further find that the ratio of broad to narrow Ly$\alpha$ absorbers is higher at $z \la 0.4$ than
at $1.5 \la z\la 3.6$, implying that a larger fraction of the low redshift universe is hotter
and/or more kinematically disturbed.
We base these conclusions on an analysis of 7 QSOs observed
with both {\em FUSE}\ and the {\em HST}/STIS E140M ultraviolet echelle spectrograph.
Our sample has 341 \hi\ absorbers with a total unblocked redshift path of
2.064. The observed absorber population is complete for $\log N_{\rm
H I} \ga 13.2 $, with a column density distribution $f(N_{\rm H I})
\propto N^{-\beta}_{\rm H I}$. For narrow 
($b\le 40$ \km) absorbers $\beta = 1.76 \pm 0.06$. The distribution of the Doppler 
parameter $b$ at low redshift implies two populations: narrow ($b\le 40$ \km) 
and broad ($b>40$ \km) Ly$\alpha$ absorbers (referred to as NLAs and BLAs, 
respectively). Both the NLAs and some BLAs probe the cool ($T\sim 10^4$ K) 
photoionized IGM. The BLAs also probe the highly-ionized  gas of the warm-hot IGM ($T\simeq 10^5$--$10^6$~K). 
The distribution of $b$ has a more prominent high velocity tail at $z\la 0.4$ than at $1.5 \la z\la 3.6$, which results 
in median and mean $b$-values that are 15--30\% higher at low $z$ than at 
high $z$. The ratio of the number density of BLAs to NLAs at $z\la 0.4$ is 
a factor of $\sim$3 higher than at $1.5 \la z\la 3.6$.
\end{abstract}


\section{Introduction}
Observations of Ly$\alpha$ absorption lines in the spectra of QSOs
provide a sensitive probe of the evolution and the distribution of the 
gas in the universe from high to low redshift. A forest of \hi\ absorption 
lines occurs at different redshifts, $z$, along QSO sightlines
with  $\log N_{\rm H I} < 17 $. 
Understanding the evolution of the Ly$\alpha$ forest with redshift
is critical to understanding the evolution and formation of 
structures in the universe. At $z\gtrsim 1.5$,  observations of 
the Ly$\alpha$ forest are obtained from ground-based telescopes at a 
spectral resolution of 7--8 \km\ using 8--10 m class telescopes 
\citep[e.g.,][]{hu95,lu96,kirkman97,kim97,kim02}, but at $z\le1.5$ they
require UV space-based instruments. Space-based UV astronomy has produced 
remarkable results, including the discovery itself of the Ly$\alpha$ forest
at low redshift \citep{bahcall91,morris91}. However, most 
of the UV studies of the IGM at low $z$ have lacked the spectral resolution 
and wavelength coverage of the higher redshift studies 
\citep{weymann98,impey99,penton00,penton04}, requiring 
assumptions for the Doppler parameter to derive the column density. 
The situation at $z \la 0.5$ has dramatically improved in the last few years with 
high quality observations of several low redshift QSOs obtained with 
the {\em Hubble Space Telescope (HST)}\ and its Space Telescope Imaging Spectrograph (STIS).
In its E140M echelle mode, STIS provides a spectral resolution of $\sim$7 \km, comparable
to the resolution of the high redshift observations of the Ly$\alpha$ forest.
The high spectral resolution has allowed the derivation of accurate Doppler parameters, $b$,
and column densities, $N$, using techniques similar to those used at high redshift. 
The Doppler parameter is important because it is related to the temperature 
of the gas via $b^2 = b^2_{\rm th} + b^2_{\rm nt} $, where $b_{\rm nt}$ is 
the non-thermal broadening of the absorption line and $ b_{\rm th} = \sqrt{2 kT/m} = 0.129 \sqrt{T} $
is the thermal broadening of the \hi\ absorption line; therefore the measured $b$ 
directly provides  an upper limit to the temperature of the observed gas. 

In parallel, hydrodynamic cosmological simulations  of the local universe 
have quickly evolved in recent years  \citep{dave99,cen99,dave01a,cen06a,cen06b}. These models
predict that at low redshift roughly 30--50\% of the baryons are in a hot (\dex5--\dex7 K) and highly ionized
intergalactic medium (IGM) known as the warm-hot intergalactic medium (WHIM), 
30--40\% are in a cooler medium ($\la 10^4$ K)  photoionized by the UV background, 
and the remaining baryons are in galaxies. Observationally, \citet{penton04} 
found with moderate spectral resolution (FWHM\,$\sim$20 \km) UV observations 
that the cool phase of the IGM may contain 29\% of the baryon budget. 
The observational detection of the WHIM came first from an intensive search of collisionally-ionized
\ovi\ systems and other highly ionized species such as \neviii\ 
using space-based observatories such as the {\em Far Ultraviolet
Spectroscopic Explorer}\ (\fuse) and {\em HST}/STIS
\citep[e.g.,][]{tripp00b,danforth05,savage05}. The baryon 
content of the \ovi\ absorbers suggests that $\Omega_{\rm b}($\ovi$)\ga 0.0022 h_{70}^{-1}$
or at least 5\% of the total baryon budget \citep{sembach04,tripp04,danforth05}, but 
this estimate relies critically on the assumed oxygen abundance and the 
number of collisionally ionized \ovi\ systems versus photoionized \ovi\ systems 
\citep{prochaska04,lehner06}. The \ovi\ systems probe the lower end of the WHIM temperature
range ($T < 10^{6}$ K). Higher temperatures can be traced with more highly ionized oxygen  ions (\ovii, \oviii) 
that can be observed in principle with X-ray observatories such as {\em Chandra}\ \citep{nicastro05,fang06}
and {\em XMM-Newton}, but the IGM detections of \ovii\ and \oviii\ are still controversial 
\citep{kaastra06,rasmussen06} for $z>0$.

The WHIM can also be detected  through broad \hi\ absorption lines. The high 
temperature of the WHIM will broaden the Ly$\alpha$ absorption line resulting in a large Doppler 
parameter ($b>40$ \km). A very small fraction of \hi\ (typically $<10^{-5}$) 
is expected to be found in the highly ionized plasma of the WHIM, so that
the broader the \hi\ absorption line is, the shallower it should be.
Recent observations, both with moderate spectral resolution and with
the higher-resolution STIS E140M echelle mode, have revealed the
presence of broad \hi\ absorption lines that could be modeled by smooth
and broad Gaussian components \citep{tripp01,bowen02,richter04,sembach04,lehner06}.
\hi\ absorbers with  $b<40$ \km\ imply that the temperature of these absorbers must be  $T < 10^5$ K. Since the 
nominal lower temperature of the WHIM is $T\sim 10^5$ K, it is {\em a priori}\ natural to consider 
two different physical populations of \hi\ absorbers:
the narrow Ly$\alpha$ absorbers (NLAs) as \hi\ systems having $b<40$ \km\ and the 
broad Ly$\alpha$ absorbers (BLAs) as \hi\ systems having $b \ge 40$ \km. We will 
see that the separation between the NLAs and BLAs is not so evident observationally,
with an important overlap between the two populations, particularly in the range
$b=40$--50 \km. While it is clear that the NLAs are tracing mostly the cool photoionized 
IGM, the situation is less clear for the BLAs. There is a fuzziness in the separation of the 
BLAs from the NLAs because non-thermal broadening can be important and 
unresolved components can hide the true structure of the \hi\ absorbers. 
Recent simulations show that non-thermal broadening could be particularly 
important for the \hi\ absorbers with $40\la b\la 60$ \km\ \citep{richter06a}. Therefore BLAs may
probe photoionized gas (that can be either cool ($T\la 10^4$ K) or hot ($T\ga 10^5$ K) 
but with a very low density  ($\log n_{\rm H} < -5.3$)) {\em and}\  collisionally ionized, hot ($T\ga 10^5$ K) gas. 
In this paper, we define a BLA as an absorber that can be fitted with a single Gaussian component with $b > 40$ \km\ 
(a NLA has $b \le 40$ \km\ and can have multiple components). BLAs probe the IGM 
that is either hotter ($b_{\rm th} \gg b_{\rm nt}$) or is more kinematically disturbed ($b_{\rm nt} \gg b_{\rm th}$)
than the IGM probed by the NLAs.  
Studying in detail the NLA and BLA populations is also important for
estimating the baryon density because an accurate inventory of the
baryon distribution must separate the fraction of the baryons that are
located in the cool photoionized IGM versus those that are in the
substantially hotter shock-heated WHIM phase.  The narrow and broad
Ly$\alpha$ lines provide a means to discriminate between the cool and
shock-heated gas clouds.  Therefore to make a reliable assessment of
the baryonic content of the Ly$\alpha$ forest at low z,
it is necessary to investigate the frequency and properties 
of the NLAs and BLAs.  The current estimate of the baryon budget residing in the 
BLAs over a redshift path $\Delta z = 0.928$ yields $\Omega_{\rm b}({\rm BLA})\ga 0.0027 h_{70}^{-1}$ 
or at least 6\% of the total baryon budget \citep{richter06}, assuming  
that the observed broadening is mostly thermal and collisional ionization equilibrium
applies. 

In this paper, we will address these issues using a sample of 7 QSO sightlines that have been 
observed with both STIS E140M and {\fuse}\ and for which the 
data have been fully analyzed using similar techniques and are in press or to be submitted soon.
While the spectral resolution of {\fuse}\ is only $\sim$20 \km\ compared to $\sim$7 \km\ for STIS E140M, 
{\fuse}\ gives access to several Lyman series and metal lines 
making the line identification more reliable and providing an unprecedented insight of the
physical conditions and metallicity. 
The fundamental parameters of the Ly$\alpha$ forest lines (redshift $z$, 
column density $N$,  and Doppler parameter $b$) were accurately determined using
profile fitting of all the observed Lyman series lines. Because $b$ could be derived,
this is the first time with a large sample that the properties of the Ly$\alpha$ forest
in the low redshift universe can be studied as a function of $b$. 
The main aims of this work are   1) to study the distribution and evolution of the Doppler 
parameter of the Ly$\alpha$ forest, and 2) to determine the baryon density of the Ly$\alpha$ forest
in the NLA and BLA populations.  
The organization of this paper is as follows: \S\ref{datasample} describes the sample and its 
completeness; in \S\ref{bdistr} we study the distribution of the Doppler parameter $b$
and the Ly$\alpha$ density number as a function of $b$; in \S\ref{sec-evol} we study the evolution 
of $b$ with redshift by comparing the low-$z$ ($z \la 0.4$) Ly$\alpha$ forest with the
mid-$z$ ($0.5\la z \la 1.5$) and high-$z$ ($1.5 \la z \la 3.6$) Ly$\alpha$ forest;
in \S\ref{dddf} we estimate the column density distribution; and
finally, in \S\ref{bd} we estimate the baryon content of the cool photoionized IGM probed by the NLAs 
and the photoionized IGM and the WHIM probed by the BLAs,  and discuss the uncertainties in estimating 
the baryon budget. We summarize our results in \S\ref{sum}.

\begin{deluxetable}{lccccc}
\tabcolsep=3pt
\tablecolumns{6}
\tablewidth{0pt} 
\tabletypesize{\scriptsize}
\tablecaption{Line of Sight Properties \label{t1}} 
\tablehead{\colhead{Sightline}    &  \colhead{S/N}&\colhead{$z$} &  \colhead{$\Delta z$}&  \colhead{$\Delta X$}&  \colhead{Refs.}
	\\
	\colhead{(1)}    &  \colhead{(2)}&\colhead{(3)} &  \colhead{(4)}&  \colhead{(5)}&  \colhead{(6)}}
\startdata
H\,1821+643	&   15--20	&  0.297   &  0.238    &  0.266  & (1) \\		  
HE\,0226--4110 	&   5--11   	&  0.495   &  0.401    &  0.481  & (2) \\		  
HS\,0624+6907 	&   8--12   	&  0.370   &  0.329    &  0.383  & (3) \\		  
PG\,0953+415   	&   7--11   	&  0.239   &  0.202    &  0.222  & (4) \\		  
PG\,1116+215   	&   10--15   	&  0.176   &  0.126    &  0.134  & (5) \\		  
PG\,1259+593   	&   9--17   	&  0.478   &  0.355    &  0.418  & (6) \\		  
PKS\,0405--123	&   5--10    	&  0.574   &  0.413    &  0.498  & (7) 		   	  
\enddata		
\tablecomments{Column 2: Range of signal-to-noise (S/N) per resolution element in STIS E140M mode. 
Column 3: QSO redshift. Column 4: Unblocked redshift
path for Ly$\alpha$. Column 5:  Absorption distance (see \S\ref{datasample}). 
Column 6: References: (1) Sembach et al. (2007, in prep.), (2) Lehner et al. (2006), 
(3) Aracil et al. (2006), (4) T.M. Tripp (2006, priv. comm.), (5) Sembach et al. (2004), 
(6) Richter et al. (2004), (7) Williger et al. (2006) and appendix of this paper. 
}
\end{deluxetable}

\section{The Low $z$ Sample}\label{datasample}
\subsection{Description of the Sample and Completeness}
Our low-$z$ sample consists of 7 QSO sightlines that were
observed with  {\em HST}/STIS E140M and {\fuse}: H\,1821+643 (Sembach et al. 2007, in preparation; 
see also Tripp, Savage, \& Jenkins 2000 and Oegerle et al. 2000 for the metal-line systems), 
HE\,0226--4110 	\citep{lehner06}, HS\,0624+6907 \citep{aracil06}, PG\,0953+415	
(T.M. Tripp 2006, private communication; see also Savage et
al. 2002 for the metal-line systems), PG\,1259+593 \citep{richter04}, PG\,1116+215 \citep{sembach04},    	
and PKS\,0405--123 (Williger et al. 2006; see also Prochaska et al. 2004
for the metal-line systems). Note that for HS\,0624+6907, we used the results summarized
in the erratum produced by \citet{aracil06a}. The data handling and analysis 
are described in detail in the above papers. For PKS\,0405--123, 
we adopt the  new measurements and a new line
list that we describe in the Appendix. The motivation to revisit Williger et
al.'s analysis was first driven by differentiating a real detection from a noise feature 
for the BLAs since these authors noted that several of their BLAs could be just noise. 
This re-analysis also provides an overall coherent 
data sample that was analyzed following the same methodology. Signal-to-noise (S/N) where Ly$\alpha$
can be observed, redshift ($z$), unblocked redshift path ($\Delta z$),
and absorption distance ($\Delta X$) are summarized in Table~\ref{t1} for each sightline. 
The absorption distance was computed assuming a Friedman cosmology, 
$\Delta X = 0.5[(1+\Delta z)^2 - 1]$ with $q_0 = 0 $.
For the lines of sight to QSOs with $z_{\rm QSO} < 0.42$, we have
excluded absorption systems within 5000 \km\ of the QSO
redshift.  Some observations of low-redshift QSOs have provided
evidence that ``intrinsic'' absorption lines can arise in clouds that
are spatially close to the QSO and yet, due to the cloud kinematics,
are substantially offset in redshift from the QSO (e.g., Yuan et
al. 2002; Ganguly et al. 2003).  To avoid contaminating the sample
with these intrinsic absorbers, we do not use systems detected with 5000
\km\ of the QSO.
The Voigt profile fitting method was employed for each line of sight 
to measure the column densities ($N_{\rm H I}$), Doppler parameters ($b$), 
and redshifts of the absorbers and we adopt those results
for our analysis. Table~\ref{t1a} lists these parameters. 
For PG\,1259+593 we did not include in our sample the systems marked 
uncertain (UC) in Table~5 of \citet{richter04}. A colon in Table~\ref{t1a} indicates
that there is uncertainty in the determination of the physical parameters, which 
are not accounted for in the formal errors produced by the profile fitting. These systems are
not taken into account when we use an error cutoff (see below).  If we had, it would not have
changed the results presented in this paper in a statistically significant manner. 

The total sample consists of 341 \hi\ systems, with a total unblocked redshift path 
$\Sigma \Delta z = 2.064$ and a total absorption distance $\Sigma \Delta X = 2.404$. 
There are 201 systems 
at $z\le0.2$ and 131 systems at $0.2 < z \le 0.4$. The remaining 9 systems lie at 
$0.4 < z < 0.44$. Therefore, our sample mostly probes the universe at $z\le 0.4$.
More absorbers are found at $z<0.2$ because several lines of sight do not 
extend out to $z=0.3$--0.4 and to a lesser extent because the S/N
at $\lambda > 1650$ \AA\ decreases rapidly (see below). 

{\LongTables
\begin{deluxetable}{lcc}
\tabcolsep=6pt
\tablecolumns{3}
\tablewidth{0pt} 
\tabletypesize{\scriptsize}
\tablecaption{\hi\ measurements \label{t1a}} 
\tablehead{\colhead{$z$}    &  \colhead{$\log N_{\rm H I}$}$$&\colhead{$b$} 
	\\
	\colhead{}    &  \colhead{(dex)}&\colhead{(\km)}}
\startdata
  \cutinhead{H\,1821+643 (Sembach et al. 2006, in prep.)}
    0.02438  & $  14.28 \,^{+0.06}_{-0.05}  $   & $   26.9 \pm   1.7  $ \\
    0.02642  & $  13.26 \,^{+0.07}_{-0.08}  $   & $   48.6 \pm   6.0  $ \\
    0.06718  & $  13.72 \,^{+0.01}_{-0.01}  $   & $   20.2 \pm   0.6  $ \\
    0.07166  & $  13.87 \,^{+0.02}_{-0.02}  $   & $   34.0 \pm   1.0  $ \\
    0.08911  & $  13.01 \,^{+0.06}_{-0.07}  $   & $   23.3 \pm   2.8  $ \\
    0.11133  & $  12.95 \,^{+0.10}_{-0.13}  $   & $   88.0 \pm  14.0  $ \\
    0.11166  & $  12.99 \,^{+0.08}_{-0.09}  $   & $   28.6 \pm   4.4  $ \\
    0.11961  & $  13.15 \,^{+0.05}_{-0.06}  $   & $   36.8 \pm   3.4  $ \\
    0.12055  & $  12.64 \,^{+0.13}_{-0.18}  $   & $   22.5 \pm   5.9  $ \\
    0.12112  & $  13.93 \pm  0.37  $   & $   26.0 \,^{+13.0}_{- 9.0}  $ \\
    0.12125  & $  14.04 \pm  0.36  $   & $   40.0 \,^{+44.0}_{-21.0}  $ \\
    0.12147  & $  13.78 \pm  0.17  $   & $   85.0 \,^{+37.0}_{-26.0}  $ \\
    0.12221  & $  13.19 \,^{+0.06}_{-0.06}  $   & $   41.7 \pm   4.4  $ \\
    0.14754  & $  13.51 \,^{+0.03}_{-0.03}  $   & $   44.6 \pm   2.5  $ \\
    0.14776  & $  13.30 \,^{+0.04}_{-0.05}  $   & $   19.2 \pm   1.6  $ \\
    0.15731  & $  13.11 \,^{+0.05}_{-0.06}  $   & $   22.9 \pm   2.1  $ \\
    0.16127  & $  12.74 \,^{+0.10}_{-0.12}  $   & $   21.5 \pm   4.0  $ \\
    0.16352  & $  13.17 \,^{+0.06}_{-0.07}  $   & $   52.2 \pm   5.9  $ \\
    0.16966  & $  13.93 \,^{+0.02}_{-0.02}  $   & $   35.7 \pm   2.0  $ \\
    0.17001  & $  13.65 \,^{+0.02}_{-0.02}  $   & 	\nodata	 	\\
    0.17051  & $  13.40 \,^{+0.04}_{-0.04}  $   & $   21.1 \pm   1.4  $ \\
    0.17926  & $  12.87 \,^{+0.09}_{-0.12}  $   & $   18.4 \pm   3.4  $ \\
    0.18047  & $  13.14 \,^{+0.07}_{-0.08}  $   & $   50.7 \pm   6.6  $ \\
    0.19662  & $  12.98 \,^{+0.07}_{-0.08}  $   & $   21.1 \pm   2.7  $ \\
    0.19904  & $  12.74 \,^{+0.12}_{-0.17}  $   & $   22.1 \pm   5.4  $ \\
    0.20957  & $  13.10 \,^{+0.03}_{-0.04}  $   & $   16.2 \pm   0.9  $ \\
    0.21161  & $  13.16 \,^{+0.06}_{-0.06}  $   & $   22.3 \pm   2.3  $ \\
    0.21668  & $  12.88 \,^{+0.08}_{-0.10}  $   & $   19.7 \pm   3.0  $ \\
    0.21326  & $  14.41 \,^{+0.04}_{-0.04}  $   & $   43.0 \pm   2.4  $ \\
    0.22497  & $  15.53 \,^{+0.05}_{-0.05}  $   & $   25.0 \pm   7.0  $ \\
    0.22616  & $  13.51 \,^{+0.04}_{-0.04}  $   & $   54.7 \pm   3.8  $ \\
    0.22786  & $  13.26 \,^{+0.03}_{-0.04}  $   & $   35.3 \pm   2.0  $ \\
    0.23869  & $  12.86 \,^{+0.10}_{-0.12}  $   & $   20.2 \pm   3.8  $ \\
    0.24142  & $  13.12 \,^{+0.03}_{-0.04}  $   & $   23.4 \pm   1.4  $ \\
    0.24531  & $  13.06 \,^{+0.08}_{-0.10}  $   & $   34.2 \pm   5.1  $ \\
    0.25689  & $  12.80 \,^{+0.11}_{-0.15}  $   & $   22.6 \pm   5.1  $ \\
    0.25814  & $  13.38 \,^{+0.10}_{-0.13}  $   & $   60.3 \pm   8.8  $ \\
    0.25816  & $  12.98 \,^{+0.11}_{-0.15}  $   & $   14.5 \pm   3.4  $ \\
    0.26152  & $  13.70 \,^{+0.02}_{-0.03}  $   & $   37.7 \pm   1.5  $ \\
    0.26659  & $  13.64 \,^{+0.03}_{-0.03}  $   & $   44.5 \pm   2.1  $ \\
  \cutinhead{HE\,0226--4110 \citep{lehner06}}
    0.01746  & $  13.22 \pm  0.06  $   & $   17.9 \pm   4.3  $ \\
    0.02679  & $  13.22 \pm  0.08  $   & $   41.6 \pm  11.0  $ \\
    0.04121  & $  12.82 \pm  0.14  $   & $   23.6 \pm  18.6  $ \\
    0.04535  & $  12.71 \pm  0.13  $   & $   16.2 \pm  12.8  $ \\
    0.04609  & $  13.66 \pm  0.03  $   & $   25.0 \pm   2.1  $ \\
    0.06015  & $  13.19 \pm  0.06  $   & $   35.5 \pm   7.6  $ \\
    0.06083  & $  14.65 \pm  0.02  $   & $   44.5 \pm   1.0  $ \\
    0.07023  & $  13.81 \pm  0.11  $   & $   26.0 \pm  12.4  $ \\
    0.08375  & $  13.67 \pm  0.05  $   & $   29.6 \pm   4.5  $ \\
    0.08901  & $  13.33 \pm  0.05  $   & $   23.8 \pm   3.7  $ \\
    0.08938  & $  12.59 \pm  0.22  $   & $    8.2 :		  $ \\
    0.08950  & $  12.72 \pm  0.20  $   & $   22.0 :		  $ \\
    0.09059  & $  13.71 \pm  0.03  $   & $   28.3 \pm   2.0  $ \\
    0.09220  & $  12.94 \pm  0.11  $   & $   40.2 \pm  18.0  $ \\
    0.10668  & $  13.09 \pm  0.08  $   & $   32.7 \pm   9.2  $ \\
    0.11514  & $  12.90 \pm  0.09  $   & $   10.4 \pm   4.1  $ \\
    0.11680  & $  13.27 \pm  0.05  $   & $   23.7 \pm   3.9  $ \\
    0.11733  & $  12.64 \pm  0.15  $   & $   15.0 :		  $ \\
    0.12589  & $  13.01 \pm  0.09  $   & $   29.2 \pm  10.1  $ \\
    0.13832  & $  13.19 \pm  0.06  $   & $   25.9 \pm   5.3  $ \\
    0.15175  & $  13.42 \pm  0.05  $   & $   48.6 \pm   6.7  $ \\
    0.15549  & $  13.13 \pm  0.08  $   & $   34.7 \pm   9.8  $ \\
    0.16237  & $  13.04 \pm  0.08  $   & $   29.7 \pm   8.6  $ \\
    0.16339  & $  14.36 \pm  0.04  $   & $   46.3 \pm   1.9  $ \\
    0.16971  & $  13.35 \pm  0.05  $   & $   25.3 \pm   3.9  $ \\
    0.18619  & $  13.26 \pm  0.08  $   & $   53.9 \pm  16.2  $ \\
    0.18811  & $  13.47 \pm  0.05  $   & $   22.4 \pm   3.3  $ \\
    0.18891  & $  13.34 \pm  0.07  $   & $   22.2 \pm   4.0  $ \\
    0.19374  & $  13.20 \pm  0.06  $   & $   28.7 \pm   6.0  $ \\
    0.19453  & $  12.89 \pm  0.12  $   & $   26.1 \pm  14.0  $ \\
    0.19860  & $  14.18 \pm  0.04  $   & $   37.0 \pm   2.0  $ \\
    0.20055  & $  13.38 \pm  0.05  $   & $   38.9 \pm   6.4  $ \\
    0.20698  & $  13.31 \pm  0.34  $   & $   97.0 :	     $ \\
    0.20700  & $  15.06 \pm  0.04  $   & $   17.4 \pm   1.4  $ \\
    0.20703  & $  14.89 \pm  0.05  $   & $   35.9 \pm   1.1  $ \\
    0.22005  & $  14.40 \pm  0.04  $   & $   27.7 \pm   1.1  $ \\
    0.22099  & $  12.99 \pm  0.12  $   & $   34.1 \pm  18.1  $ \\
    0.23009  & $  13.69 \pm  0.04  $   & $   67.9 \pm   7.5  $ \\
    0.23964  & $  13.13 \pm  0.08  $   & $   28.8 \pm   8.8  $ \\
    0.24514  & $  14.20 \pm  0.03  $   & $   34.5 \pm   1.6  $ \\
    0.25099  & $  13.17 \pm  0.08  $   & $   37.9 \pm  11.5  $ \\
    0.27147  & $  13.85 \pm  0.07  $   & $   25.7 \pm   4.2  $ \\
    0.27164  & $  13.33 \pm  0.28  $   & $   26.2  :	  $ \\
    0.27175  & $  12.88 \pm  0.38  $   & $   11.2  :	 $ \\
    0.27956  & $  13.22 \pm  0.14  $   & $   36.3 \pm  24.6  $ \\
    0.28041  & $  13.03 \pm  0.11  $   & $   13.9 \pm   6.7  $ \\
    0.29134  & $  13.53 \pm  0.07  $   & $   27.0 \pm   6.2  $ \\
    0.29213  & $  13.19 \pm  0.12  $   & $   33.4 \pm  17.8  $ \\
    0.30930  & $  14.26 \pm  0.03  $   & $   43.8 \pm   2.3  $ \\
    0.34034  & $  13.68 \pm  0.06  $   & $   33.4 \pm   4.9  $ \\
    0.35523  & $  13.60 \pm  0.07  $   & $   27.1 \pm   6.8  $ \\
    0.37281  & $  13.16 \pm  0.12  $   & $   25.9 \pm  13.3  $ \\
    0.38420  & $  13.91 \pm  0.04  $   & $   62.0 \pm   7.1  $ \\
    0.38636  & $  13.36 \pm  0.09  $   & $   38.1 \pm  12.7  $ \\
    0.39641  & $  13.59 \pm  0.10  $   & $   62.8 \pm  22.7  $ \\
    0.39890  & $  13.50 \pm  0.16  $   & $  151.7 :	  $ \\
    0.40034  & $  13.39 \pm  0.11  $   & $   60.7 \pm  26.1  $ \\
    0.40274  & $  14.13 \pm  0.04  $   & $   45.7 \pm   4.2  $ \\
  \cutinhead{HS\,0624+6907 \citep{aracil06a}}
    0.01755  & $  12.96 \pm  0.05  $   & $   29.0 \pm   4.3  $ \\
    0.03065  & $  13.36 \pm  0.03  $   & $   22.0 \pm   1.7  $ \\
    0.04116  & $  13.33 \pm  0.03  $   & $   41.0 \pm   3.0  $ \\
    0.05394  & $  13.26 \pm  0.04  $   & $   24.0 \pm   2.3  $ \\
    0.05437  & $  13.09 \pm  0.11  $   & $   60.0 \pm  19.2:  $ \\
    0.05483  & $  14.50 :	   $   & $   35.0 :	     $ \\
    0.05515  & $  13.68 \pm  0.17  $   & $   84.0 \pm  30.7:  $ \\
    0.06188  & $  13.77 \pm  0.03  $   & $   21.0 \pm   1.4  $ \\
    0.06201  & $  12.63 \pm  0.17  $   & $    8.0 \pm   4.7  $ \\
    0.06215  & $  12.41 \pm  0.22  $   & $   10.0 \pm   7.9  $ \\
    0.06234  & $  13.45 \pm  0.05  $   & $   30.0 \pm   4.0  $ \\
    0.06265  & $  13.31 \pm  0.14  $   & $   35.0 \pm  12.3  $ \\
    0.06276  & $  12.95 \pm  0.28  $   & $    8.0 \pm   3.7  $ \\
    0.06285  & $  13.42 \pm  0.14  $   & $   20.0 \pm   7.0  $ \\
    0.06304  & $  13.33 \pm  0.13  $   & $   27.0 \pm   8.8  $ \\
    0.06346  & $  14.46 \pm  0.30  $   & $   48.0 \pm   8.4:  $ \\
    0.06348  & $  15.27 \pm  0.13  $   & $   24.0 \pm   5.5  $ \\
    0.06362  & $  14.29 \pm  0.38  $   & $   10.0 \pm   5.6  $ \\
    0.06475  & $  13.87 \pm  0.04  $   & $   33.0 \pm   3.0  $ \\
    0.06502  & $  13.97 \pm  0.04  $   & $   31.0 \pm   2.7  $ \\
    0.07573  & $  14.18 \pm  0.03  $   & $   24.0 \pm   0.8  $ \\
    0.09023  & $  13.29 \pm  0.08  $   & $   76.0 \pm  13.7  $ \\
    0.13076  & $  13.34 \pm  0.04  $   & $   34.0 \pm   3.6  $ \\
    0.13597  & $  13.33 \pm  0.10  $   & $   57.0 \pm  10.7  $ \\
    0.16054  & $  13.08 \pm  0.21  $   & $   34.0 \pm  10.3  $ \\
    0.16074  & $  13.66 \pm  0.05  $   & $   30.0 \pm   2.4  $ \\
    0.19975  & $  13.24 \pm  0.05  $   & $   17.0 \pm   2.0  $ \\
    0.19995  & $  13.17 \pm  0.06  $   & $   26.0 \pm   4.6  $ \\
    0.20483  & $  13.72 \pm  0.02  $   & $   24.0 \pm   1.0  $ \\
    0.20533  & $  14.12 \pm  0.03  $   & $   25.0 \pm   0.8  $ \\
    0.20754  & $  13.48 \pm  0.02  $   & $   27.0 \pm   1.5  $ \\
    0.21323  & $  13.22 \pm  0.05  $   & $   45.0 \pm   5.6  $ \\
    0.21990  & $  13.39 \pm  0.05  $   & $   60.0 \pm   8.6  $ \\
    0.22329  & $  13.86 \pm  0.02  $   & $   25.0 \pm   0.9  $ \\
    0.23231  & $  13.33 \pm  0.08  $   & $   44.0 \pm   7.7:  $ \\
    0.23255  & $  12.86 \pm  0.21  $   & $   24.0 \pm   7.3  $ \\
    0.24060  & $  13.33 \pm  0.04  $   & $   20.0 \pm   2.0  $ \\
    0.25225  & $  12.96 \pm  0.06  $   & $   24.0 \pm   4.2  $ \\
    0.26856  & $  13.03 \pm  0.05  $   & $   51.0 \pm   7.2  $ \\
    0.27224  & $  12.80 \pm  0.06  $   & $   12.0 \pm   2.2  $ \\
    0.27977  & $  13.50 \pm  0.06  $   & $   34.0 \pm   4.9  $ \\
    0.28017  & $  14.32 \pm  0.02  $   & $   43.0 \pm   1.9  $ \\
    0.29531  & $  13.80 \pm  0.02  $   & $   42.0 \pm   2.0:  $ \\
    0.29661  & $  13.54 \pm  0.02  $   & $   52.0 \pm   2.9  $ \\
    0.30899  & $  13.49 \pm  0.03  $   & $   28.0 \pm   1.8  $ \\
    0.30991  & $  13.61 \pm  0.10  $   & $   66.0 \pm  12.3:  $ \\
    0.31045  & $  13.43 \pm  0.33  $   & $   62.0 \pm  40.3:  $ \\
    0.31088  & $  13.13 \pm  0.43  $   & $   51.0 \pm  27.7  $ \\
    0.31280  & $  13.65 \pm  0.10  $   & $   54.0 \pm   9.3  $ \\
    0.31303  & $  13.09 \pm  0.24  $   & $   17.0 \pm   6.8  $ \\
    0.31326  & $  13.62 \pm  0.10  $   & $   55.0 \pm  10.9  $ \\
    0.31790  & $  13.37 \pm  0.04  $   & $   34.0 \pm   3.6  $ \\
    0.32089  & $  13.97 \pm  0.02  $   & $   31.0 \pm   1.2  $ \\
    0.32724  & $  13.73 \pm  0.32  $   & $   69.0 \pm  15.6:  $ \\
    0.32772  & $  13.61 \pm  0.43  $   & $  115.0 \pm  62.1:  $ \\
    0.33267  & $  13.55 \pm  0.04  $   & $   38.0 \pm   3.4  $ \\
    0.33976  & $  14.45 \pm  0.03  $   & $   42.0 \pm   1.3  $ \\
    0.34682  & $  13.59 \pm  0.02  $   & $   39.0 \pm   1.9  $ \\
    0.34865  & $  12.78 \pm  0.06  $   & $   18.0 \pm   3.0  $ \\
  \cutinhead{PG\,0953+415 (T.M. Tripp, 2006, priv. comm.)}
    0.01558  & $  13.19 \pm  0.08  $   & $   35.0 \,^{+10.0}_{- 8.0}  $ \\
    0.01587  & $  12.86 \pm  0.21  $   & $   19.0 \,^{+15.0}_{- 8.0}  $ \\
    0.01606  & $  13.54 \pm  0.05  $   & $   32.0 \pm   5.0  $ \\
    0.01655  & $  13.51 \pm  0.04  $   & $   26.0 \pm   3.0  $ \\
    0.02336  & $  13.21 \pm  0.08  $   & $   56.0 \,^{+14.0}_{-11.0}  $ \\
    0.04416  & $  12.72 \pm  0.09  $   & $   15.0 \pm   6.0  $ \\
    0.04469  & $  12.77 \pm  0.06  $   & $   12.0 \pm   3.0  $ \\
    0.04512  & $  13.42 \pm  0.02  $   & $   38.0 \pm   2.0  $ \\
    0.05876  & $  13.82 \pm  0.07  $   & $   25.0 \pm   3.0  $ \\
    0.05879  & $  13.41 \pm  0.16  $   & $   63.0 \,^{+19.0}_{-14.0}  $ \\
    0.06808  & $  14.47 \pm  0.03  $   & $   21.0 \pm   2.0  $ \\
    0.09228  & $  13.08 \pm  0.07  $   & $   27.0 \pm   6.0  $ \\
    0.09315  & $  13.66 \pm  0.03  $   & $   39.0 \pm   3.0  $ \\
    0.10940  & $  13.68 \pm  0.03  $   & $   23.0 \pm   2.0  $ \\
    0.11558  & $  13.47 \pm  0.03  $   & $   24.0 \pm   2.0  $ \\
    0.11826  & $  13.68 \pm  0.02  $   & $   30.0 \pm   2.0  $ \\
    0.11871  & $  12.81 \pm  0.10  $   & $   16.0 \pm   6.0  $ \\
    0.12558  & $  12.77 \pm  0.09  $   & $    8.0 \,^{+ 4.0}_{- 3.0}  $ \\
    0.12784  & $  12.83 \pm  0.35  $   & $   44.0 \,^{+76.0}_{-28.0}  $ \\
    0.12804  & $  13.27 \pm  0.12  $   & $   21.0 \pm   5.0  $ \\
    0.14178  & $  12.68 \pm  0.10  $   & $   10.0 \,^{+ 5.0}_{- 3.0}  $ \\
    0.14233  & $  13.58 \pm  0.03  $   & $   28.0 \pm   2.0  $ \\
    0.14263  & $  13.45 \pm  0.04  $   & $   31.0 \pm   4.0  $ \\
    0.14294  & $  12.63 \pm  0.27  $   & $   17.0 \,^{+23.0}_{-10.0}  $ \\
    0.14310  & $  13.05 \pm  0.11  $   & $   22.0 \,^{+ 9.0}_{- 6.0}  $ \\
    0.14333  & $  12.77 \pm  0.14  $   & $   15.0 \,^{+ 9.0}_{- 6.0}  $ \\
    0.17985  & $  13.27 \pm  0.07  $   & $   48.0 \,^{+11.0}_{- 9.0}  $ \\
    0.19072  & $  13.04 \pm  0.08  $   & $   26.0 \pm   6.0  $ \\
    0.19126  & $  13.08 \pm  0.55  $   & $   48.0 \,^{+95.0}_{-32.0}  $ \\
    0.19147  & $  13.33 \pm  0.34  $   & $   30.0 \,^{+12.0}_{- 9.0}  $ \\
    0.19210  & $  13.14 \pm  0.07  $   & $   28.0 \pm   6.0  $ \\
    0.19241  & $  12.94 \pm  0.12  $   & $   35.0 \,^{+15.0}_{-11.0}  $ \\
    0.19361  & $  13.94 \pm  0.02  $   & $   40.0 \pm   2.0  $ \\
    0.20007  & $  13.24 \pm  0.09  $   & $   66.0 \,^{+17.0}_{-14.0}  $ \\
    0.20104  & $  13.16 \pm  0.16  $   & $   71.0 \,^{+41.0}_{-26.0}  $ \\
    0.20136  & $  12.91 \pm  0.19  $   & $   19.0 \,^{+10.0}_{- 7.0}  $ \\
    0.20895  & $  12.94 \pm  0.10  $   & $   28.0 \,^{+10.0}_{- 7.0}  $ \\
    0.21514  & $  13.30 \pm  0.04  $   & $   27.0 \pm   3.0  $ \\
    0.22526  & $  12.74 \pm  0.23  $   & $    2.0 \,^{+ 2.0}_{- 1.0}  $ \\
    0.22527  & $  13.19 \pm  0.05  $   & $   34.0 \pm   5.0  $ \\
  \cutinhead{PG\,1116+215 \citep{sembach04}}
    0.00493  & $  13.36 \,^{+0.07}_{-0.06}  $   & $   34.2 \pm   3.6  $ \\
    0.01635  & $  13.39 \pm 0.06	    $   & $   48.5 \pm   5.1  $ \\
    0.02827  & $  13.80 \pm 0.02	    $   & $   31.4 \pm   1.1  $ \\
    0.03223  & $  13.33 \,^{+0.06}_{-0.05}  $   & $   31.6 \pm   2.9  $ \\
    0.04125  & $  13.25 \,^{+0.11}_{-0.09}  $   & $  105.0 \pm  18.0  $ \\
    0.04996  & $  12.72 \,^{+0.10}_{-0.08}  $   & $   16.5 \pm   3.2  $ \\
    0.05895  & $  13.56 \pm 0.05	    $   & $   25.0 \pm   5.0  $ \\
    0.05928  & $  12.41 \,^{+0.18}_{-0.13}  $   & $   10.0 :		  $ \\
    0.06072  & $  13.28 \,^{+0.06}_{-0.05}  $   & $   55.4 \pm   5.8  $ \\
    0.06244  & $  13.18 \,^{+0.07}_{-0.06}  $   & $   77.3 \pm   9.0  $ \\
    0.07188  & $  12.79 \,^{+0.08}_{-0.07}  $   & $    9.6 \pm   2.0  $ \\
    0.08096  & $  13.45 \,^{+0.03}_{-0.02}  $   & $   24.9 \pm   1.0  $ \\
    0.08587  & $  12.90 \,^{+0.19}_{-0.13}  $   & $   52.0 \pm  14.0  $ \\
    0.08632  & $  12.66 \,^{+0.32}_{-0.18}  $   & $   36.0 \pm  15.0  $ \\
    0.09279  & $  13.39 \,^{+0.09}_{-0.08}  $   & $  133.0 \pm  17.0  $ \\
    0.10003  & $  12.73 \,^{+0.07}_{-0.06}  $   & $   23.2 \pm   2.2  $ \\
    0.11895  & $  13.44 \,^{+0.04}_{-0.03}  $   & $   31.5 \pm   1.8  $ \\
    0.13151  & $  13.41 \pm 0.03	    $   & $   28.8 \pm   1.3  $ \\
    0.13370  & $  13.27 \,^{+0.08}_{-0.07}  $   & $   83.6 \pm  10.4  $ \\
    0.13847  & $  16.20 \,^{+0.05}_{-0.04}  $   & $   22.4 \pm   0.3  $ \\
 \cutinhead{PG\,1259+593 \citep{richter04}}
    0.00229  & $  13.57 \pm  0.10  $   & $   42.1 \pm   4.4  $ \\
    0.00760  & $  14.05 \pm  0.05  $   & $   34.6 \pm   2.0  $ \\
    0.01502  & $  13.21 \pm  0.06  $   & $   22.6 \pm   4.4  $ \\
    0.02217  & $  13.67 \pm  0.04  $   & $   30.2 \pm   2.2  $ \\
    0.03924  & $  12.94 \pm  0.05  $   & $   15.3 \pm   2.8  $ \\
    0.04606  & $  15.58 \pm  0.21  $   & $   47.6 \pm  12.4  $ \\
    0.05112  & $  13.62 \pm  0.07  $   & $   34.4 \pm   2.5  $ \\
    0.05257  & $  12.75 \pm  0.06  $   & $   20.7 \pm   3.0  $ \\
    0.05376  & $  13.44 \pm  0.04  $   & $   30.5 \pm   1.9  $ \\
    0.06644  & $  13.65 \pm  0.05  $   & $   28.7 \pm   3.3  $ \\
    0.08041  & $  12.97 \pm  0.10  $   & $   42.0 \pm   4.5  $ \\
    0.08933  & $  14.04 \pm  0.03  $   & $   28.9 \pm   1.7  $ \\
    0.09591  & $  12.97 \pm  0.03  $   & $   21.5 \pm   2.3  $ \\
    0.12188  & $  13.03 \pm  0.07  $   & $   26.9 \pm   4.2  $ \\
    0.12387  & $  13.47 \pm  0.06  $   & $   28.2 \pm   3.0  $ \\
    0.14852  & $  13.91 \pm  0.06  $   & $   42.1 \pm   2.4  $ \\
    0.15029  & $  13.25 \pm  0.11  $   & $   25.7 \pm   4.3  $ \\
    0.15058  & $  13.45 \pm  0.13  $   & $   32.0 \pm   5.1  $ \\
    0.15136  & $  13.32 \pm  0.09  $   & $   65.3 \pm   5.5  $ \\
    0.15435  & $  13.22 \pm  0.04  $   & $   25.2 \pm   1.9  $ \\
    0.17891  & $  13.29 \pm  0.10  $   & $   98.5 \pm   9.1  $ \\
    0.18650  & $  13.02 \pm  0.03  $   & $   19.6 \pm   1.4  $ \\
    0.19620  & $  13.65 \pm  0.05  $   & $   32.7 \pm   3.2  $ \\
    0.19775  & $  13.33 \pm  0.06  $   & $   23.9 \pm   2.8  $ \\
    0.21949  & $  15.08 \pm  0.08  $   & $   32.3 \pm   1.4  $ \\
    0.22313  & $  13.92 \pm  0.04  $   & $   34.8 \pm   1.1  $ \\
    0.22471  & $  13.59 \pm  0.06  $   & $   28.9 \pm   1.7  $ \\
    0.22861  & $  13.47 \pm  0.05  $   & $   40.3 \pm   2.9  $ \\
    0.23280  & $  13.50 \pm  0.07  $   & $   37.4 \pm   3.2  $ \\
    0.23951  & $  13.40 \pm  0.03  $   & $   16.3 \pm   1.3  $ \\
    0.24126  & $  13.41 \pm  0.09  $   & $   89.1 \pm   6.9  $ \\
    0.25642  & $  13.74 \pm  0.03  $   & $   25.0 \pm   0.9  $ \\
    0.25971  & $  13.84 \pm  0.12  $   & $   40.5 \pm   4.9  $ \\
    0.28335  & $  13.59 \pm  0.10  $   & $   37.0 \pm   5.2  $ \\
    0.29236  & $  14.65 \pm  0.09  $   & $   24.3 \pm   2.8  $ \\
    0.29847  & $  13.09 \pm  0.10  $   & $   33.3 \pm   2.9  $ \\
    0.30164  & $  13.26 \pm  0.14  $   & $   31.7 \pm   4.7  $ \\
    0.30434  & $  13.76 \pm  0.14  $   & $   64.5 \pm   9.6  $ \\
    0.31070  & $  13.40 \pm  0.07  $   & $   22.8 \pm   2.9  $ \\
    0.31978  & $  13.98 \pm  0.06  $   & $   74.4 \pm   8.7  $ \\
    0.32478  & $  13.24 \pm  0.15  $   & $   46.1 \pm  10.2  $ \\
    0.33269  & $  13.88 \pm  0.08  $   & $   25.9 \pm   3.2  $ \\
    0.34477  & $  14.02 \pm  0.08  $   & $   34.3 \pm   4.4  $ \\
    0.34914  & $  13.36 \pm  0.09  $   & $   31.3 \pm   4.8  $ \\
    0.35375  & $  13.41 \pm  0.06  $   & $   16.4 \pm   1.8  $ \\
    0.37660  & $  13.45 \pm  0.13  $   & $   36.4 \pm   6.2  $ \\
    0.38833  & $  13.02 \pm  0.19  $   & $   14.1 \pm   3.8  $ \\
    0.41081  & $  13.57 \pm  0.08  $   & $   32.8 \pm   5.0  $ \\
    0.41786  & $  13.25 \pm  0.08  $   & $   50.7 \pm   3.9  $ \\
    0.43148  & $  14.10 \pm  0.06  $   & $   20.9 \pm   4.3  $ \\
    0.43569  & $  14.22 \pm  0.10  $   & $   44.0 \pm   3.9  $ \\
  \cutinhead{PKS\,0405--123 (this paper, see Appendix)}
\enddata		
\end{deluxetable}
}

\begin{figure*}
\epsscale{1} 
\plotone{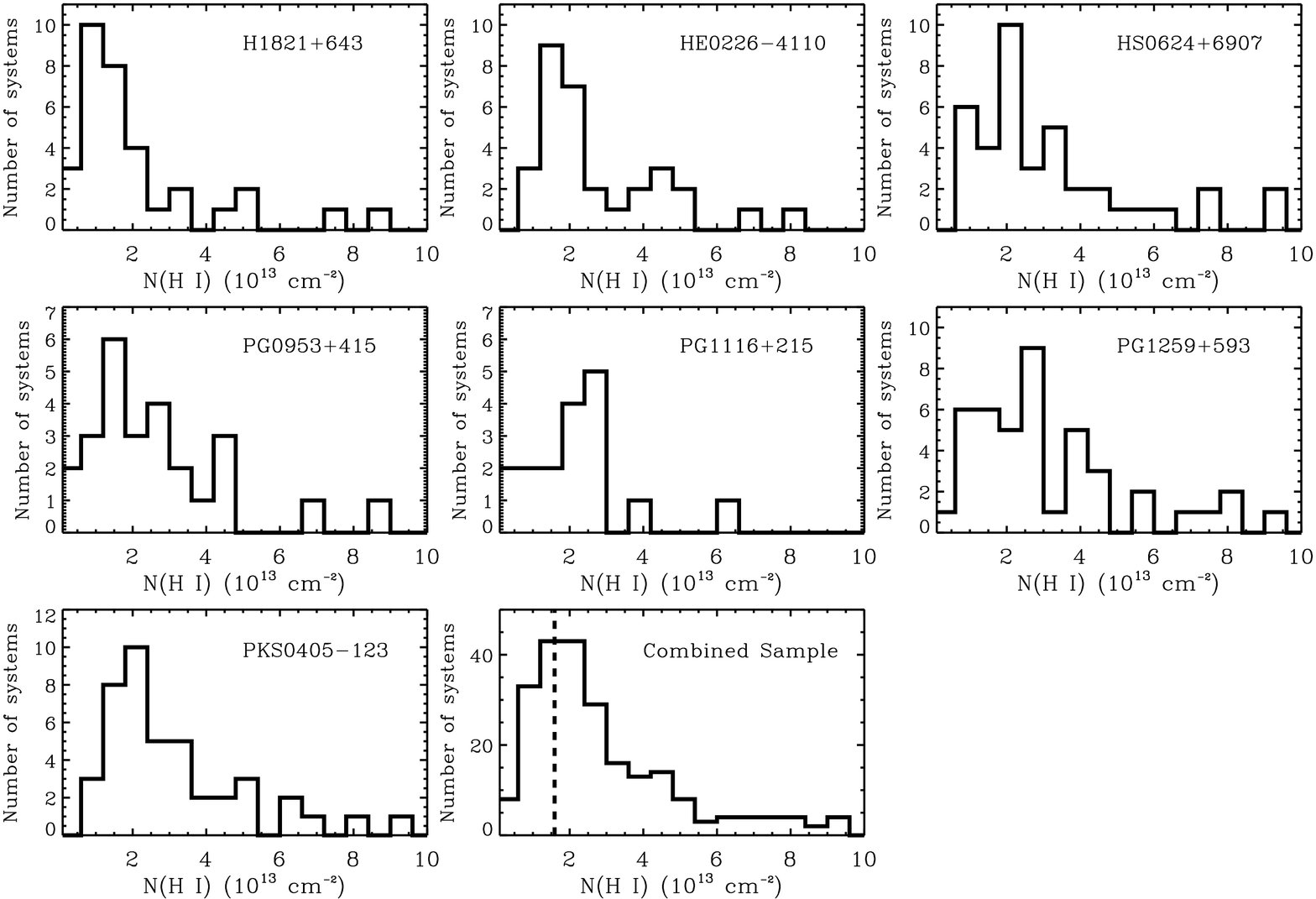}
\medskip
\caption{Distribution of the column density of \hi\ for each sightline and the combined sample
with a column density interval of $6\times 10^{12}$ cm$^{-2}$.  Only data with $N_{\rm H I} \le 10^{14}$ cm$^{-2}$
and less than 40\% errors in $b$ and $N_{\rm H I}$
are considered. The vertical dashed line on the complete sample marks the completeness limit of 
$N_{\rm H I} = 1.6 \times 10^{13}$ cm$^{-2}$.
\label{histall}}
\end{figure*}

The sample is not homogeneous with respect to the achieved S/N
in the STIS E140M observations toward the 7 sightlines considered. The detection limit
depends on the S/N and the breadth over which the spectrum is 
integrated. The matter is complicated by the S/N not being constant over the full 
wavelength range from $\sim$1216 \AA\ to $\sim$1730 \AA\
available with STIS E140M where Ly$\alpha$ can be observed; 
in particular, it deteriorates rapidly at $\lambda > 1650$ \AA. 
Only lines of sight with $z \ga 0.32 $ can reach the last 100 \AA\ of the STIS E140M 
wavelength coverage. Note that  297 systems with $z\le 0.32$ are observed at wavelengths 
$\lambda \la 1650$ \AA. For example, in one of the lowest 
S/N spectra in our sample (HE\,0226--4110), we estimate that at $\sim$1700 \AA,
a $3\sigma$ limit is  $\sim$75 m\AA\ for the profile integrated over $\delta v= [-50,50]$ \km\ 
and $\sim$100 m\AA\  over $\delta v= [-90,90]$ \km.
We estimate that our sample is complete for  $\log N($\hi$)\gtrsim 13.20 $ (corresponding to 
a rest frame equivalent width $W_{1215} \simeq 88$ m\AA) at a 3$\sigma$
level for $b\la 80$ \km. For the high S/N lines of sight (H\,1821+643, PG\,1259+593, and 
PG\,1116+215), this limit is quite conservative. 
For systems with $b>80$ \km\ and $\log N($\hi$)\gtrsim 13.20 $, our sample is incomplete, especially at 
$z\ge0.32$  and for the lowest S/N spectra. 

Since our sample is not homogeneous with respect to the achieved S/N, 
it is useful to have a sample in which the cloud parameters are relatively
well determined so that scatter due to noise is reduced. We therefore
consider systems with errors on $b$ and $N_{\rm H I}$ that are less
than 40\%. The sample has 270 \hi\ systems with 
errors on $b$ and $N_{\rm H I}$ that are less than 40\%. In this case, 
there are 155 systems at $z\le0.2$, 107 systems at $0.2 < z \le 0.4$, and 8 systems at 
$0.4 < z < 0.44$. At the completeness level $\log N($\hi$)\ge 13.20 $ and with
errors on $b$ and $N_{\rm H I}$ less than 40\%, there is a total of 202 \hi\ systems with
109 systems at $z\le0.2$, 85 systems at $0.2 < z \le 0.4$, and 8 systems at $0.4 < z < 0.44$.

\subsection{Overview of the Distributions of $N_{\rm H I}$, $b$, $z$}
In Fig.~\ref{histall}, we show the distribution of the column density for each line of sight
and the total sample (last panel, where we show the completeness limit of the sample) 
for systems with $N_{\rm H I} \le 10^{14}$ cm$^{-2}$ and $\sigma_b/b, \sigma_N/N \le 0.4$.
About 86\% of the absorbers have   $N_{\rm H I} \le 10^{14}$ cm$^{-2}$ and $\sim$94\% of them have
$N_{\rm H I} \le 10^{14.5}$ cm$^{-2}$. The column density distribution 
peaks near $N_{\rm H I} \sim 2 \times 10^{13}$ cm$^{-2}$ and drops sharply at smaller 
column densities. Since the peak of the distribution corresponds to about our completeness limit, 
the observed decrease of the number of systems at $N_{\rm H I} \la 1.6 \times 10^{13}$ cm$^{-2}$ 
can be understood from the reduction in sensitivity. 
Higher S/N spectra would be needed to further understand the distribution of 
the smaller column densities.

\begin{figure*}
\epsscale{1} 
\plotone{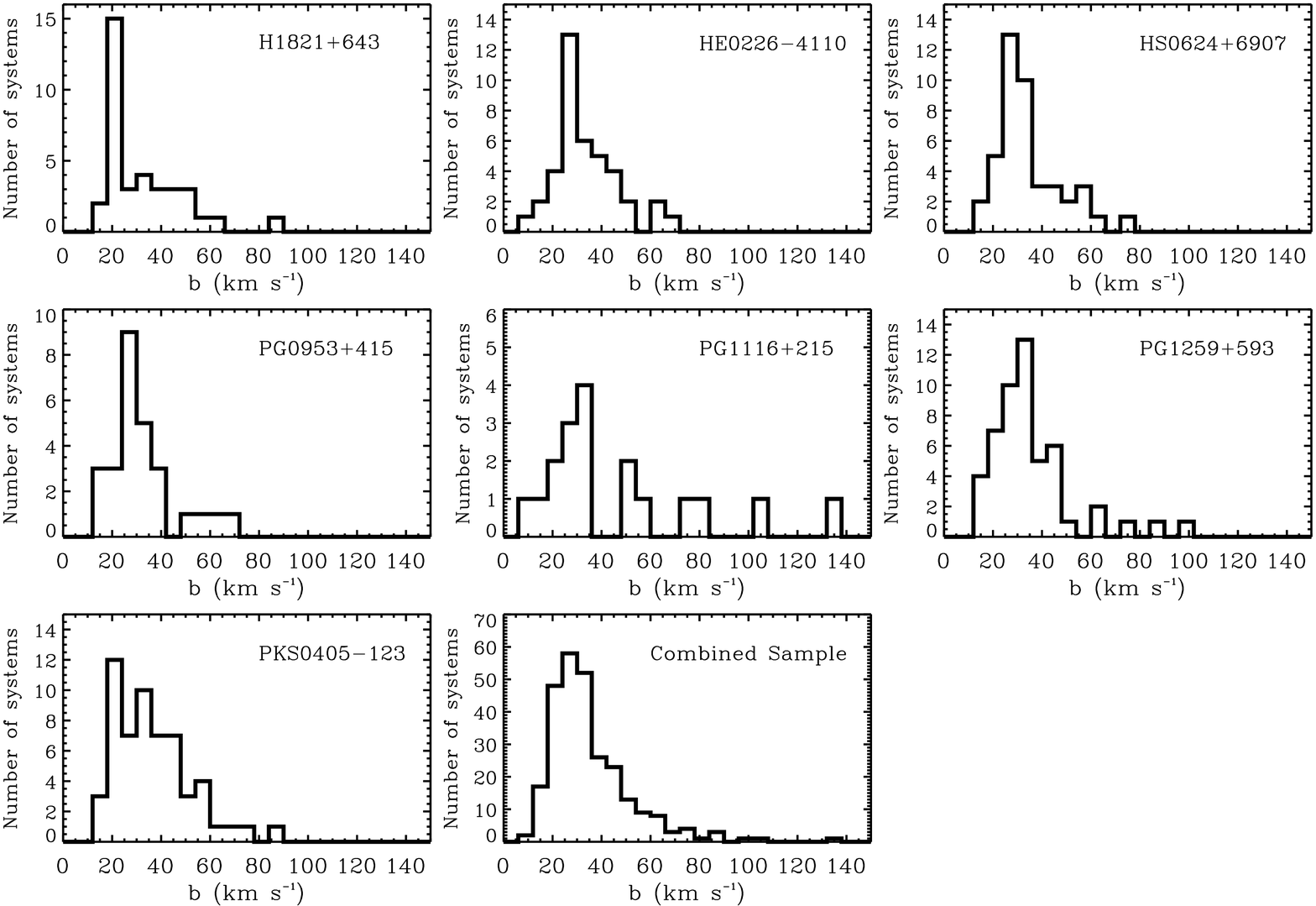}
\medskip
\caption{Distribution of the Doppler parameter of \hi\ for each sightline and the combined sample. 
The observations are plotted in 6 \km\ bins.  Only data with less than 40\% errors in $b$ and $N_{\rm H I}$
are considered.
\label{histallb}}
\end{figure*}

\begin{figure*}
\epsscale{1} 
\plotone{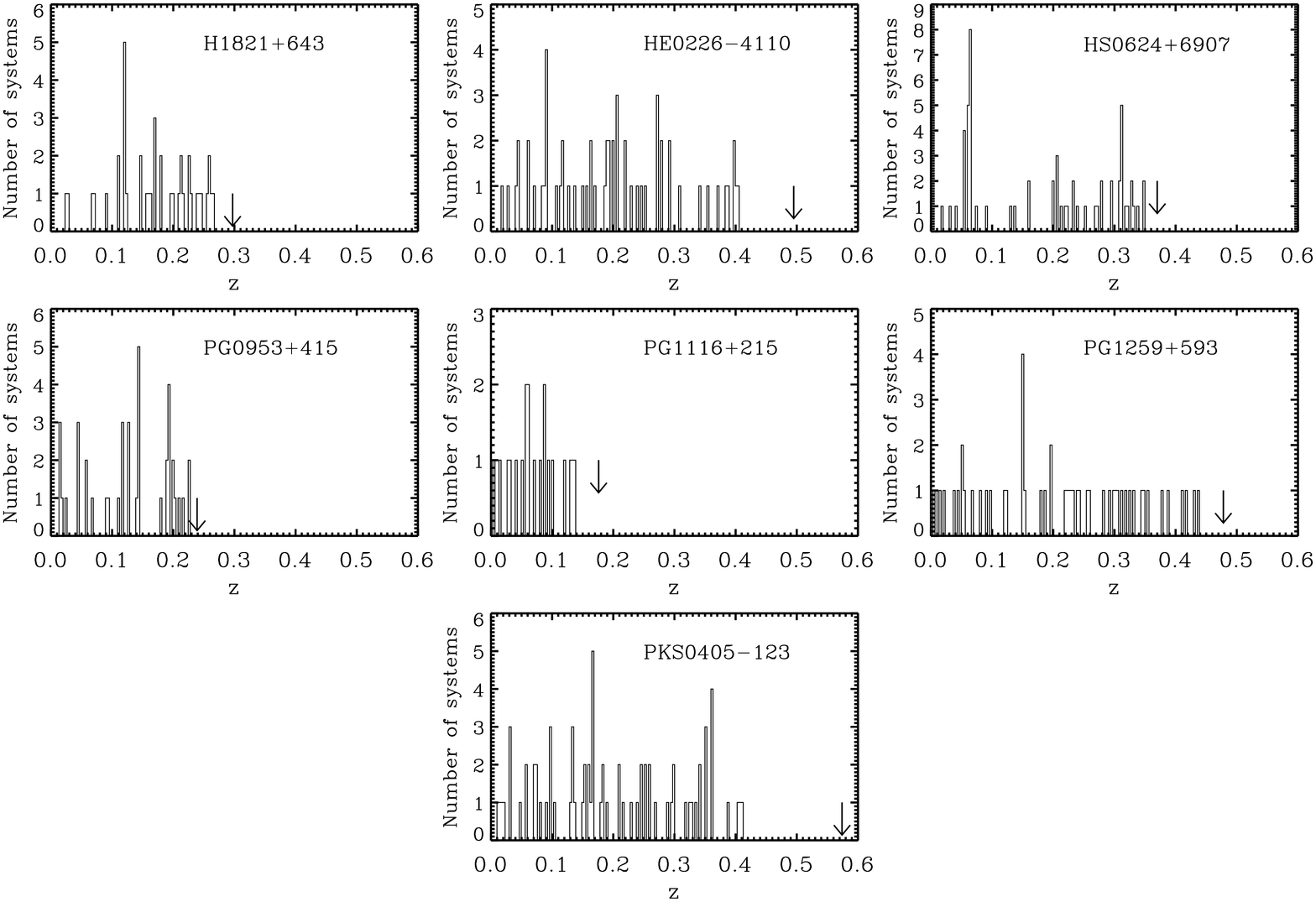}
\medskip
\caption{Distribution of the redshift of \hi\ for each sightline
with a redshift interval of $\Delta z = 0.0033$. The arrows show the redshifts of the QSOs.
The wavelength coverage of the E140M mode of STIS limits Ly$\alpha$ measurements to $z\la 0.42$.  
\label{histallz}}
\end{figure*}

In Fig.~\ref{histallb}, we show the distribution of the Doppler parameters for each line of sight
and for the entire sample (last panel) with $\sigma_b/b, \sigma_N/N \le 0.4$. 
For every sightline, the $b$-distribution peaks between about 20 and 30 \km. Yet, it is apparent that the 
distribution is not Gaussian around these values, and, in particular,
there are many systems with $b \ga 40$--50 \km\ that produce a tail 
in the $b$ distribution. This is clearly
observed in the histogram of the combined sample where
$b$ peaks  around 20--30 \km\ with an excess of systems with $b > 40$ 
\km\ compared to the number of systems with small $b$. This preview 
already shows clearly that a mean $b$-value does not provide an adequate description
of the Ly$\alpha$ forest. 

In Fig.~\ref{histallz}, we show the distribution of the redshift 
for each line of sight with a redshift bin of 0.0033 (corresponding to 1000 \km\ intervals). 
There is no striking difference between the different lines of sight, but 
there are clearly voids and clustering in the distribution, 
consistent with the Ly$\alpha$ forest tracing filaments of matter in 
the universe. An example of clustering of absorbers is found near the system at $z= 0.06352$ 
toward HS\,0624+6907 where at least 8 absorbers are found in a single bin and 17  
absorbers are only separated by about 3000 \km. \citet{aracil06} attribute
this clustering to the absorption of intragroup  gas, possibly from a filament viewed along its long axis.

\section{Distribution of the Doppler Parameter}\label{bdistr}
Earlier low-redshift UV studies of the \hi\ forest have low or 
moderate spectral resolutions and smaller wavelength 
coverage and did not allow access to several Lyman series transitions. Therefore,
the Doppler parameter $b$ generally had to be assumed and a study of the evolution 
and distribution of $b$ was not possible 
\citep[see for example the studies of][]{weymann98,penton00,penton04}. 
With STIS E140M observations (spectral resolution of 6.5 \km), $b$ can 
be derived from profile fitting analysis. Furthermore, combining STIS E140M 
and \fuse\ observations allows further constraints on $b$ by using 
several Lyman series lines and reducing possible misidentifications. 
Here, we review the frequency and properties of the narrow Ly$\alpha$ absorbers 
(NLAs, $b\le 40$ \km) and the broad Ly$\alpha$ absorbers (BLAs, $b > 40$ \km).

\subsection{Distribution of $b$ and Other Low $z$ Studies}

The median $b$-value of 31 \km\ for our combined sample is larger than the medians 
found in the low redshift IGM studies by \citet{dave01} and \citet{shull00}.
\citet{dave01} used automated software to derive
$b$ and $N$, but their criteria did not allow a search for broad components.  
\citet{shull00} went only after the  Ly$\beta$ absorption lines in the {\fuse}\ wavelength 
range to combine with known  Ly$\alpha$ absorption lines observed with GHRS
and were therefore less likely to find the broad \hi\ absorbers.  
\citet{dave01} found median and mean $b$-values of 
22 \km\ and 25 \km, respectively.  We find that the distribution of $b$ is not gaussian, 
making the mean and dispersion less useful quantities.
If  we restrict our sample to data with $b\le 40$ \km, 
the median and mean (with 1$\sigma$ dispersion) are 27 \km\ and $27 \pm 7$ \km. If
only data with $\log N($\hi$) \ga 13.2$ are considered, we obtain 29 \km\ and $28 \pm 6 $ \km. 
If we set the cutoff at $b\le 50$ \km, the median and the mean increase by 2 \km. 
The median, mean, and dispersion of $b$ in our sample (with $b\le 40$ or 50 \km) 
compare well to those derived by \citet{shull00}: 28 \km\ and $31 \pm 7$ \km\ for the median 
and mean, respectively.

\begin{figure}
\epsscale{1} 
\plotone{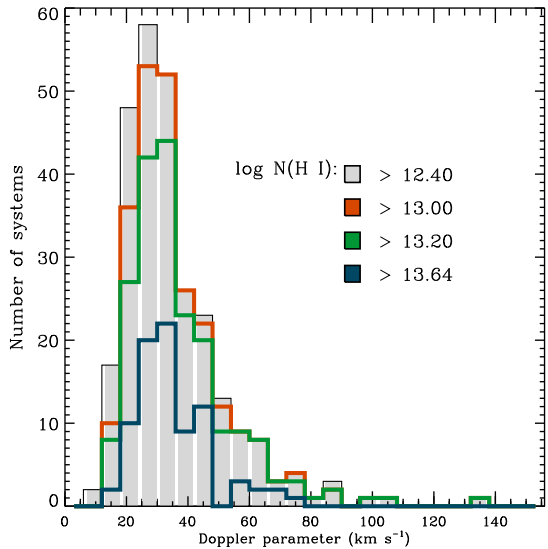}
\caption{Distribution of the Doppler parameter of \hi\ for a series of minimum $\log N($\hi) for $z\la 0.4$
for the seven sightlines.
Only data with less than 40\% errors in $b$ and $N_{\rm H I}$
are considered.
\label{bhist}}
\end{figure}

In Fig.~\ref{bhist}, we show the distribution of $b$ for samples with various $N_{\rm H I}$ 
cutoffs. For any $N_{\rm H I}$ cutoff, 
the maximum of the distribution always peaks 
near 25--30 \km\ and in all cases there is clearly an asymmetry in the distribution with
the presence of a tail in the distribution that develops at $b>40 $--60 \km. 
For the BLAs, the number of systems with \hi\ column between 13.0 and 13.2 dex is the largest, 
although this could be in part an observational bias since weaker column density absorbers 
with $b>40$ \km\ would require higher S/N to be detected. 
Although the effects of line blending can contaminate the measurements of
the observed tails, the various papers describing
the data \citep[see also][]{richter06} show that a large fraction of these 
broad absorbers can be described with a single Gaussian within the S/N. For 
the stronger of these absorbers, several Lyman series lines were also used 
to derive the physical parameters.

\begin{figure*}
\epsscale{0.8} 
\plotone{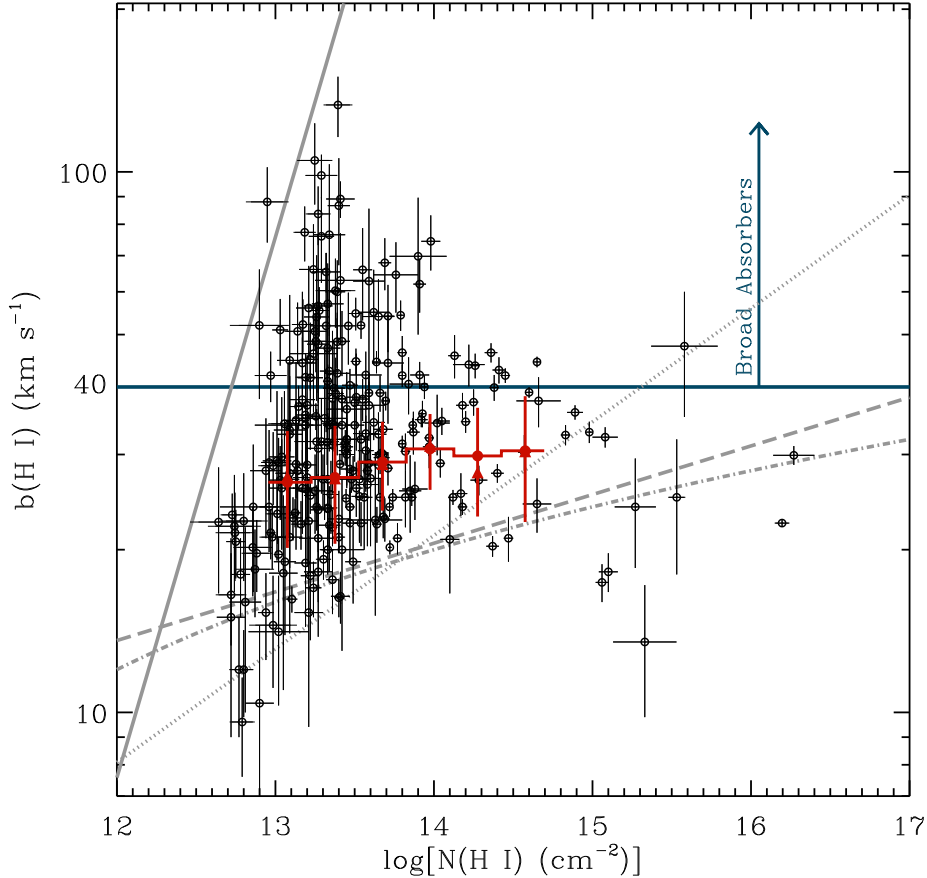}
\caption{Line width versus \hi\ column density for the Ly$\alpha$ absorbers along the seven
sightlines. Only data with less than 40\% errors in $b$ and $N_{\rm H I}$
are considered. The solid gray line shows $b$ vs $N$(\hi) for a Gaussian line
shape with 10\% central optical depth -- absorption lines to the left of this line are not detectable.
The dotted gray line shows the predicted minimum $b$-value from hydrodynamic cosmological simulations
at low redshift \citep{dave99}. The dot-dashed and long-dashed lines shows the fits to lower cutoff in the $b$ 
distribution of high redshift systems \citep{kirkman97,kim02a} (see \S\ref{sec-bn} for more details).
The red histogram and red symbols show the median $b$ (triangles), the mean and dispersion of $b$ (circles and vertical bars) 
in six bins of $N_{\rm H I}$ {\em only} for data with $b \le 40 $ \km\ 
(see Table~\ref{t3a}). 
\label{bcomp}} 
\end{figure*}

\subsection{The $b$--$N_{\rm H I}$ distribution}\label{sec-bn}
The $b$--$N_{\rm H I}$ distribution of the low redshift sample is shown in Fig.~\ref{bcomp}.
The solid gray line shows the threshold detection of \hi\ absorbers.  
This curve is the relationship between $b$ and $N_{\rm H I}$ for a Gaussian line profile with 10\%
central optical depth -- absorption lines to the left of this line are not detectable.
The lack of data with low $b$ and $\log N_{\rm H I} \la 12.7$ is most likely because
our sample is not complete at these low column densities. When absorbers with 
$\log N_{\rm H I} \ga 13.2$  and $b>0$ \km\ are considered, the 
$b$--$N_{\rm H I}$ plot reveals mostly a scatter diagram. Most of the 
absorbers are present in the column density range 13.2--14 dex.  In Table~\ref{t3a}, 
we list the median, mean, and dispersion of $b$ for the entire sample, NLAs, and BLAs. 
For the entire sample, as expected from  Fig.~\ref{bcomp}, it is not clear if $b$ increases
or decreases as $N_{\rm H I}$ increases. A Spearman rank-order
correlation test on the entire sample with $\log N_{\rm H I} \ge 13.2$ shows a 
very marginal negative correlation between $b$ and $N_{\rm H I}$ with 
a rank-order correlation coefficient $r = -0.08$ and a
statistical significance $t = 0.25$. 
We note that when systems with  $\log N_{\rm H I} \la 13.2$ 
are considered, it creates an apparent correlation ($r= 0.21$ and $t=4.7\times 10^{-4}$) 
between $N_{\rm H I}$ and $b$, which can be understood in terms of measurement biases, since weak broad systems 
are more difficult to detect than weak narrow absorbers. There is no 
clear separation between the NLAs and BLAs in this figure, although
we note that most BLAs are found at $\log N_{\rm H I} \la 14.0$ and 
no BLAs with $b>50$ \km\ have $\log N_{\rm H I} \ga 14.0$ (see below). 
This scatter and absence of clear separation between NLAs and BLAs are expected if 
the \hi\ lines trace systems with different temperatures {\em and}\ turbulent velocities. 

For NLAs with $b\le 40$ \km\ and  $\log N_{\rm H I} \ge 13.2$, 
we show in Fig.~\ref{bcomp} the median, mean, and dispersion of $b$ (red curves 
and symbols) derived in six intervals of $N_{\rm H I}$. These estimates are summarized
in Table~\ref{t3a}. This shows evidence of an increase of $b$ with increasing $N_{\rm H I}$
for the NLAs, at least for the weak absorbers with $\log N_{\rm H I} \la 14.1$. 
The Spearman rank-order correlation test for the NLA sample
shows a weak correlation between $b$ and $N_{\rm H I}$  with $r = 0.12$ and  $t = 0.15$. 
Since the sample is complete for the NLAs, the increase of $b$ with increasing $N_{\rm H I}$
must be real. The large scatter is again expected if the \hi\ lines are broadened as a result 
of different temperatures and turbulent velocities. 

\begin{figure*}[tbp]
\epsscale{1} 
\plotone{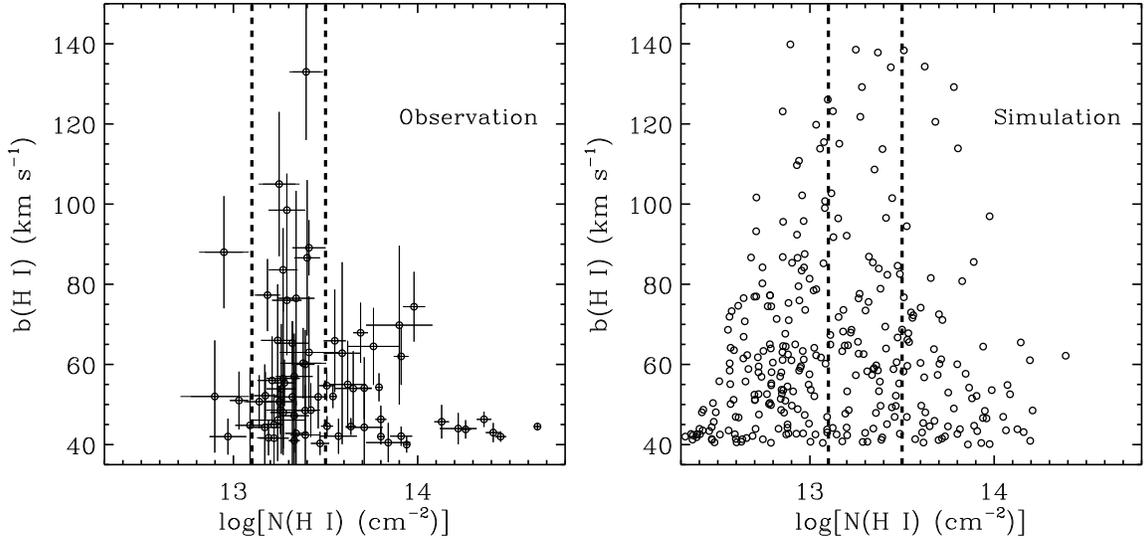}
\caption{{\em Left}: Line width versus \hi\ column density for the BLAs ($b>40$ \km) along the seven
sightlines. Only data with less than 40\% errors in $b$ and $N_{\rm H I}$
are considered. {\em Right:} Line width versus \hi\ column density for the high-quality
sample of BLAs simulated by \citet{richter06a}.  The dashed vertical lines highlight
the column density range [13.1,13.5] dex where the broadest absorbers with $b>80$ \km\ are detected 
in the observational sample. 
\label{bcomp1}} 
\end{figure*}

In Fig.~\ref{bcomp1}, we present the $b$--$N_{\rm H I}$ distribution only for the BLAs (left panel),
which shows that $b$ appears to decrease  with increasing  $N_{\rm H I}$: For  
$13.1 \la \log N_{\rm H I} \la 13.5$ (this range is highlighted in Fig.~\ref{bcomp1} by 
the vertical dotted lines), $b$ is distributed between about 40 and 130 \km; 
for $13.5 \la \log N_{\rm H I} \la 14.0$, $b$ is mostly distributed between about 40 and 80 \km; 
and for $\log N_{\rm H I} \ga 14.0$, $b$ is always lower than 50 \km. 
This trend is also confirmed in the last two columns of Table~\ref{t3a}  
(note that in the [13.8,14.1] interval there are only 6 systems, with 3 of them having $b>60$ 
\km\ and the other 3 having $b\sim 40$ \km). The Spearman rank-order correlation  for the BLA sample 
with $\log N_{\rm H I}\ge 13.2 $ confirms a negative correlation between 
$b$ and $N_{\rm H I}$ ($r = -0.30$ and  $t = 0.016$). 

In the right panel of Fig.~\ref{bcomp1}, we show the recent simulation of BLAs 
undertaken by \citet{richter06a}, in which artificial spectra were generated
from the hydrodynamical simulation. Their numerical model was part of an earlier investigation of the \ovi\
absorption arising in WHIM filaments \citep{fang01}, and they include collisional ionization 
and photoionization processes.  The simulated sample presented in  Fig.~\ref{bcomp1} corresponds 
to their high-quality sample that includes 321 BLAs with almost perfect Gaussian profiles 
(note that 58\% of the sample has $\log N_{\rm H I} \la 13.2$). 
Since our observations are complete only to  $\log N_{\rm H I} \ga 13.2$,
it is not surprising that we are missing absorbers below this limit. It is, however, interesting
to note that the simulation and observations have a similar trend: (i) most of the broadest absorbers are found 
at $\log N_{\rm H I} \la 13.5$, and (ii) most of the strong absorbers ($\log N_{\rm H I} \ga 14$) have 
$b\la 50$ \km\ (although we note that in the simulation a few systems have $b$  up to 65 \km, and 
the simulation does not produce BLAs with $\log N_{\rm H I} \ga 14.4$). 
This is also in general agreement with the simulation of the WHIM produced by \citet{dave01a}
where they show  the WHIM fraction peaks for an overdensity $\rho/\bar{\rho}$
of $\sim$$5$--30. If Eq.~\ref{e-nrho} (see below) applies for the BLAs, 
the \hi\ column density range $13.2$ to 14.0 dex corresponds to 
$\rho/\bar{\rho} \sim 5$--17, in general agreement with the
hydrodynamical simulations of \citet{dave01a} if the BLAs trace mostly the WHIM. 
We note that the simulation of \citet{richter06a} generally produces larger $b$ than currently observed: 
for systems with $\log N_{\rm H I} \ga 13.2$,  the median, mean, and standard deviation are 
59, 69, 31 \km\ for the simulation, while they are 52, 57, 18 \km\ for the observations. 
The current S/N of the observed data limits the detection of the broader absorbers with $b>80$ \km,
as discussed in \S\ref{datasample}. We also note that the broader systems ($b>80$ \km) are likely 
to be more uncertain, especially since they are detected in the lowest column density range 
($\log N_{\rm H I} \la 13.4$).

\citet{schaye99} and others demonstrated that the temperature-density
relation, which is well described by a power law $T = T_0(\rho/\bar{\rho})^{\gamma-1}$
(where $\bar{\rho}$ is the average density, $\rho/\bar{\rho}$ is the overdensity
of the IGM), implies a lower envelope to the $b$--$N_{\rm H I}$ distribution.
This lower envelope is not clearly observed in Fig.~\ref{bcomp}. We further explore
this by considering the numerical cosmological simulations of the low-redshift Ly$\alpha$ forest  
that predict that the temperature $T_4 = T/(10^4\,{\rm K})$  is a power law of the overdensity 
$\rho/\bar{\rho}$, with
\begin{equation}\label{e-trho}
T_4 \approx 0.5 \left(\frac{\rho}{\bar{\rho}}\right)^{0.6}\, , 
\end{equation}
for the coolest systems at any given density. The overdensity is connected to the \hi\ column density 
through \citep{dave99},
\begin{equation}\label{e-nrho}
\frac{\rho}{\bar{\rho}} \approx 20 \left(\frac{N_{\rm H I}}{10^{14}\, 
{\rm cm}^{-2}}\right)^{0.7} \, 10^{-0.4 z} \,.
\end{equation}
\begin{deluxetable*}{lcc|cc|cc}[!t]
\tabcolsep=3pt
\tablecolumns{7}
\tablewidth{0pt} 
\tabletypesize{\scriptsize}
\tablecaption{Median, mean, dispersion of $b$ at low $z$ in a given column density interval for the 7 QSO sample \label{t3a}} 
\tablehead{\colhead{Column}    &  \colhead{median}&\colhead{mean\,$\pm \sigma$ }  &  \colhead{median}&\colhead{mean\,$\pm \sigma$ } &  \colhead{median}&\colhead{mean\,$\pm \sigma$ } \\ 
\colhead{Density interval} & \multicolumn{2}{c}{$b>0$ \km} & \multicolumn{2}{c}{$b\le 40$ \km} & \multicolumn{2}{c}{$b>40$ \km} }
\startdata
$ [12.9,13.2]$   &    28.7    &  $  31.5  \pm	13.9 $   &     27.2   &$   26.7 \pm  6.5 $  &  50.7	& $  54.7 \pm	16.6 $   \\
$ [13.2,13.5]$   &    34.0    &  $  39.9  \pm	21.7 $   &     27.0   &$   27.2 \pm  6.7 $  &  55.4	& $  61.5  \pm  21.1 $   \\
$ [13.5,13.8]$   &    32.0    &  $  36.3  \pm	12.9 $   &     28.7   &$   29.1 \pm  5.4 $  &  54.1	& $  53.2  \pm  8.4  $   \\	 
$ [13.8,14.1]$   &    34.0    &  $  36.9  \pm	14.1 $   &     31.0   &$   30.7 \pm  4.9 $  &  62.0	& $  57.8  \pm  15.7 $   \\
$ [14.1,14.4]$   &    37.0    &  $  34.1  \pm	9.1  $   &     27.7   &$   29.8 \pm  6.8 $  &  45.7	& $  45.0  \pm  1.2  $   \\
$ [14.4,14.7]$   &    39.1    &  $  35.2  \pm	8.9  $   &     30.3   &$   30.5 \pm  8.0 $  &  43.0	& $  43.2  \pm  1.3  $  \\ 
\enddata		
\tablecomments{Only data with $\sigma_b/b, \sigma_N/N \le 0.4$ are included in the various samples.}
\end{deluxetable*}
If we combine these two equations, we have: 
\begin{equation}\label{e-tn}
T_4 \approx 3 \left(\frac{N_{\rm H I}}{10^{14}\, {\rm cm}^{-2}}\right)^{0.42} \, 10^{-0.24 z} \,.
\end{equation}
The pure thermal Doppler broadening for H is $b_{\rm th} = 0.129 \sqrt{T}$, so 
for photoionized hydrogen absorbers we can write
\begin{equation}\label{e-bn}
b_{\rm th} \approx 22 \left(\frac{N_{\rm H I}}{10^{14}\, {\rm cm}^{-2}}\right)^{0.21} 
\, 10^{-0.12 z} \, {\rm km\,s}^{-1} \,.
\end{equation}
This latter relation is shown in Fig~\ref{bcomp} with the dotted line where we set $z=0.2$, which is
about the mean and median $z$ in our sample. Since the $\rho$--$T$
fit to photoionized absorbers is for the coolest systems, this relation should provide
a lower envelope to the $b$--$N_{\rm H I}$ distribution.  The lower envelope to the 
observed $b$--$N_{\rm H I}$ distribution roughly agrees with the numerical simulations, at least 
as long as $\log N_{\rm H I} \la 14.2$. But the low redshift sample does not provide yet 
as sharp a lower envelope to the $b$--$N_{\rm H I}$ distribution 
as high redshift samples do \citep[see for example][]{kirkman97,kim02} because
there are still too few systems at the completeness level. 
In Fig.~\ref{bcomp}, we show with the dot-dashed line the relation 
$b_{\rm min} = 20 + 4 \log [N_{\rm H I}/(10^{14} {\rm cm^{-2}})]$ \km\
found by \citet{kirkman97} at $\bar{z} = 2.7$ and with the long dashed line the relation 
$\log b_{\rm min} = 1.3 + 0.090 \log [N_{\rm H I}/(10^{14} {\rm cm^{-2}})]$ \km\ found 
by \citet{kim02a} at $\bar{z} = 2.1$ (smoothed power-law fit). These power laws provide 
a good approximation to the observed lower envelope of $b$ at high redshift. 
At low redshift at $\log N_{\rm H I} \ga 13.2$, a few absorbers lie below
these fits, especially at $\log N_{\rm H I} \ga 15$. For systems with $\log N_{\rm H I} \la 15$
it is not clear if the lower $b$ cutoff evolves with redshift.

Finally, following \citet{dave01}, if we compare the median in the various intervals with 
$b \le 40$ \km\ (red curve in Fig.~\ref{bcomp}) and $b_{\rm th}$ defined in Eq.~\ref{e-bn}, we find that 
$b_{\rm th} \sim  (0.6-0.7) b_{\rm obs}$ for a typical absorber with $\log N_{\rm H I} \la 14.2$.  
Therefore, the contribution from thermal broadening is substantial for the low redshift NLAs.
However, if the some BLAs actually trace cool photoionized gas, the non-thermal broadening
will be dominant in these absorbers.  

\subsection{Ly$\alpha$ Line Density}\label{lined}
This sample provides the first opportunity to investigate 
the relative number of systems as a function of the Doppler parameter in the low-$z$ IGM. 
In Table~\ref{t2}, we summarize $d{\mathcal N}/dz$ for each sightline and for 
the combined sample where our sub-samples have either $b \le 40$ \km\ or $b\le 150$ \km. 
For both $b$-samples, we choose three different column density ranges: (a) $[13.2,14.0]$ dex, 
(b) $[13.2,16.5]$ dex and (c) $[13.64,16.5]$ dex. The lower limit of samples (a) 
and (b) corresponds to our threshold of completeness. 
The largest observed column  density in our sample is about $10^{16.5}$ cm$^{-2}$. 
Hence sample (b) corresponds to the combined sample with   $W \ga 90$ m\AA, while 
sample (a) only covers the weaker Ly$\alpha$ lines which may evolve differently 
than the stronger lines since the weaker lines may arise from tenuous gas
in the IGM while the stronger lines may mostly trace the gas in the outskirts of galaxies.   
The threshold of sample (c) was chosen to be comparable to the equivalent
width threshold of 0.24 \AA\ from  the {\em HST}\ QSO absorption line key project \citep{weymann98}.  
The last rows of the subtables in Table~\ref{t4} show the mean $d{\mathcal N}/dz$ for the 7 sightlines.   
We considered only absorbers with $\sigma_b/b, \sigma_N/N \le 0.4$. Note 
that the average values would have increased only by $\sim$5\% if we did not 
apply this cutoff.   

\begin{deluxetable}{lcc|cc}
\tabcolsep=4pt
\tablecolumns{5}
\tablewidth{0pt} 
\tablecaption{Ly$\alpha$ Line Density \label{t2}} 
\tablehead{\colhead{Sightline}    &  \colhead{${\mathcal N}_{\rm H I}$} & \colhead{$d{\mathcal N}_{\rm H I}/dz$}    &  \colhead{${\mathcal N}_{\rm H I}$}&\colhead{$d{\mathcal N}_{\rm H I}/dz$}  \\
\colhead{} & \multicolumn{2}{c}{$b \le 40$ \km} & \multicolumn{2}{c}{$b \le 150$ \km}}
\startdata
\cutinhead{(a) $13.2 \le \log N_{\rm H I} \le 14.0$}
H\,1821+643    &   7   &  $  29 \pm    11 $	 &  12  &  $ 50 \pm	15 $	   \\
HE\,0226--4110 &  16   &  $  40 \pm    10 $	 &  22  &  $ 55  \pm	12 $	   \\
HS\,0624+6907  &  22   &  $  67 \pm    14 $	 &  31  &  $ 94 \pm	17 $	   \\
PG\,0953+415   &  13   &  $  64 \pm    18 $	 &  17  &  $ 84 \pm	20 $	   \\
PG\,1116+215   &   7   &  $  56 \pm    21 $	 &  12  &  $ 95 \pm	28 $	  \\
PG\,1259+593   &  24   &  $  68 \pm    14 $	 &  35  &  $ 99 \pm	17 $	\\
PKS\,0405--123 &  21   &  $  51 \pm    11 $	 &  35  &  $ 85 \pm     14 $	   \\
Mean	       &     	& $  54 \pm    5 (15) $	 &      &  $  80 \pm    6 (20) $ \\
\cutinhead{(b) $13.2 \le \log N_{\rm H I} \le 16.5$}
H\,1821+643    &   9   &  $  38 \pm	13 $	  &  15   &  $ 63  \pm    16 $      \\
HE\,0226--4110 &  21   &  $  52 \pm	11 $	  &  31   &  $ 77  \pm    14 $      \\
HS\,0624+6907  &  25   &  $  76 \pm	15 $	  &  35   &  $ 106 \pm    18 $      \\
PG\,0953+415   &  14   &  $  69 \pm	19 $	  &  18   &  $ 89  \pm    21 $     \\
PG\,1116+215   &  8    &  $  64 \pm	23 $	  &  13   &  $ 103  \pm   29 $        \\
PG\,1259+593   &  30   &  $  85 \pm	15 $	  &  43   &  $ 121 \pm    19 $     \\
PKS\,0405--123 &  33   &  $  80 \pm	14 $	  &  47   &  $ 114 \pm    17 $      \\
Mean	       &     	& $  66 \pm     6 (17) $      &     & $   96 \pm    7 (21) $   \\
\cutinhead{(c) $13.64 \le \log N_{\rm H I} \le 16.5$}
H\,1821+643    &   6    &  $  25 \pm	10 $	  &   7   &  $  29 \pm    11 $   \\
HE\,0226--4110 &  10    &  $  25 \pm	8  $	  &  16   &  $  40 \pm    10 $   \\
HS\,0624+6907  &  10    &  $  30 \pm    10 $	  &  13   &  $  40 \pm    11 $  	 \\
PG\,0953+415   &   6    &  $  30 \pm	12 $	  &   6   &  $  30 \pm    11 $ \\
PG\,1116+215   &   2    &  $  16 \pm	11 $	  &   2   &  $  16 \pm    11 $  	\\
PG\,1259+593   &  12    &  $  34 \pm	10 $	  &  18   &  $  51 \pm    13 $  \\
PKS\,0405--123 &  16    &  $  39 \pm	10 $	  &  21   &  $  51 \pm    12 $	\\
Mean	       &     	&  $  28 \pm	4 (8) $      &   	  &  $  37 \pm    4 (13) $   
\enddata		
\tablecomments{Only data with $\sigma_b/b, \sigma_N/N \le 0.4$ are included in the various samples.
Errors are from Poisson statistics.  The number between parentheses in the 
row showing the mean corresponds to the standard deviation around the mean for 
the different lines of sight.}
\end{deluxetable}

The average value of $d{\mathcal N}/dz$ in sample (c) ($W \ga 0.24$ \AA) 
of $28 \pm 4$ for $b \le 40 $ \km\ 
is slightly smaller than the estimates of \citet{weymann98}  at $\log(1+z)< 0.15$ 
and \citet{impey99} at $0<z<0.22$, because the broader absorbers and uncertain absorbers 
are not included in our sample. Indeed, the estimates of \citet{weymann98} and \citet{impey99} appear
intermediate between our two $b$ samples. 
\citet{penton04}  found $d{\mathcal N}/dz = 25 \pm (4, 5) $ for a sample with  $z < 0.069$ which
overlaps with our estimate within the 1$\sigma$ dispersion, which 
implies little or no evolution of  $d{\mathcal N}/dz$ at $z \la 0.4$ for the strong \hi\ absorbers. For 
$b \le 150 $ \km, the value of  $d{\mathcal N}/dz$ is about 1.3 times larger than for the NLAs. 
We find that the broad Ly$\alpha$ lines with $40<  b \le 150$
\km\ have $ d{\mathcal N}({\rm BLA})/dz \approx 9$, implying
that for the stronger lines of the Ly$\alpha$ forest, the number of BLAs
per unit redshift may be important. 

Comparison of samples (a) and (b) shows that the weak systems are far more frequent than
the strong systems. For these samples, we note  that $d{\mathcal N}/dz$ is systematically
smaller toward H\,1821+643 and HE\,0226--4110 than $d{\mathcal N}/dz$ toward the other sightlines 
for both $b$ sub-samples. $d{\mathcal N}/dz$ toward PG\,1116+215 and PKS\,0405--123 is 
intermediate for the NLAs. The 3 other lines of sight have similar $d{\mathcal N}/dz$. 
These trends do not appear related to the S/N of the data
since the S/N is the highest toward H\,1821+643 and comparatively low for HE\,0226--4110.
The redshift paths do not seem to explain all the differences. For example,
while the redshift paths are comparable between HE\,0226--4110 and
PG\,1259+593, $\Delta z$ is significantly smaller toward PG\,1116+215.  $d{\mathcal N}/dz$ is 
very similar for either column density range toward the sightlines that cover small and large redshift paths, 
implying no redshift evolution of $d{\mathcal N}/dz$ between  $z > 0.$ and $z\la 0.4$. 
Therefore, some of the observed variation in $d{\mathcal N}/dz$ must be cosmic variance 
between sightlines.

We explore the effect of the S/N on the $d{\mathcal N}({\rm BLA})/dz$ estimate in Table~\ref{t2a}
by considering data with $\sigma_b/b, \sigma_N/N \le 0.4$, 0.3, and 0.2. 
As expected, the spectra with the highest S/N are less affected by these cutoffs than 
the spectra with the lowest S/N. Decreasing the error thresholds has an effect mostly 
on the weak systems ($\log N_{\rm H I}\la 13.40 $).
Toward PG\,0953+415, the four BLAs have $ 13.2 < \log N_{\rm H I}\le 13.40 $ and 
$0.23 \le \sigma_b/b \le 0.30$, which explains why there is no BLA at the threshold $\sigma_b/b \le 0.2$.
Although the HS\,0624+6207 spectrum has several weak BLAs, only 
if $\sigma_b/b, \sigma_N/N < 0.15$ is applied do we observe a significant 
drop in $d{\mathcal N}({\rm BLA})/dz$ by a factor 2.5.  
On average, there is a decrease in $d{\mathcal N}({\rm BLA})/dz$
by a factor  $\sim$1.4 between the cutoff at 0.4 and 0.2. 
If only BLAs with $40<b\le 100$ \km\ are considered,  $d{\mathcal N}({\rm BLA})/dz = 28, 25,20$
for $\sigma_b/b, \sigma_N/N \le 0.4, 0.3,0.2$, respectively. These estimates 
are in agreement with those of \citet{richter06} (see also \S\ref{sec-descbevol}). 
The BLA density is about twice smaller than $d{\mathcal N}({\rm NLA})/dz$ for samples
(a) and (b).

\begin{deluxetable}{lcccc}
\tabcolsep=0pt
\tablecolumns{5}
\tablewidth{0pt} 
\tabletypesize{\scriptsize}
\tablecaption{Effect of the Signal-to-Noise on the Broad Ly$\alpha$ Density Estimate \label{t2a}} 
\tablehead{\colhead{Sightline}    &\colhead{S/N}    &  \colhead{$d{\mathcal N}_{\rm H I}/dz$}&\colhead{$d{\mathcal N}_{\rm H I}/dz$} &\colhead{$d{\mathcal N}_{\rm H I}/dz$}  \\
\colhead{}    & \colhead{}    &  \colhead{$\sigma_b/b \le 0.4$}&\colhead{$\sigma_b/b\le 0.3$} &\colhead{$\sigma_b/b \le 0.2$} \\ 
\colhead{}    & \colhead{}    &  \colhead{$\sigma_N/N \le 0.4$}&\colhead{$\sigma_N/N \le 0.3$} &\colhead{$\sigma_N/N \le 0.2$}  }
\startdata
H\,1821+643     &   15--20    & $ 25 \pm 10  $&  $ 25 \pm 10 $ &  $ 25 \pm 10 $   \\	  
HE\,0226--4110  &   5--11     & $ 25 \pm 8   $&  $ 20 \pm 7  $ &  $ 18 \pm 7  $   \\
HS\,0624+6907   &   8--12     & $ 30 \pm 10  $&  $ 30 \pm 10 $ &  $ 30 \pm 10 $   \\
PG\,0953+415    &   7--11     & $ 20 \pm 10  $&  $ 15 \pm 9  $ &   0	 \\	 
PG\,1116+215    &   10--15    & $ 40 \pm 18  $&  $ 40 \pm 18 $ &  $ 32 \pm 16 $   \\	
PG\,1259+593    &   9--17     & $ 37 \pm 10  $&  $ 34 \pm 10 $ &  $ 25 \pm 9 $   \\
PKS\,0405--123  &   5--10     & $ 34 \pm  9  $&  $ 24 \pm 8  $ &  $ 14 \pm 6  $   \\ 
Mean		& 	      & $ 30 \pm 4 (7)	$&  $ 27 \pm 4 (8)  $ &  $ 21 \pm 3 (11)  $  
\enddata		
\tablecomments{S/N is measured per resolution element. 
Sample with $40 < b \le 150$ \km\ and $13.2 \le \log N_{\rm H I} \le 16.5$.
Errors are from Poisson statistics. The number between parentheses in the 
row showing the mean corresponds to the standard deviation around the mean for the different lines
of sight. }
\end{deluxetable}

\begin{figure}[t]
\epsscale{1} 
\plotone{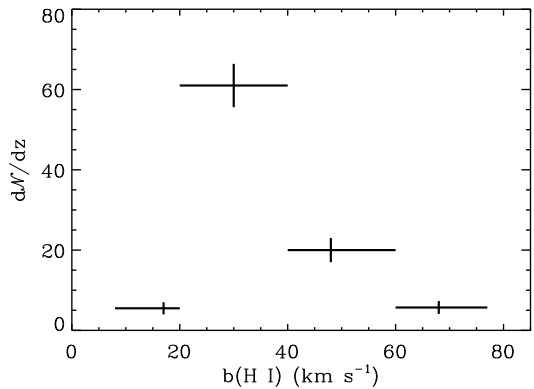}
\caption{Mean number density per unit redshift with Poissonian errors (vertical bars) in four intervals of $b$ 
($[0,20]$, $[20,40]$, $[40,60]$, $[60,80]$ \km) using data with  $\sigma_b/b, \sigma_N/N \le 0.4$ and 
$\log N_{\rm H I} \ge 13.2$. The mean number density is shown at the mean $b$-value of each 
$b$-interval sample. The horizontal bars show the $b$-intervals delimited by the minimum and maximum 
$b$-values of each $b$-interval sample.  
\label{bdndz}}
\end{figure}

Finally, we summarize in Fig.~\ref{bdndz} the frequency of the Ly$\alpha$ absorbers 
with various $b$-values in the low redshift universe. In this figure, $d{\mathcal N}/dz$ 
represents the mean of the number density of \hi\ absorbers obtained toward
each line of sight in  four intervals of $b$ ($[0,20]$, $[20,40]$, $[40,60]$, $[60,80]$ \km) 
using data with  $\sigma_b/b, \sigma_N/N \le 0.4$ and 
$\log N_{\rm H I} \ge 13.2$. The vertical error bars assume Poissonian errors. 
The mean $d{\mathcal N}/dz$ is shown at the mean $b$-value of each $b$-interval sample.
The horizontal bars show the $b$-intervals delimited by the minimum and maximum $b$-values 
of each $b$-interval sample.  
The NLAs are mostly found between 20 and 40 \km. 
Very narrow  Ly$\alpha$ absorbers with $b<20$ \km\ are rare, and we note that in the $[0,20]$ 
\km\ interval, most of the absorbers have $b\ga 15$ \km. The paucity of very narrow absorbers
with $\log N_{\rm H I} \ge 13.2$ is real since the resolution of STIS E140M ($b_{\rm inst}\sim 4$ \km) 
allows to fully resolve absorbers with $b \ga 5$ \km. The broad absorbers 
are more frequent in the $b=[40,60]$ \km\ range: there are about 3.5 times more absorbers in
$[40,60]$ \km\ interval than in the $b=[60,80]$ \km\ interval. Ly$\alpha$ absorbers are therefore
more frequent for low column densities ($\log N($\hi$) \la 14$) and for $b$-values
between 20 and 40 \km. 

\section{Evolution of the Doppler Parameter}\label{sec-evol}

\subsection{Higher Redshift Samples}

Recently, the analysis of the Ly$\alpha$ absorbers in the redshift range $0.5 < z < 2.0 $ 
from \citet{jank06} has become available at the Centre de Donn\'ees de Strasbourg (CDS).
Their data that sample the mid-$z$ IGM at $0.5 < z \la 1.5$ were obtained with 
 {\em HST}/STIS E230M: PG\,1634+706 ($z=0.534$--1.295), 
PKS\,0232--04 ($z=0.876$--1.419), PG\,1630+377 ($z=0.875$--1.451), PG\,0117+213 ($z=0.875$--1.475),
HE\,0515--4414 ($z=0.874$--1.475) and HS\,0747+4259 ($z=0.760$--1.443). The redshifts between
parentheses indicate the redshift interval probed by the observations.
The spectral resolution of E230M data ($R \sim 30000$) is lower than that of the low-$z$
sample obtained with the E140M grating ($R \sim 44000$). The S/N of the 
mid-$z$ sample is comparable to the lowest S/N of the low-$z$ sample, except for
PG\,1634+706, which has a S/N per resolution element of 5--40. The high redshift sample ($z>1.5$)
consists of data obtained with VLT/UVES ($R\sim 40000$): 
 HE\,0515--4414 (1.515--1.682), HE\,0141--3932 ($z=1.518$--1.784), HE\,2225--2258 ($z=1.515$--1.861), and 
HE\,0429--4901 ($z=1.662$--1.910). The S/N is typically higher than 30--40, except for HE\,0429--4901, 
which is about 15. The spectrum of  HS\,0747+4259 ($z=1.562$--1.866)
was also obtained with Keck/HIRES ($R\sim 50000$) with a S/N in the range of 6--24.

For the high redshift sample, we also consider the spectra of QSOs at $z>1.5$  presented by Kim et al. (2002a).
Their profile fitting results are available at the CDS. 
These data were obtained with the VLT/UVES at resolution of $R\sim 45000$ and typical S/N\,$\simeq 40$--50. 
The QSOs considered are: HE\,0515--4414 ($z=1.53$--1.69), Q\,1101--264 ($z=1.66$--2.08), 
J\,2233--606 ($z=1.80$--2.20), HE\,1122-1648 ($z=1.88$--2.37), 
HE\,2217--2818 ($z=1.89$--2.37), HE\,1347--2457 ($z=2.09$--2.57), Q\,0302--003 
($z=2.96$--3.24), Q\,0055--269 ($z=2.99$--3.60). Note that the HE\,0515--4414 VLT spectrum in 
\citet{jank06} has a different wavelength coverage and exposure time than the HE\,0515--4414 VLT spectrum
presented by \citet{kim02}. We finally consider the \hi\ parameters measured toward HS\,1946+7658 
($z=2.43$--3.05) by \citet{kirkman97}. The spectrum of HS\,1946+7658 was obtained with Keck/HIRES. 
The S/N per resolution element varies from about 30 to 200 and the spectral resolution is 
similar to UVES. The spectral resolution of the high redshift sample is similar to the low redshift sample, 
but the S/N is generally much higher in the $z>1.5$ sample than in the low
or mid $z$ samples

\subsection{Conditions for Comparing Various Samples}\label{sec-descbevol}
The definition of a BLA that was followed in recent papers presented by our group is an 
absorber that can be fitted with a single Gaussian component with $b > 40$  \km\ 
and for which the reduced-$\chi^2$ does not improve statistically by adding more components in the model. 
Low S/N data can have, however, treacherous effects that can mask a BLA or confuse narrow multi-absorbers
with a BLA (see Figure~2 in Richter et al. 2006b). To overcome the effects of noise,
the sample of BLAs defined by \citet{richter06} has also to satisfy the following rules: 
(i) the line does not show any asymmetry, (ii) the line is not blended, (iii) there is no 
evidence of multiple components in the profile; (iv) the S/N must be high enough 
(i.e. $\log (N_{\rm H I}/b) \ga \log [3 \times 10^{12}/({\rm S/N})] \ga 11.3 $, where
$N_{\rm H I}$ and $b$ are in cm$^{-2}$ and \km, respectively).
For the low redshift sample presented here, the BLAs strictly follow criteria
(iv) when the condition $\sigma_b/b, \sigma_N/N \le 0.4$ is set.. 
Following these rules, \citet{richter06}
found $d{\mathcal N}({\rm BLA})/dz = 22 \pm 5$ for their secure detections (53 for the entire candidate
sample). Within $1\sigma$, our BLA number density estimate overlaps with the result 
of \citet{richter06} when the cutoff  $\sigma_b/b, \sigma_N/N \le 0.4$ is applied to our sample. 
Therefore, this shows that by applying an error cutoff  without scrutinizing  
each profile for the conditions listed above, we find a similar average 
$d{\mathcal N}({\rm BLA})/dz$. 
This is crucial because for a comparison with other samples at
higher $z$, we can not examine each absorber individually. For the sample presented
by \citet{jank06}, no spectra or fits are shown. \citet{kim02} and \citet{kirkman97}
present their spectra and fits, but over too broad wavelength ranges to study in detail
the conditions listed above. Furthermore, as $z$ increases 
absorbers are more often blended due to the higher redshift line density, so we can not reject
the blended systems, otherwise we would introduce a strong bias in the comparison.
At high, mid, and low redshift, the  $\chi^2$ of the profile fit governs the number of components 
allowed in the model of an absorber; therefore a similar methodology was applied in each sample, 
allowing a direct comparison of the various measurements. 
We apply the same cutoff  $\sigma_b/b, \sigma_N/N \le 0.4$ to  the $z>0.5$
redshift samples to remove the uncertain profile fit results in a similar manner in 
each sample. Such cutoff, however, introduces a systematic effect:  more BLAs will be rejected
in low S/N spectra (low and some mid $z$ data) than in high S/N spectra (high $z$ data). 
But such systematics should underestimate the number of BLAs in low S/N data, and therefore
this should strengthen the differences observed at low $z$ compared to the 
higher $z$ samples. 

For our comparison, we also consider only absorbers with $\log N_{\rm H I} \ge 13.2$,
the completeness level of the low redshift sample, which also corresponds to about the 
completeness of the lowest S/N spectra of the mid-$z$ sample. If the S/N of the high
redshift spectra is solely considered, absorbers with $\log N_{\rm H I} \ge 13.2$
would be far above the completeness level of the high redshift sample, which is
$\log N_{\rm H I} \ga 12.5$ for the data presented by \citet{kim02} and \citet{kirkman97}. 
However, line blending and blanketing reduce the completeness threshold, especially
at $z\ga 2.5$ \citep{kirkman97,lu96}.  At $z\sim 4$, 
\citet{lu96} show using simulated spectra that line blanketing was not important as long as 
$\log N_{\rm H I} \ga 13.5$, while at $z\sim 2.7$, \citet{kirkman97} show following 
a similar methodology that their sample is likely to be complete  at 
$\log N_{\rm H I} \ga 12.8$--13.0. Since broad and shallow absorbers are more uncertain,
considering only absorbers  with $\log N_{\rm H I} \ge 13.2$ also reduces the problem of creating a higher 
proportion of wide lines than is really present in the intrinsic distribution. 

For our comparison, we only consider  BLAs with $40<b \le 100$ \km. The choice of $b\le 100$ \km\ 
reduces the incompleteness at the high $b$ end of the low redshift sample. Furthermore,
at $z\sim4$ (which is at higher redshift than any absorbers considered here), \citet{lu96}
argue that absorbers with $b>100$ \km\ are caused essentially by heavily blended
forest lines since they are systematically found in the high line density region
of the spectrum. This effect should diminish greatly as $z$ decreases. At $z \sim 2.7$, 
\citet{kirkman97} have produced simulated spectra to better understand the intrinsic properties
of their observational data. They observe a tail at high $b$ in both the simulated
and observed $b$ distributions. In the simulated spectra, the tail at $b>80$ \km\ is only 
due to line blending because their simulation did not allow such broad lines. They
find more lines at $b>80$ \km\  in the observed distribution (3.6\% compared to 0.5\%
in the simulated spectra), suggesting that many of these absorbers could 
be intrinsically broad. Refined simulations, so that simulated observations
match exactly the real observations, would be needed to be entirely conclusive
on this point \citep{kirkman97}. Yet,  at high $z$, if BLAs with $b>80$ \km\ that
are in fact blended narrow lines were frequent, this effect should
be more important as $z$ increases. We will see that this effect, if present,
is not statistically significant (see next two sections). By considering only absorbers 
with $40<b \le 100$ \km, $\log N_{\rm H I} \ge 13.2$, and  $\sigma_b/b, \sigma_N/N \le 0.4$, 
we reduce significantly the risk of including spurious broad absorbers. 
We also note that, in any samples considered, the majority of BLAs are actually 
found in the $b$-range  $40<b \le 60$ \km, not in the $b$-range $b> 80$ \km.

\begin{figure*}[tbp]
\epsscale{1} 
\plottwo{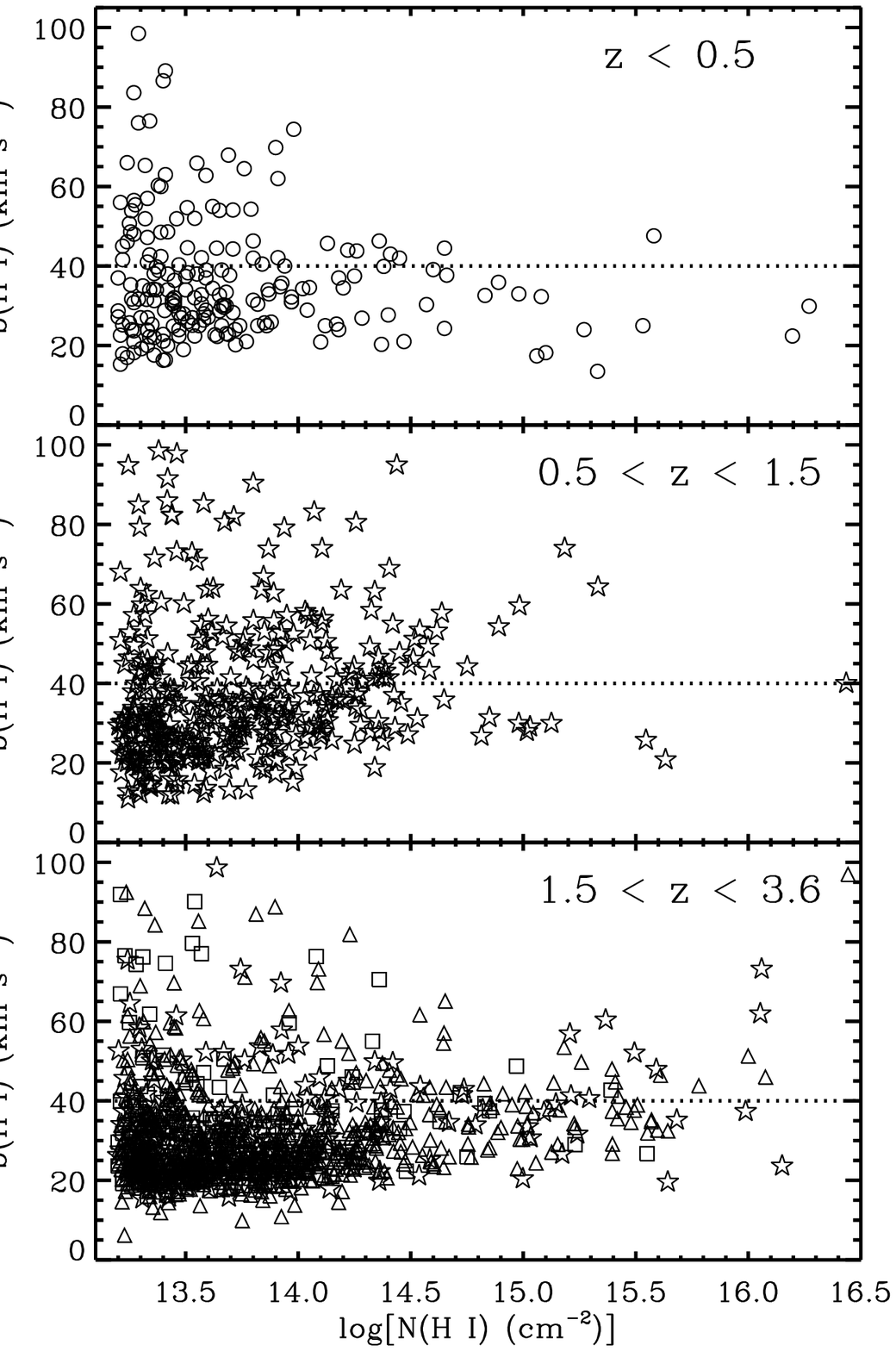}{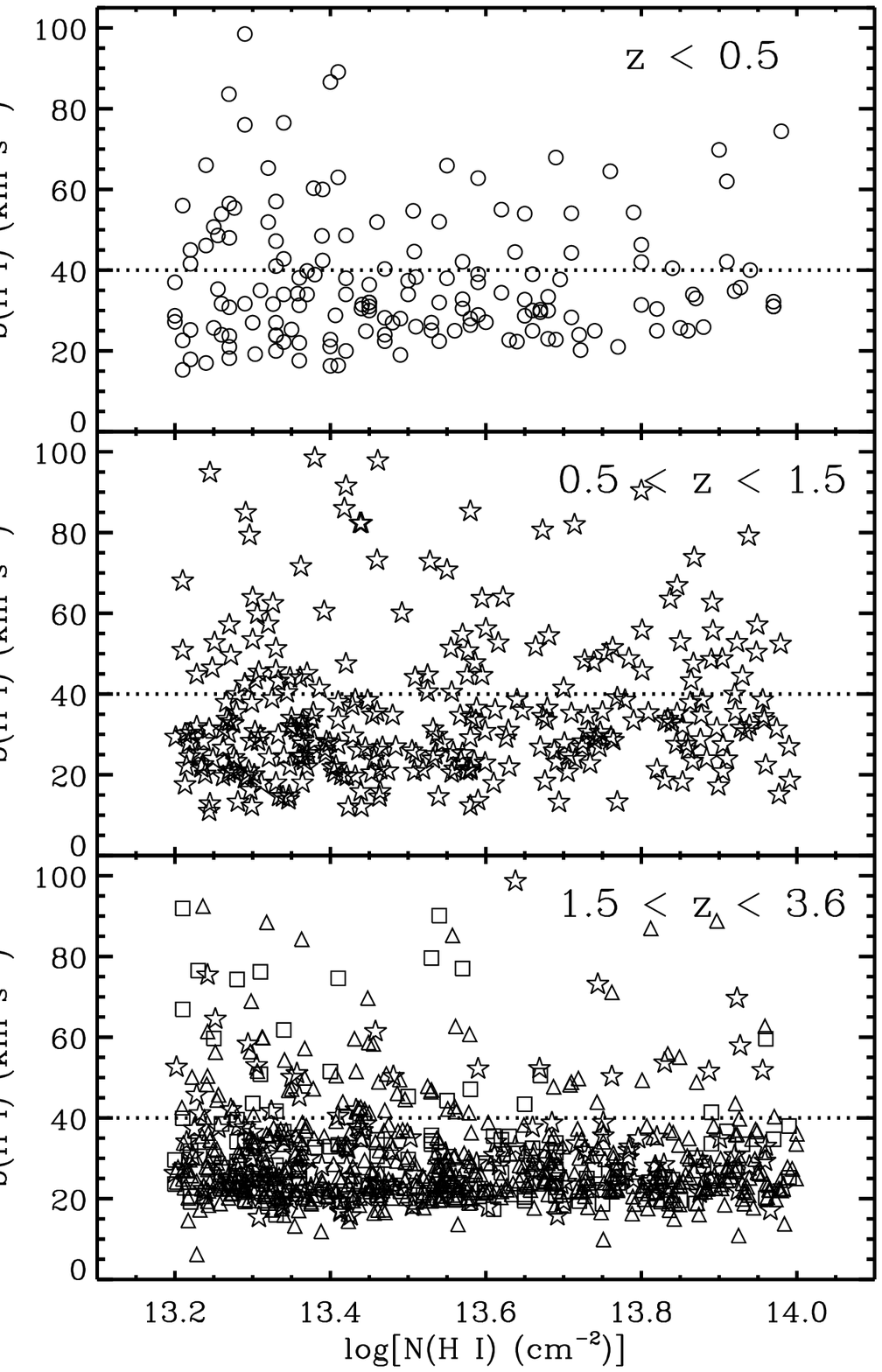}
\caption{{\em Left panel}: Line width versus \hi\ column density for the Ly$\alpha$ absorbers 
for $13.2 \le \log N_{\rm H I} \le 16.5$ and $0<b \le 100$ \km\  
in different redshift intervals indicated in the
upper-right corner. Only data with less than 40\% errors in $b$ and $N_{\rm H I}$
are considered.  The symbols have the following meaning: 
data represented by circles were estimated from the low redshift sample (see Table~\ref{t1} for the references); 
data represented by stars are from the sample of \citet{jank06};
data represented by triangles are from the sample of  Kim et al. (2002a); 
data represented by squares are from the sample of \citet{kirkman97}. 
{\em Right panel}: Same as the left-hand side, but only for absorbers with  
$13.2 \le \log N_{\rm H I} \le 14.0$ 
\label{bcomp2}} 
\end{figure*}

In Fig.~\ref{bcomp2}, we show the $b$--$N_{\rm H I}$ distribution
for three redshift intervals, $z<0.5$ (top panel), $0.5<z<1.5$ (middle panel), 
and $1.5 < z \la 3.6$ (bottom panel). Only absorbers that satisfy the conditions
$0<b \le 100$ \km,  $\log N_{\rm H I} \ge 13.2$,  $\sigma_b/b, \sigma_N/N \le 0.4$
are shown in the figure. 
At $z>0.5$, there are many weak systems with $b<40$ \km, which darkens part of the 
middle and bottom panels. At $z<0.5$, there are very few absorbers with $b\ga 40$ 
\km\ and $\log N_{\rm H I} > 14$, and none of them have $b>50$ \km. This contrasts
remarkably with what is observed at higher redshifts: many absorbers at $z > 0.5$
have $b>40$ \km\ and  $\log N_{\rm H I} \ge 14.0$. 
This effect must be due to strong, saturated Ly$\alpha$ lines for which the errors for $b$ and $\log N_{\rm H I}$
are far too optimistic. At $z<0.5$,  saturation can be dealt with because 
higher Lyman series lines are systematically used in the profile  fitting, reducing the possibility 
of non-uniqueness in the profile fitting solution and of finding unrealistic large $b$-values for strong 
lines.  At $z>0.5$, Ly$\alpha$ is generally the only transition available, although 
we note  \citet{kim02} analyzed  the Ly$\beta$ forest at $z<2.5$ with a small 
sample and their results suggested that line blending and saturation are not an important issue.
Yet,  Fig.~\ref{bcomp2} shows a system with $b = 97.0 \pm 3.8 $ \km\ and 
$\log N_{\rm H I} = 16.44 \pm 0.14$. This absorber is observed in the spectrum of J\,2233--606 
at $z=1.869$, and the parameters that are presented here were estimated by \citet{kim02}. 
The spectrum of J\,2233--606 is shown in \citet{cristiani00} and at the wavelength corresponding 
to this absorber, there is a strong, saturated line that \citet{kim02} model with a single 
component. \citet{dodo01} actually found that this absorber has $\sim$11--16 components
using higher Lyman series lines and metal lines. While this example is extreme, most of the
absorbers with $\log N_{\rm H I} > 14$ and $b > 40$ \km\ are likely to be strong 
saturated Ly$\alpha$ lines at $z>0.5$.  Fig.~1 in \citet{kirkman97}, which shows the whole HS\,1946+7658 spectrum 
with  the values of $\log N_{\rm H I}, b$, corroborates this conclusion. The strong, saturated
Ly$\alpha$ lines are unlikely to probe a single broad line, but are likely composed of several 
unresolved components.  Therefore, for our comparison, 
we will consider absorbers with $13.2 \le \log N_{\rm H I} \le 14.0$. This sample is 
summarized in the right-hand side of Fig.~\ref{bcomp2}.  

As we have just illustrated, the Voigt profile fitting method adopted for deriving $z,b,N_{\rm H I}$  
may not be unique, especially for highly blended regions, strong lines, and low S/N spectra.
We believe that some of this effect is reduced by considering absorbers with 
$13.2 \le \log N_{\rm H I} \le 14.0$, $b<100$ \km,  and  $\sigma_b/b, \sigma_N/N \le 0.4$. 
At low $z$, line blending and strong lines are not a major issue because often 
higher Lyman series lines can be used. Futhermore, several persons analyzed these data independently, 
and other methods (curve-of-growth and/or optical depth method) were also used 
for the low-$z$  sample, yielding consistent results (see the references 
given in Table~\ref{t1}), although the Appendix highlights some differences between 
various groups. Even in this latter case, there is roughly a 1$\sigma$ consistency
between the various results for absorbers with $13.2 \le \log N_{\rm H I} \le 14.0$.
 At $z \sim 2.7$, \citet{kirkman97} show that profile fitting
may not provide a unique solution even in very high S/N data. Yet, their comparison with simulated
spectra show that the intrinsic distribution of $b$ is very similar to the observed
$b$ distribution. Therefore, in a statistical sense, the observed $b$ distribution at high $z$ 
can be used for our comparison. 
We also note that three different research groups have worked on the 
absorbers at $z\ga 1.5$ presented here, with the spectra obtained from different telescopes with
different S/N, and we did not find any major differences between their results, at least within
the conditions listed above.

Continuum placement may also be a problem for finding or defining a BLA. At low $z$, BLAs may be 
confused with continuum undulations. With our constraints, most of these ill-defined BLAs are rejected 
\citep[see also][]{richter06}. We believe that
this conclusion should apply to the mid-$z$ sample, but since \citet{jank06} did not present any of the 
spectra they analyze, we treat this sample with more caution 
(see also below). At $z>1.5$, the continuum is more difficult  because each 
order must be fitted with a high order Legendre polynomial, and during the Voigt profile fitting
process, the continuum level is often adjusted to give an optimum fit between the data and the calculated 
Voigt model. Furthermore, as $z$ increases the continuum placement becomes more difficult
because the line density increases, and therefore the line blending and blanketing effects 
increase. High order polynomials and adjusted continua can therefore mask broad and shallow BLAs. 
However, such effects may be counterbalanced by a possible spurious increase of BLAs caused by the 
same line blending and blanketing effects (see above). 
The high S/N of the data at high redshifts and considering only the absorbers with 
$13.2 \le \log N_{\rm H I} \le 14.0$ and  $b \le 100$ \km\ reduce greatly the risk of losing 
many BLAs because of the line blending and blanketing at high $z$. The problem of not finding BLAs is 
potentially more important in the highest $z$ spectra, but we do not notice significant 
differences between $z\sim 2$ and $z>3$.  We also note above that simulated spectra show that line blanketing 
was not important at $\log N_{\rm H I} \ga 13.2$. Therefore, 
for the samples considered here using the limits $\log N_{\rm H I} \ge 13.2$ and 
$b \le 100$ \km, they must almost be complete. Ultimately, one would like to produce 
simulated spectra of various $z$ with realistic inputs to fully consider the
impacts of S/N, continuum placement, and line blending and blanketing on finding BLAs 
and deriving their intrinsic properties. 

\begin{deluxetable*}{lcccccc}
\tabletypesize{\scriptsize}
\tablecaption{Evolution of $b$  \label{t3}} 
\tabcolsep=7pt
\tablecolumns{7}
\tablewidth{0pt} 
\tablehead{\colhead{} & \colhead{$z < 0.5$}  &  \colhead{$0.5 < z < 1.0$}&  \colhead{$1.0 < z < 1.5$}&  \colhead{$1.5 < z < 2.0$}&  \colhead{$1.5 < z < 3.6$}  &  \colhead{$2.4 < z < 3.1$} \\
		\colhead{} & \colhead{(this paper)} &  \colhead{(J06)} &  \colhead{(J06)} &  \colhead{(J06)}&  \colhead{(K02)} &  \colhead{(KT97)}   }
\startdata
\cutinhead{$0 <b \le 100$ \km}
$b$ median ($m_{\rm tot}$)  &  30.5 (336)  	    &  27.9 (263)	  & 28.3 (588) 		&   28.0 (450)        & 24.8 (2305)	    &	27.1 (452)	    \\
$b$ mean\,$\pm \sigma$      &  $  34.0 \pm 16.9$&  $  32.9 \pm 17.8$  & $31.8 \pm 17.0$	&   $31.4 \pm 14.5$   & $27.2 \pm 13.5$     &	$31.2 \pm 16.2$ 	    \\
$m(b>40)/m_{\rm tot}$   & 0.277		    & 0.274		  &  0.240		&    0.204	      &  0.131 	    &	 0.210 		    \\
\cutinhead{$0 <b\le 100$  \km\ and $0 < (\sigma_b/b, \sigma_N/N) \le 0.4$ and $13.2 \le \log N_{\rm H I} \le 14.0 $}
$b$ median ($m_{\rm tot}$)  &  32.7 (162)  	    &  31.5 (82)	  & 30.8 (209) 		&   28.2 (154)       &  25.5 (509)	   &   27.1 (123)		       \\
$b$ mean\,$\pm \sigma$      &  $ 37.4 \pm 16.1$ &  $ 36.0 \pm 18.3$  & $35.5 \pm 17.3$	&   $31.7 \pm 12.9$  &  $29.2 \pm 11.9$    &   $32.5 \pm 16.0$  	       \\
$m(b>40)/m_{\rm tot}$   & 0.321		    & 0.317		  &  0.272		&    0.169	     &   0.132	 	   &	0.179			      \\
\enddata		

\tablecomments{Median and mean values of $b$ are listed for the different
redshift intervals for the entire samples  with $0 <b \le 100$ \km\ and 
the higher quality samples restricted to  absorbers with $13.2 \le \log N_{\rm H I} \le 14.0 $.
$m_{\rm tot}$ is the total number of absorbers in the sample.
$m(b>40)$ is the total number of absorbers with $40 < b \le 100$ \km\ in the sample.
The values of $\sigma$ listed are the standard deviation around the mean.
Source of profile fitting measurements: J06: \citet{jank06}; K02: \citet{kim02}; KT97: \citet{kirkman97}.
}
\end{deluxetable*}

The redshift range $z=0.5$--1.5 is more problematic than the other redshift ranges 
because most of the data have low S/N, all have lower resolution than the higher 
or lower redshift spectra, and no spectra or fits were presented by \citet{jank06}
making it more difficult to assess some of the issues discussed above. Ly$\beta$
lines are only available for about half the wavelength coverage of this redshift range and
lower spectral resolution and S/N of  these data further worsen the problems
of saturation and non-uniqueness in the profile fitting results.
Setting the conditions  $0<b \le 100$ \km, $13.2 \le \log N_{\rm H I} \le 14.0$, 
and $\sigma_b/b, \sigma_N/N \le 0.4$ make this sample stronger, but we nonetheless treat
the results from this sample with caution. We will consider two sub-samples 
for the estimate of the redshift density based on the \citet{jank06} sample (see \S\ref{sec-bdndz}): one with all
their data, and one with only their highest S/N data (i.e. PG\,1634+706, HE\,0515--4144, 
HE\,0141--3932, HE\,2225--2258). Among the high S/N data in the mid-$z$ sample are E230M spectra of
PG\,1634+706 and  HE\,0515--4144.

For the reasons aforementioned, with the conditions
$0<b \le 100$ \km, $13.2 \le \log N_{\rm H I} \le 14.0$, and
$\sigma_b/b, \sigma_N/N \le 0.4$,  we greatly reduce some pitfalls in comparing
data analyzed by various groups, with different S/N,
obtained with various instruments, with different line blending issues.
A BLA is therefore defined here as an absorber that is fitted with a single Gaussian for 
which the $\chi^2$ does not drop significantly by adding more components
(this condition was adopted by the various groups who analyzed the data
used here) and that has $40 \le b \le 100$ \km, $13.2 \le \log N_{\rm H I} \le 14.0$,
and $\sigma_b/b, \sigma_N/N \le 0.4$. We believe that we are statistically
comparing the intrinsic properties of the broadening of \hi\ absorbers, although
we note that simulated spectra probing various $z$, with realistic inputs may be
the only way to fully understand some of the effects discussed above and
how these effects balance each other at various $z$.

\subsection{Comparison of the Distributions of $b$ at Low and Higher $z$}\label{sec-bevold}

Table~\ref{t3} summarizes the $b$-value median, mean, dispersion, and fraction
of BLAs (i.e. systems with $40 <b \le 100$ \km). In this table, we consider two sub-samples
with $0<b \le 100$ \km: (a) the entire sample, (b) the higher quality sample with 
$13.2 \le \log N_{\rm H I} \le 14.0$ and errors in $b$ and $N_{\rm H I}$ less than 40\%. 
We note that the number of systems in each sample is roughly similar, except for the \citet{kim02} sample,
which is noticeably larger, and the $0.5<z<1.0$ sample, which is somewhat smaller. In sample (a) there
appears to be an increase in the median, mean, and the fraction of BLAs as $z$ decreases.
The same trend is observed in sample (b), but the differences are better revealed, where
the median and mean are always larger by 15--30\% in the low redshift sample ($z<0.5$) 
than in the high redshift sample ($z>1.5$).  
For the sample (b), the fraction of BLAs at $z\la 0.4$ is 1.9--2.4 times 
larger than the fraction of BLAs at  $z \ga 1.5$. 
Therefore, $b$ is larger on average at $z\la 1.0$ than at $z\ga 1.5$, and the fraction of BLAs is larger 
at $z<1.5$ than at $z>1.5$. 

\begin{figure}[tbp]
\epsscale{1} 
\plotone{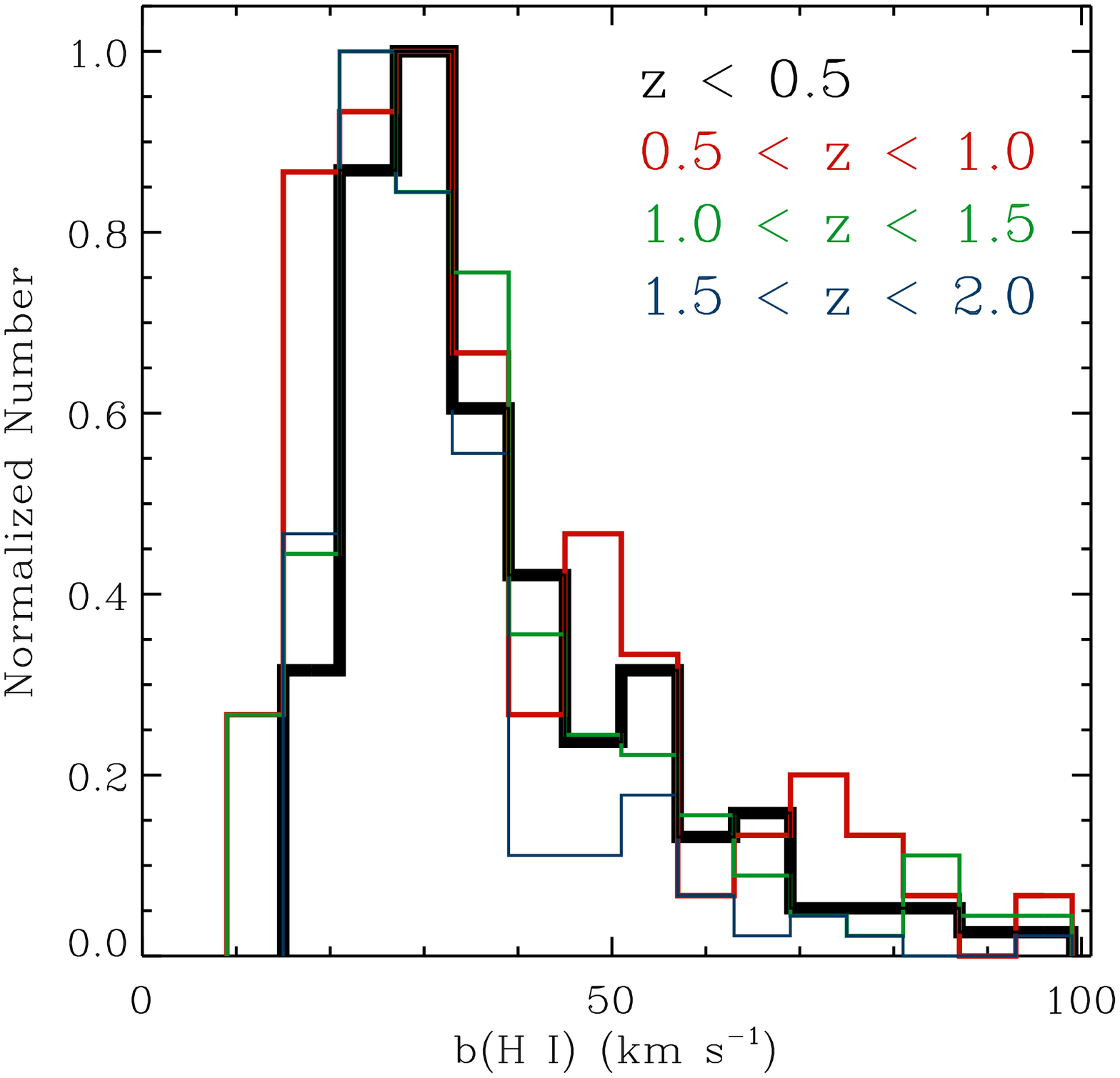}
\caption{Comparison of the normalized distributions of $b$(\hi) for the low $z$
sample  (black histogram) and for the higher $z$
sample (colored  histograms) of \citet{jank06} for the weak absorbers $13.2 \le \log N_{\rm HI} \le 14.0$. 
The sample in the redshift range $z\simeq0.5$--1.5 has a spectral resolution of 
about 10 \km, while the other samples have  about 7--8~\km.
Only data with less than 40\% errors in $b$ and $N_{\rm H I}$
are considered.
\label{bzm}}
\end{figure}

\begin{figure*}[tbp]
\epsscale{1} 
\plottwo{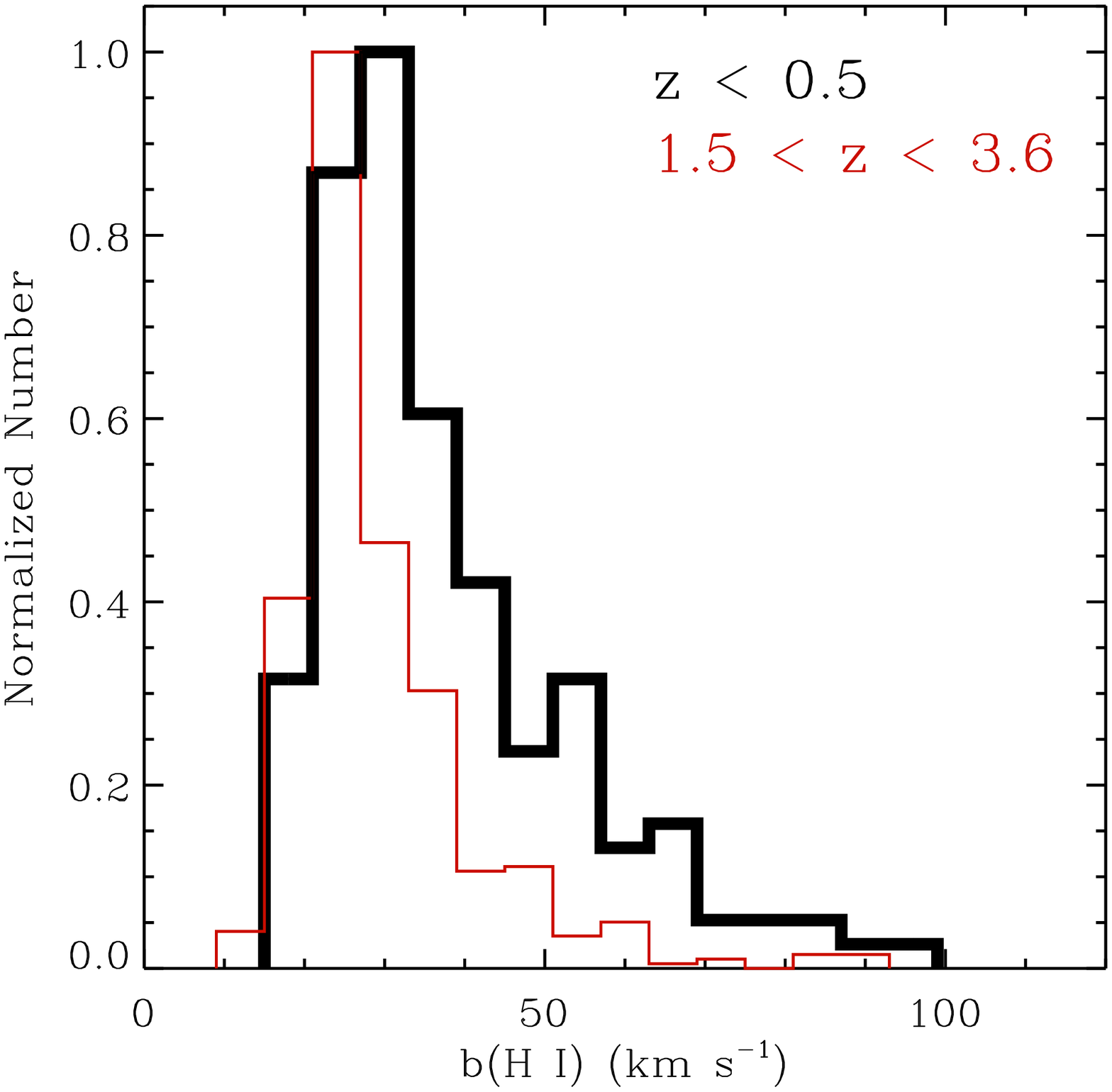}{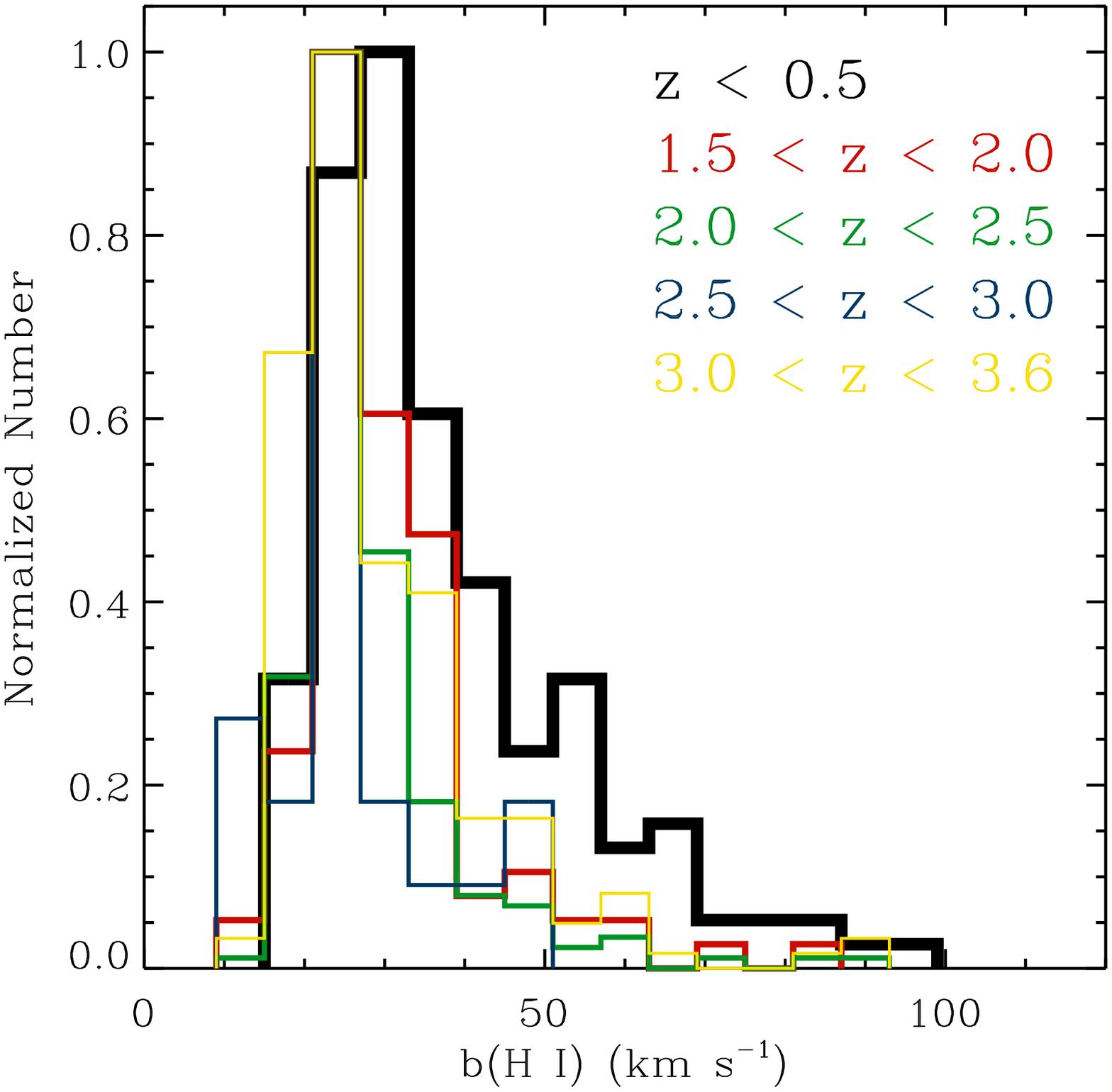}
\caption{{\em Left panel}: Comparison of the normalized distributions of $b$(\hi) for the low $z$
sample of data points  (black histogram) and for the
high redshift sample (red histogram) from \citet{kim02}  for the weak absorbers $13.2 \le \log N_{\rm HI} \le 14.0$.
{\em Right panel}: Same as the left panel but the sample of \citet{kim02} is 
separated in various redshift intervals.
 Each sample has a similar spectral resolution of about 7--8~\km.
Only data with less than 40\% errors in $b$ and $N_{\rm H I}$
are considered for the different samples.
\label{bz}}
\end{figure*}

\begin{figure}[tbp]
\epsscale{1} 
\plotone{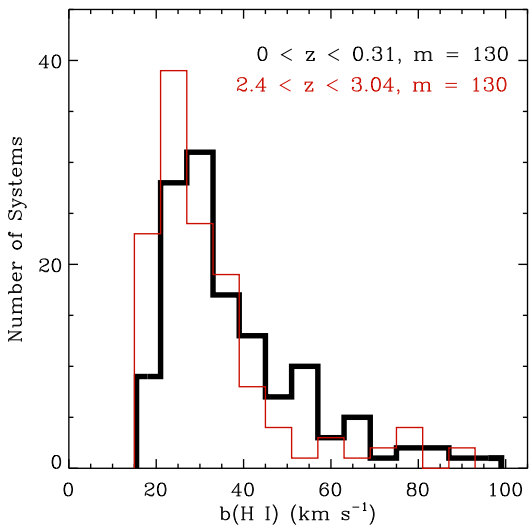}
\caption{Comparison of the distributions of $b$(\hi) for the low $z$ STIS E140M
sample  (black histogram) and for the high $z$ Keck sample  (red histogram) of \citet{kirkman97}
 for the weak absorbers $13.2 \le \log N_{\rm HI} \le 14.0$. The number of absorbers in each sample
 is the same: $m = 130$.
Only data with less than 40\% errors in $b$ and $N_{\rm H I}$
are considered for both samples. 
\label{bz2}}
\end{figure}

\begin{figure}[tbp]
\epsscale{1} 
\plotone{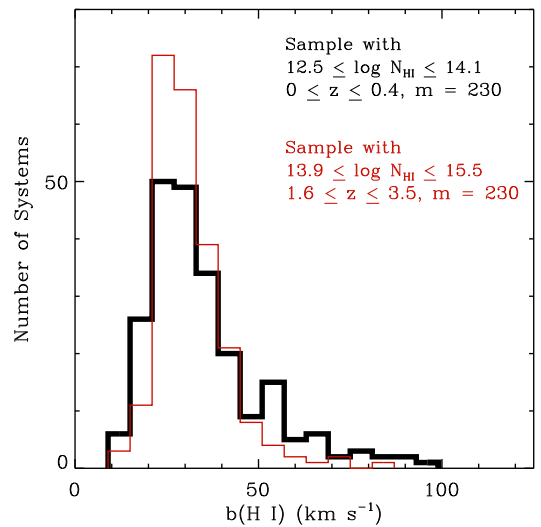}
\caption{Comparison of the distribution of $b$(\hi) for the low $z$ STIS E140M
sample of data points  (black histogram) and for the
high redshift sample  (red histogram). The samples cover different ranges in $\log N_{\rm H I}$.
Higher column density systems at high redshift
are expected to be physically analogous to the low density systems
at low redshift in an expanding universe.  But note that in the high redshift
sample, BLAs are often likely to be blends of narrower lines (see \S4.2). 
The high redshift data are from \citet{kim02}. The number of absorbers in each sample
 is the same: $m = 230$.
Only data with less than 40\% errors in $b$ and $N_{\rm H I}$
are considered in both samples.
\label{bz3}}
\end{figure}

In Figs.~\ref{bzm}--\ref{bz3} we compare the distributions of $b($\hi). In all comparisons 
we restrict the observations to the higher quality measurements with less than
40\% errors in $b$ and $N_{\rm H I}$.

In Fig.~\ref{bzm}, we compare the normalized number distribution of $b$(\hi) for the 
low redshift sample with the normalized number distribution of $b$(\hi) for 
the \citet{jank06} sample, which is subdivided in 3 $z$ sub-samples. The distributions
peak at about $b \sim 20$--30 \km. Each distribution shows a tail at 
higher $b$. The tail in the $1.5 < z < 2.0$ distribution is always weaker
than in the lower $z$-interval distributions, especially for $40<b<55$ \km. 
The  tail in the $0.5 < z < 1.0$ distribution is generally stronger than any
other $z$-interval distribution presented in this figure, which is possibly due to a combination 
of blending effects, low S/N spectra, and lower resolution spectra in the redshift 
range $0.5 < z < 1.0$.  We note that data probing the redshift ranges  $0.5 < z < 1.0$ 
and  $1.0<z<1.5$ were obtained with the STIS E230M grating, but the sensitivity of E230M is lower 
for Ly$\alpha$ in the redshift range $0.5 \la z \la 0.9$ than  $0.9 \la z \la 1.4$. 
A Kolmogorov-Smirnov test does not reveal a  significant difference between the samples $z<0.5$ and $0.5 < z< 1.5$ 
(the maximum deviation between the two cumulative distributions is $D = 0.118$ with 
a level of significance  $P = 0.104$), but suggests a difference between the samples 
$z<0.5$ and $1.5 < z< 2.0$. Therefore,  BLAs appear more frequent at $z<1.5$ than at $z>1.5$.

In Fig.~\ref{bz}, we compare the normalized number distribution of $b$(\hi) for the low redshift sample 
with the normalized number distribution of $b$(\hi) for the high redshift sample ($1.5 \le  z \le 3.6 $)
from Kim et al. (2002a). The left panel shows the entire sample at $1.5 \le  z \le 3.6 $, 
while the right panel shows the distribution of $b$ in various redshift intervals for
the high $z$ sample. The peak of the high redshift normalized distribution in
Fig.~\ref{bz} is not only shifted by $\sim -5$ \km\ with respect to the peak of 
the low redshift distribution, but also the width of the distribution is smaller for
the high redshift sample. Moreover, the low redshift sample shows a tail of high 
$b$(\hi) absorbers that is much weaker in the high redshift sample, showing that BLAs are more frequent 
at $z<0.5$ than at $z>1.5$.  
The right-hand side of Fig.~\ref{bz} verifies this conclusion for the various redshift intervals, 
i.e.  BLAs appear more frequent at $z<0.5$ than for any redshift intervals at $z>1.5$.
There is, however, no major difference between the various redshift intervals at $z>1.5$. 
We also note  that the $b$ distribution at $z>1.5$  shows little effect
of line blending as $z$ increases since there is scant evidence of a larger of fraction 
of BLAs at $z\ga 3$ than at $1.5 \la z\la 2$. 

We compare in Fig.~\ref{bz2} our low $z$ sample to the high redshift absorbers observed 
toward the QSO HS\,1946+7658 \citep{kirkman97}. The median and mean are somewhat intermediate
between our sample and the \citet{kim02} sample (see Table~\ref{t3}). We slightly 
adjusted the redshift of the low $z$ sample so that
the low and high redshift samples have exactly the same number of systems ($m=130$);
hence the two distributions can be directly compared.  
The peak of the high redshift distribution in
Fig.~\ref{bz} is again shifted by $\sim -5$ \km\ with respect to the peak of 
the low $z$ distribution.
A higher number of systems with $40< b \le 70$ \km\ is found in the low redshift sample. 
At $b>70$ \km\ the high redshift sample appears to have a larger number of systems. 
However, we note that the data of \citet{kirkman97} have S/N\, up to 200, and 
as we discussed in \S\ref{datasample} our sample is not complete for $b > 80$ \km. 
A Kolmogorov-Smirnov test yields a maximum deviation between the two 
cumulative distributions of the low and high redshift samples $D = 0.210$ with a level of significance  $P = 0.006$;
the null hypothesis of no difference between the two datasets is therefore rejected.

So far, we have ignored the effects of evolution of column density in the absorbers in 
our comparison. According to the numerical simulation of \citet{dave99}, 
the dynamical state of an absorber depends mainly
on its overdensity, so that a column density range traces
absorbers of progressively higher overdensity as the universe expands.
Therefore, according to Eq.~\ref{e-nrho}, 
a low redshift  \hi\ absorber is physically analogous to a high redshift \hi\ absorber 
not with the same column density but to an absorber with column 
density $10^{0.4\times (z_{\rm high} - z_{\rm low})/0.7}$ times higher \citep{dave99}. 
With the average redshifts of 2.6 and 0.2 of the high redshift sample
of \citet{kim02} and our sample, this corresponds to a column density
roughly 25 times higher, i.e. a sample with $12.5 \le \log N_{\rm H I} \le 14.1$ 
at low $z$ is physically analogous to the  high redshift sample with 
$13.9 \le \log N_{\rm H I} \le 15.5$. At such high column density, 
BLAs are likely to trace in large part narrower, strong absorbers that
are blended together (see \S\ref{sec-descbevol}). We, nonetheless, make this comparison
in Fig.~\ref{bz3}, where both samples have the same number
of systems.   The high redshift sample has again a larger fraction of systems
in the range $b=[20,40]$ \km\ than the low redshift sample. 
The fraction of BLAs at low redshift is again larger than the fraction 
of BLAs at high redshift with $b > 50$ \km. This is remarkable since 
the low redshift sample is far from complete for $\log N_{\rm H I} \la 13.2$
and the high redshift sample likely overestimates the number of true BLAs (i.e. BLAs 
that are not blends of NLAs).  We again apply  a Kolmogorov-Smirnov test 
on the two cumulative distributions of the low and high redshift samples and find 
$D = 0.130$ with a level of significance  $P = 0.029$; again 
the null hypothesis of no difference between the two datasets is rejected.

\subsection{Evolution of $d{\mathcal N}/dz$ as a Function of $b$}\label{sec-bdndz}
\begin{figure*}[tbp]
\epsscale{1} 
\plottwo{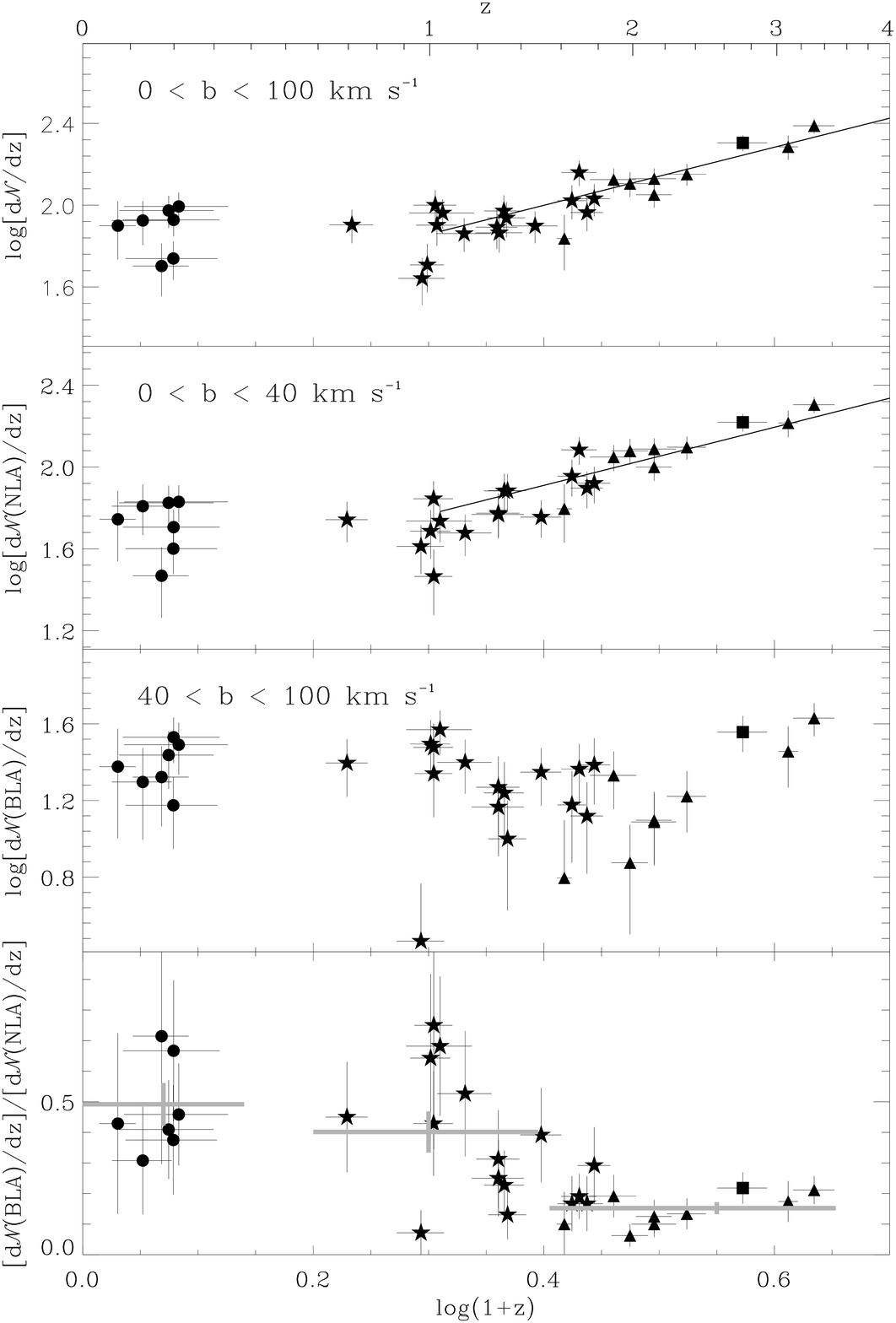}{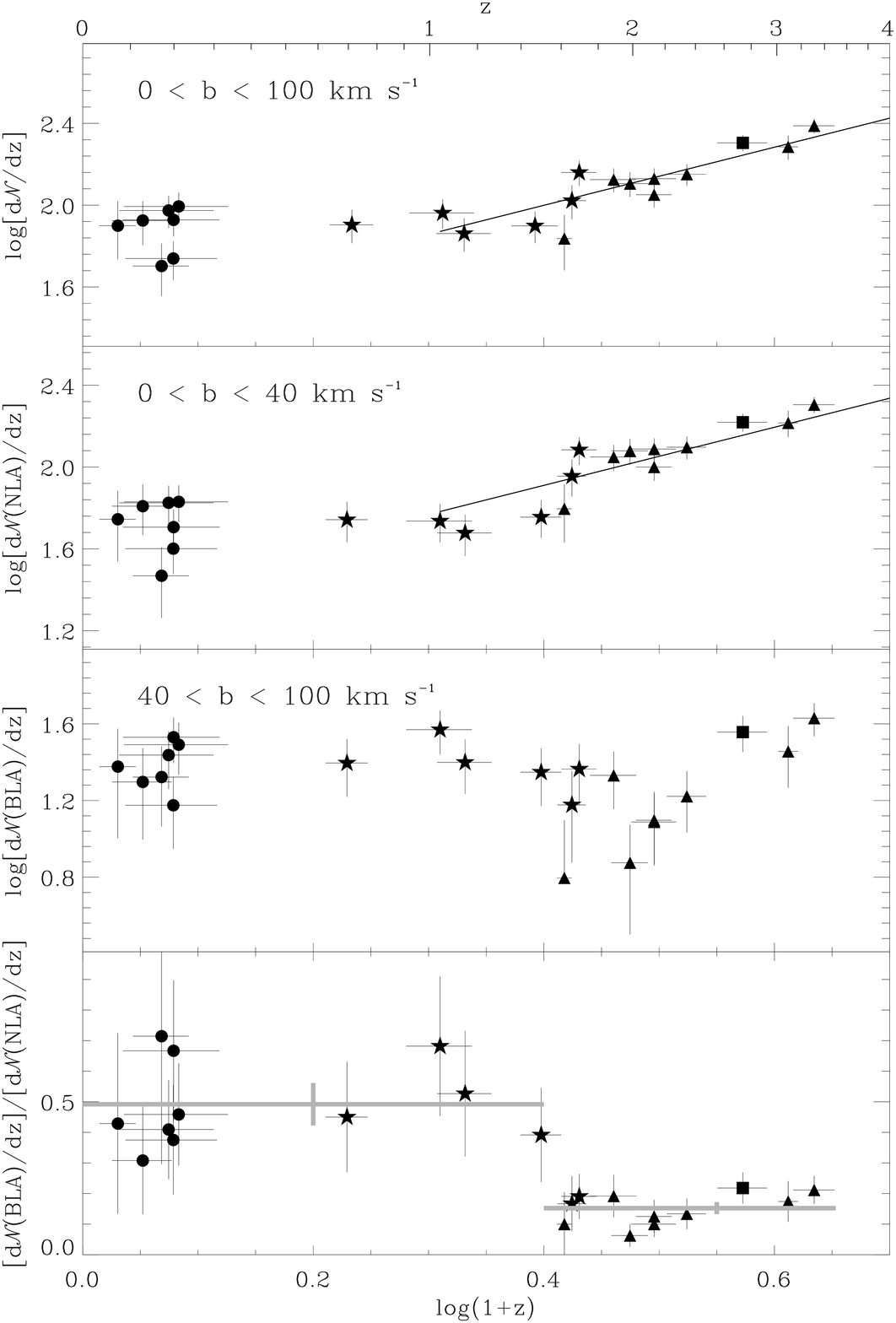}
\caption{The Ly$\alpha$ redshift density of the entire sample ($0<b \le 100$ \km),
the NLAs ($b< 40$ \km), the BLAs ($40 \le b \le 100$ \km), and the ratio 
of the Ly$\alpha$ redshift density fraction of the BLAs to the NLAs (bottom panel). 
We only consider absorbers with $13.2 \le \log N_{\rm H I} \le 14.0$ and with errors
less than 40\% in $b$ and $N_{\rm H I}$. In the top two panels, 
the solid line shows $d{\mathcal N}/dz \propto (1+z)^{1.42}$ adopted from \citet{kim02}.
The solid gray crosses 
in the bottom panels are the mean of $(d{\mathcal N}({\rm BLA})/dz)/(d{\mathcal N}({\rm NLA})/dz)$
where the horizontal bars indicate the redshift range over which the mean is estimated. 
The vertical bars are Poissonian  errors. The symbols have the following meaning: 
data represented by circles are estimated from the low redshift sample (see Table~\ref{t1} for the references); 
data represented by stars are estimated from the sample of \citet{jank06};
data represented by triangles are estimated from the sample of  \citet{kim02}; 
data represented by the square are estimated from the sample of \citet{kirkman97}. 
The left panel considers all the data with the conditions listed above from these various samples,
while in the right panel we remove the lowest quality measurements of \citet{jank06} (see \S4.2). 
\label{dndzcomp}}
\end{figure*}

In Fig.~\ref{dndzcomp}, we show $d{\mathcal N}/dz$ ($0<b\le 100$ \km),
$d{\mathcal N}({\rm NLA})/dz$ ($b \le 40$ \km),
$d{\mathcal N}({\rm BLA})/dz$ ($40 < b\le 100$ \km), and
$(d{\mathcal N}({\rm BLA})/dz)/(d{\mathcal N}({\rm NLA})/dz)$
as a function of the redshift. Only systems with 
$13.2 \le \log N_{\rm H I} \le 14.0$  and $\sigma_b/b, \sigma_N/N \le 0.4$
are considered in this figure.  $d{\mathcal N}/dz$ was estimated for
each sightline over the redshift interval available along a given
sightline.  The redshift at which  $d{\mathcal N}/dz$ is plotted in Fig.~\ref{dndzcomp}
corresponds to the mean redshift interval in a given sightline. The vertical bars are Poissonian 
errors, while the horizontal bars represent the standard deviation
around the mean of the observed redshifts of the Ly$\alpha$ absorbers. 
For the sample of \citet{jank06}, several lines of sight have a redshift path larger than 0.6 and 
we divided those in two redshift intervals. The left diagram in Fig.~\ref{dndzcomp} 
includes all the sightlines available in the various samples, 
while the right diagram includes all the sightlines available in  the
low $z$ sample and high $z$ samples of \citet{kim02} and \citet{kirkman97}, 
but only the highest quality data in the \citet{jank06} sample 
(see \S\ref{sec-descbevol}).

The top two panels of Fig.~\ref{dndzcomp} show the usual number density
evolution with little or no evolution between redshifts 0 and $\sim$1.0--1.6, 
and an evolution of $d{\mathcal N}/dz$ with $z$ at higher 
redshift (see for example Kim et al. 2002a). At high redshift,
$d{\mathcal N}/dz$ decreases with decreasing $z$ 
($d{\mathcal N}/dz \propto (1+z)^\gamma$, $\gamma>0$)
according to the expansion of the universe, which forces 
any initial baryon overdensity to thin out \citep[e.g.,][ and references therein]{dave99}. 
The expansion  also results in a decrease of recombinations of the free electrons 
with the protons and thus in an additional decline in $d{\mathcal N}/dz$. The break in 
$d{\mathcal N}/dz$ near $z\sim 1.6$ ($\log (1+z) \sim 0.4$)
is believed to be primarily caused by the 
drop in the UV background  because of the declining 
quasar population \citep{theuns98,dave99}. In the top panels, 
the solid line shows $d{\mathcal N}/dz = (d{\mathcal N}/dz)_0 (1+z)^\gamma$, where 
$\gamma = 1.42$ was adopted from \citet{kim02} who find this value for 
the weak absorbers with  $13.1\le \log N_{\rm H I} \le 14.0$. The value of
$(d{\mathcal N}/dz)_0$ was adjusted to match the data. For the entire sample ($0<b\le 100$ \km)
or the sample restricted to $b<40$ \km, this line represents well the 
evolution of $d{\mathcal N}/dz$ at $z> 1$, but we note that when the best 
quality data are considered (right-hand side panel), the break in the evolution 
appears to occur at $z\sim 1.6$. 

The second panels from the bottom in Fig.~\ref{dndzcomp} show that $d{\mathcal N}({\rm BLA})/dz$ is 
generally similar or higher at low and mid redshift than between redshifts 1.6 and 2. At 
$z\ga 2$, $d{\mathcal N}({\rm BLA})/dz$ increases and is larger at $z>2.5$
than $d{\mathcal N}({\rm BLA})/dz$ at $z<1.5$. The Hubble 
expansion must be the primary driver in the evolution of 
these structures at high redshift, and despite the expansion
the BLA number density is comparable to $d{\mathcal N}({\rm BLA})/dz$ at high $z$,
which contrasts remarkably from the evolution of $d{\mathcal N}({\rm NLA})/dz$ from 
low to high $z$.  This difference is even more striking when we consider  
the ratio ${\mathcal R} \equiv (d{\mathcal N}({\rm BLA})/dz)/(d{\mathcal N}({\rm NLA})/dz)$ (bottom panels),
which is higher in the low redshift sample than in the high redshift
sample. On the left-hand side diagram, at mid redshift there is a very large 
scatter in the redshift range $0.9\la z \la 1.2$. This scatter clears up 
when only the best quality data of the mid-$z$ sample are considered (see right-hand side
diagram). 
The distribution of ${\mathcal R}$ between redshifts $\sim$1.6 to 3.5 is actually nearly flat, 
showing that the slope controlling the evolution of $d{\mathcal N}({\rm NLA})/dz$ and 
$d{\mathcal N}({\rm BLA})/dz$ at $z\ga 1.6$ must be about the same, 
further strengthening that NLAs and BLAs must follow a similar evolution dictated by the 
expansion of the universe.  We note that ${\mathcal R}(z>2.5)$ is slightly larger than  ${\mathcal R}(z\sim2)$, 
possibly indicating some spurious increase of BLAs possibly due to line blending and blanketing effects
that are more serious at $z>2.5$ than at $z\sim 2$. However, in the view of the large scatter in 
the various samples, this result does not appear statistically significant. 
The grey crosses in the bottom panels show the average ${\mathcal R}$ over the redshift 
range depicted by the horizontal grey bar (the vertical grey bar assumes Poissonian errors). 
At $z\la 1.6$, ${\mathcal R}$ is larger than at $z>1.6$  as both 
the individual ${\mathcal R}$ and average ${\mathcal R}$ show. 
${\mathcal R}$ is a factor 3.2 higher at $z\la 0.4$ than at $z\ga 1.6$. 
When only the best data of \citet{jank06} are considered, ${\mathcal R}$ at mid-redshift 
($0.6 \la z \la 1.5$) appears similar to the value observed at low redshift.  

If we had considered 
systems with $b\le 80$ \km, the same conclusions would be drawn. If 
we had considered a cutoff for the NLAs of $b=50$ or 60 \km\ instead of 40 \km, similar conclusions 
would also be drawn.  If absorbers with $13.2 \le \log N_{\rm H I} \le 16.5$
are considered, we find that ${\mathcal R}(z<0.5) \sim 3 {\mathcal R}(z>1.6)$, despite
the fact many BLAs at high $z$ may actually be blends of narrower lines (see \S\ref{sec-descbevol}).

\subsection{Implications}

We find that (1) the distribution of $b$ for the BLAs has a 
distinctly more prominent high velocity tail at low and mid $z$ than at high $z$;
(2) the median and mean $b$-values at low and mid $z$ are systematically higher than 
at high $z$; and (3) the ratio $(d{\mathcal N}({\rm BLA})/dz)/(d{\mathcal N}({\rm NLA})/dz)$ 
at low and mid $z$ is a factor $\sim$3 higher than at high $z$. These conclusions 
hold for a division $b=40$, 50, or 60 \km\ between NLAs and BLAs.
For the reasons discussed in \S\ref{sec-descbevol}, we believe that these
are intrinsic properties of the evolution of the physical state of the gas that 
are not caused by comparing data with different S/N and line blending and blanketing
effects. However, simulated spectra probing various $z$, with realistic inputs would 
certainly help to unravel how exactly some of these issues (e.g., continuum placement, 
line blending effect, different S/N ratios) balance each other.
Our results strongly suggest that if the broadening is mostly thermal, 
a larger fraction of the low $z$ universe is hotter than the high $z$ universe,
and if the broadening is mostly non-thermal, the low $z$ universe is more kinematically 
disturbed than the high $z$ universe. It is likely that both possibilities are true.

\section{The Differential Column Density Distribution Function}\label{dddf}
The differential column density distribution $f(N_{\rm H I})$ is defined such that $f(N_{\rm H I},X) dXdN_{\rm H I}$
is the number of absorption systems with column density between $N_{\rm H I}$ and $N_{\rm H I}+dN_{\rm H I}$ and 
redshift path between $X$ and $X+dX$ \citep[e.g.,][]{tytler87}, 
\begin{equation}\label{eqdiff1}
f(N_{\rm H I}) dN_{\rm H I} dX =  \frac{m}{\Delta N_{\rm H I} \Sigma \Delta X  } dN_{\rm H I}dX\,,
\end{equation}
where $m$ is the observed number of absorption systems in a column density range $\Delta N_{\rm H I}$ 
centered on $N_{\rm H I}$ obtained from our sample of 7 QSOs with a total absorption 
distance coverage $ \Sigma \Delta X = 2.404 $ (see \S\ref{datasample}). 
Empirically, it has been shown that at low and high redshift,
$f(N_{\rm H I})$ is well fitted by a single power law 
\citep[e.g.,][]{tytler87,petitjean93,hu95,lu96,kim02,penton00,penton04,dave01}: 
\begin{equation}\label{eqdiff2}
f(N_{\rm H I})dN_{\rm H I}dX = C_{\rm H I} N_{\rm H I}^{-\beta}dN_{\rm H I}dX\,.
\end{equation}
In Table~\ref{t4}, we show the results from the maximum-likelihood estimate of the 
parameters for the slope $\beta$ and the normalization constant where
$C_{\rm H I} \equiv  m_{\rm tot} (1 - \beta)/ [N^{1-\beta}_{\rm max} (1 -(N_{\rm min}/N_{\rm max})^{1-\beta} )]$
where $ m_{\rm tot}$ is the total number of absorbers in the column density range 
$[N_{\rm min},N_{\rm max}]$. 
We separate our sample into absorbers with $b \le 40$ \km\ and $b\le 150$ \km. 
For both $b$-samples we also choose three different column density ranges: (a) $[13.2,16.5]$ dex, 
(b) $[13.2,14.4]$ dex and (c) $[14.4,16.5]$ dex. The lower limit of sample (a) and (b) corresponds
to our threshold of completeness.  The largest observed column  density in our sample is 
$\sim$$10^{16.5}$ cm$^{-2}$. 
Hence sample (a) corresponds to the whole sample with   $W \ga 90$ m\AA, while 
sample (b) only covers the weak Ly$\alpha$ systems, and (c) the strong Ly$\alpha$ systems. 
The cut at 14.4 was chosen because the sample analyzed by \citet{penton04} suggested
a change of slope near this value (see below). Note that for each sample, we only consider  
absorbers with $\sigma_b/b \le 0.4$ and $\sigma_N/N \le 0.4 $.

\begin{figure}[tbp]
\epsscale{1} 
\plotone{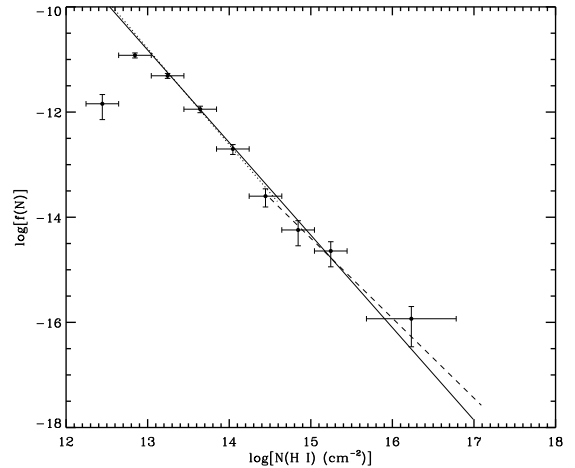}
\caption{The differential density distribution ($f(N_{\rm H I})$)
without an incompleteness correction is plotted against $\log N_{\rm H I}$. Only \hi\ absorbers with 
$b($\hi$)\le 40$ \km\ and errors less than 40\% in $b$ and $N_{\rm H I}$ are considered. The data are binned only for purpose of 
presentation. The solid line is a maximum-likelihood
fit to the data ($f(N_{\rm H I})= C_{\rm H I} N_{\rm H I}^{-\beta}$) with the slope $\beta= 1.76$ 
for \hi\ column density systems between  13.2  and 16.5 dex. The dotted and dashed
lines are maximum-likelihood fit to the data for  $13.2 \le \log N_{\rm H I} \le 14.4$ 
and $14.4 \le \log N_{\rm H I} \le 16.5$, respectively (see Table~\ref{t4} and \S\ref{dddf} for more details).
\label{max1}}
\end{figure}

In Fig.~\ref{max1}, we show the column density differential distribution of the 
identified absorbers for the combined sample with $b \le 40 $ \km\ and the maximum-likelihood fits to the data.
The fit to the sample with $b\le 150$ \km\ implies a general increase of $\beta$ (see Table~\ref{t4})
for the various column density intervals. 
For sample (a), if we vary $\log N_{\rm min}$ from 13.2 to 13.4, the results from the maximum-likelihood fit 
essentially do not change. If $\log N_{\rm min} < 13.1$, $\beta$ decreases rapidly since our 
sample is not complete anymore. If $N_{\rm max}$ is reduced, $\beta$ will increase as sample (b) shows. 
For sample (b) with either $b \le 40 $ \km\ or $b<150$ \km\, if $\log N_{\rm min}$ increases by
up to 0.4 and/or  $\log N_{\rm max}$ varies by $\pm_{0.5}^{1.0}$ dex, the results do not statistically
change. However, for sample (c), the results are more uncertain. While $\beta$ appears to 
flatten at larger column densities, there are too few absorbers with large \hi\ column density
to have a full understanding of the possible change in the slope. In particular, it is not 
clear from our sample where the break in the slope occurs since for 
$\log N_{\rm min} \approx 14.4 \pm 0.5$ dex, $\beta$ is $\sim$1.5. At the 
2$\sigma$ level, $\beta$ from sample (c) is essentially the same as for 
sample (a) and (b).

With GHRS and STIS grating moderate spectral resolution observations, 
\citet{penton04} found at $z \la 0.1$  $\beta = 1.65 \pm 0.07 $ for 
$ 12.3 \le \log N($\hi$) \le 14.5$ and $\beta = 1.30 \pm 0.30$
at $\log N($\hi$) > 14.5$ with few data points. In their work, 
$N$(\hi) was obtained from the equivalent width assuming $b = 25 $ \km.
Their results are consistent with the Key Project data
presented by \citet{weymann98} and our results. 
\citet{dave01} derived $\beta = 2.04 \pm 0.23 $ 
for $ 13 \la \log N($\hi$) \la 14$.   They used STIS E140M spectra of two QSOs and
were able to derive $b$ and $N$ independently using an automated Voigt 
profile fitting software but not allowing for BLAs. Within $1\sigma$, our results are 
the same. 

At $z\ga 1.5 $, several studies using high resolution spectra obtained with the 
Keck and the VLT have shown that $\beta \approx 1.5$ for \hi\ column
density ranging from a few times $10^{12}$ cm$^{-2}$ up to a few times 
$10^{20}$ cm$^{-2}$ \citep[e.g.,][]{tytler87,hu95,lu96}, although there may be 
some deviation from a single power law at $\log N($\hi$) > 14$  
\citep{petitjean93,kim97,kim02}. There is also some suggestion 
that for a given \hi\ interval, $\beta$ increases as $z$ decreases,
but this is statistically uncertain at $z> 2$ \citep{kim02}. 
If we compare our results to \citet{kim02}, we find an increase
of $\beta$ as $z$ decreases in the column density interval 13.2 to 14.4 
dex or 13.2 to 16.5 dex, but not in the column density interval
14.4 to 16.5 dex, although in this range the number of data points is 
too small to draw any firm conclusion. Finally, the analysis of \citet{jank06}
at  $0.5<z<2.0$ shows that $\beta $ is intermediate between  $\beta$ at $z < 0.5$
and $\beta$ at $z > 2$. Therefore, it appears that the column density distribution
steepens with decreasing $z$ in the column density range 13.2 to 16.5 dex. 
A redshift dependence was found in the  numerical results presented by \citet{theuns98},
but the observed rate of evolution of $\beta$ appears to be smaller than their models suggest. 

\begin{deluxetable}{lcc}
\tabcolsep=3pt
\tablecolumns{3}
\tablewidth{0pt} 
\tabletypesize{\scriptsize}
\tablecaption{The Power Law Fit to the Distribution Function $f(N_{\rm H I})= C_{\rm H I} N_{\rm H I}^{-\beta}$ \label{t4}} 
\tablehead{\colhead{$[\log N_{\rm min},\log N_{\rm max}$]}   &  \colhead{$\beta$}&\colhead{$\log C_{\rm H I}$}}
\startdata
\cutinhead{$b \le 40$ \km }
$[13.2,16.5]$   &  $   1.76 \pm   0.06   $  &      $ 12.1  $      	 \\	
$[13.2,14.4]$   &  $   1.83 \pm   0.06   $  &	$ 13.0$ 	      	  \\
$[14.4,16.5]$   &  $   1.52 \pm   0.10   $  &	$ 8.4 $               	  \\
\cutinhead{$b \le 150$ \km}
$[13.2,16.5]$	&  $  1.84 \pm   0.06 $ &   $13.3  $       	        		\\
$[13.2,14.4]$	&  $  1.92 \pm   0.05 $  &   14.4     	   		 	     \\
$[14.4,16.5]$	&  $  1.61 \pm   0.11 $  &   9.9     	   		 	     
\enddata		
\tablecomments{Only data with $\sigma_b/b, \sigma_N/N \le 0.4$ are included in the various samples
(for more details, see \S\ref{dddf}).}
\end{deluxetable}

\section{The Baryon Density of the IGM}\label{bd}
\subsection{Narrow Ly$\alpha$ Absorption Lines}
To estimate the baryon content of the photoionized Ly$\alpha$ forest at low $z$
we follow the method presented by \citet{schaye01}. 
The mean gas density relative to the critical density can be obtained from the 
\hi\ density distribution function:
\begin{equation}\label{omegan1}
\Omega({\rm NLA}) = \frac{\mu_{\rm H} m_{\rm H} H_0}{\rho_c\, c }\int N_{\rm H I} \frac{n_{\rm H}}{n_{\rm H I}}f(N_{\rm H I}) d N_{\rm H I}
\end{equation}
where $m_{\rm H} = 1.673\times 10^{-24}$ g is the atomic mass of hydrogen, 
$\mu_{\rm H} = 1.3 $ corrects for the presence of helium,  
$ H_0 = 100 $ \km\,Mpc$^{-1}$, $\rho_c = 3 H^2_0/(8\pi G) = 1.06 \times 10^{-29}$ g\,cm$^{-3}$ is 
the current critical density, $n_{\rm H}$ and $n_{\rm H I}$ the density 
of the total hydrogen and neutral hydrogen, respectively, $N_{\rm H I}$ the 
neutral hydrogen column density, and $f(N_{\rm H I})$ the 
differential density distribution function discussed in \S\ref{dddf}.
Assuming that the gas is isothermal and photoionized, the previous equation can be simplified to, 
\begin{eqnarray}\label{omegan2}
\Omega({\rm NLA}) \approx 2.2 \times 10^{-9} h^{-1} \Gamma_{12}^{1/3} T_4^{0.59}\left(\frac{f_g}{0.16}\right)^{1/3}~~~~~~ &&\\ 
		\int_{N_{\rm min}}^{N_{\rm max}} N^{1/3}_{\rm H I} f(N_{\rm H I}) d N_{\rm H I} \nonumber
\end{eqnarray}
where $h\equiv H_0/(100 $\,\km\,Mpc$^{-1})$, $\Gamma_{12}$ is the \hi\ photoionization 
rate in units of $10^{-12}$ s$^{-1}$, $T_4$ is the IGM temperature in units of $10^4$ K, and
$f_g$ is the fraction of mass in gas in which stars and molecules do
not contribute. \citet{schaye01} states that in cold, collapsed clumps, $f_g \approx 1$, 
but on the  scales of interest here $f_g$ is close to the ratio of the total baryon density to
the matter density, which according to \citet{spergel03} is 0.16. 
We therefore set $f_g = 0.16$. 
We use $\Gamma_{12} = 0.05 $ from \citet{haard96} for our average redshift $\bar{z} = 0.19$. 
This value of $\Gamma_{12} $ is consistent with the value derived by \citet{dave01}  at $\bar{z} = 0.17$. 
The typical temperature of the IGM is not well defined in the current simulations 
of the low redshift universe.  \citet{schaye01} noted a good agreement
between his estimate of $T_4$ and the \citet{dave99} estimate, we therefore
use Eq.~\ref{e-tn} for estimating $T_4$. 
But as these authors noted there is a very large scatter in the $T$--$\rho/\bar{\rho}$ relation
at $z \sim 0$ producing uncertainty in the baryon determination from Eq.~\ref{omegan2}
because of temperature variations.  
Note for our calculation, we set $z = \bar{z} $. 
Because $\Omega({\rm NLA}) \propto N^{4/3-\beta}_{\rm H I} $ with $\beta>4/3$,
the low column density systems dominate $\Omega({\rm NLA})$, therefore we will set the temperature
from $N_{\rm min}$ in Eq.~\ref{omegan2}. Furthermore since  $N_{\rm min}$
dominates the solution of Eq.~\ref{omegan2}, using this equation is dependent
on the data quality of the current sample.  We therefore emphasize that we only derive the 
contributions of the denser parts of the photoionized Ly$\alpha$ forest.  

To estimate the baryon content of the photoionized regions, we only use \hi\ absorbers with
$b \le 40$ \km. Broader systems are here assumed to be principally collisionally ionized. 
For the interval of \hi\ column density $10^{13.2}$--$10^{16.5}$ cm$^{-2}$, using the best fit
parameters of $f(N_{\rm H I})$ listed in Table~\ref{t4} for this range of column densities
and with $(\Gamma_{12}, T_4, f_g,h) = (0.05,1.2,0.16,0.7)$,
we find $ \Omega({\rm NLA})/\Omega_{\rm b} \simeq 0.19$ for absorbers with 
$\log N_{\rm H I} \ge 13.2$,  where $\Omega_{\rm b} = 0.044$  is the
ratio of the total baryon density to the critical density 
\citep{spergel03,burles01,omeara01}.
If  we assume that the power law described in \S\ref{dddf} fits the data extending to 
the lower observed column density (12.42 dex), we find $ \Omega({\rm NLA})/\Omega_{\rm b} = 0.29$
for the column density range 
$[12.4,16.5]$ with $(\Gamma_{12}, T_4, f_g,h) = (0.05,0.6,0.16,0.7)$. Such an assumption is not
unrealistic since, in high redshift spectra, \citet{lu96} and \citet{kirkman97} 
show using their simulation results that weak absorbers (down to 12.1--12.5 dex)
follow the same column density distribution as the stronger absorbers. 
Assuming $\beta = 1.76$ (see Table~\ref{t4}) and Eq.~\ref{e-tn}, the dependence between 
$ \Omega({\rm NLA})/\Omega_{\rm b}$ and $N_{\rm min}$ can be approximated by  
$ \Omega({\rm NLA})/\Omega_{\rm b} \simeq 0.2\, (N_{\rm min}/[10^{13.2} \, {\rm cm}^{-2}])^{-0.18}$.

With the available data, there is no indication of subcomponent structure in 
the broad Ly$\alpha$ absorption lines. However, 
broadening mechanisms other than thermal  may also be 
important. If this is the case,  the thermal broadening could decrease significantly for 
the broad Ly$\alpha$ absorption lines, implying that many of those lines would arise in the 
photoionized IGM, not the WHIM. If the complete sample with $\log N_{\rm H I} =[13.2,16.5]$
and  $b\le 150$ \km\ is considered,  $ \Omega(b<150 {\mbox \km})/\Omega_{\rm b} \ga 0.23 $ for 
$(\Gamma_{12}, T_4, f_g,h) = (0.05,1.2,0.16,0.7)$. If the sample with $\log N_{\rm H I} =[12.4,16.5]$
and  $b\le 150$ \km\ is considered,  $ \Omega(b<150 {\mbox \km})/\Omega_{\rm b} = 0.40 $ for 
$(\Gamma_{12}, T_4, f_g,h) = (0.05,0.6,0.16,0.7)$. Therefore, if the BLAs are tracing photoionized
gas, the estimate of $\Omega$ would increase by a factor $\sim$1.3. These BLAs would not then
contribute to the baryon budget in the BLAs determined in \S\ref{s-bla}.

To estimate a reliable error on $ \Omega({\rm NLA})$ remains a difficult
task with our current knowledge. The estimate of $ \Omega({\rm NLA})$ is model dependent since
the representative temperature of the IGM is not well known.
More numerical simulations of the low redshift IGM spanning a wider range of parameters 
and using the results from the current sample may tighten $\Gamma$ and $T$. 
If $T$ changes by 20\%, $ \Omega({\rm NLA})/\Omega_{\rm b}$ can change
by about 10--15\%; if $\Gamma$ changes by 20\%, $ \Omega({\rm NLA})/\Omega_{\rm b}$ can change
by about 10\%. The $\Omega$ estimate is also very sensitive to the slope $\beta$: a change of $\beta$
by $\pm 0.005$ can introduce a change in $\Omega({\rm NLA})/\Omega_{\rm b}$ by about $\mp 4$\%.
Furthermore the low column density absorbers dominate the baryon 
fraction and the low column density cutoff, $N_{\rm min}$, is  unknown.
Therefore, while the exact baryon content is uncertain in the 
photoionized IGM, it is clear that the photoionized Ly$\alpha$ forest is 
a large reservoir of baryons. The baryon fraction
in the diffuse photoionized phase traced by (narrow) Ly$\alpha$ absorbers predicted
by cosmological models in the low redshift universe varies
by a factor 2 ($\sim$20--40\%) among the various simulations \citep{dave01a},
in general agreement with our results.  We note that \citet{penton04}
derived $ \Omega({\rm Ly\alpha})/\Omega_{\rm b} = 0.29 \pm 0.04$ 
for the photoionized phase in the column density  range [12.5,17.5] dex.\footnote{It is unclear which 
parameters \citet{penton04} used with the \citet{schaye01} method (if we set the parameters to 
those given by Penton et al., i.e. $(\Gamma_{12}, T_4, f_g,h) = (0.03,0.5,0.16,0.7)$ 
we found that the values in their Table~4 (Schaye column) are a
factor $\sim$3 too high).} In view of our discussion on the various uncertainties, 
we believe that their error estimate appears optimistic.

\subsection{Broad Ly$\alpha$ Absorption Lines}\label{s-bla}
Baryons also reside in the WHIM, a shock-heated intergalactic gas with temperatures
in the range \dex5 to \dex7~K. Cosmological hydrodynamical
simulations predict that the WHIM may  contain 30--50\% of the baryons at 
low redshift (Cen \& Ostriker 1999; Dav\'e et al.\ 1999). 
BLAs may trace the \dex5 to \dex6~K WHIM if the broadening is purely 
thermal, following $T[{\rm K}] =(b[{\rm km\,s^{-1}}]/0.129)^2$.
The cosmological mass density of the broad Ly$\alpha$ absorbers 
in terms of today critical density can be written \citep{richter04,sembach04} as, 
$$
\Omega({\rm BLA}) = \frac{\mu_{\rm H} m_{\rm H} H_0}{\rho_c \,c } \frac{\Sigma f_{\rm H}(T_i) N_{\rm H I}(i)}{\Sigma \Delta X} 
$$
\begin{equation}\label{e-obla}
\Omega({\rm BLA}) \approx 1.667\times 10^{-23}  \frac{\Sigma f_{\rm H}(T_i)  N_{\rm H I}(i)}{\Sigma \Delta X}
\end{equation}
where $f_{\rm H}(T)$ being the conversion factor between \hi\ and total H  
and the other symbols have the same meaning as in Eq.~\ref{omegan1}. 
For our sample, $ \Sigma \Delta X = 2.404 $. 
In Eq.~\ref{e-obla} the sum of $f_{\rm H}(T_i)  N_{\rm H I}(i)$ over index $i$ is a 
measure of the total hydrogen column density 
in the BLAs. When collisional ionization equilibrium (CIE) is assumed,
the conversion factor between \hi\ and total H was  approximated by 
\citet{richter04} from the values given in \citet{sutherland93}
for temperatures $10^5-10^6$ K:
\begin{equation}\label{eq-fcol}
\log  f_{\rm H}(T) \approx -13.9 + 5.4 \log T -0.33 (\log T)^2 \,. 
\end{equation}

For low-density absorbers, the CIE hypothesis is, however, probably a poor approximation
because the photoionization from the UV background is important
\citep{richter06}. Recently, \citet{richter06a} used the output of the numerical
simulation, carried out by \citet{fang01} to investigate the intervening
\ovi\ absorption in the WHIM, to study the BLAs in the low redshift IGM. 
They found that a significant number of BLAs could be
photoionized, in particular systems that have relatively low $b$-values in the range 40--65 \km.
Their simulation that includes collisional ionization and photoionization suggests that  
the hydrogen ionization fraction can be approximated as: 
\begin{equation}\label{eq-fcp}
\log  f_{\rm H}(T) \approx -0.75 + 1.25 \log T  \,. 
\end{equation}
According to Fig.~5 in \citet{richter06a}, the CIE hypothesis (Eq.~\ref{eq-fcol}) 
provides a firm lower limit to $f_{\rm H}(T)$. Eqs.~\ref{eq-fcol} and \ref{eq-fcp}
converge as $b$ increases, but never intersect. 

In Table~\ref{t2a} and \S\ref{lined}, we saw that the S/N has little effect on
the estimate for $d{\mathcal N}({\rm BLA})/dz$ for the highest S/N spectra, 
but can reduce $d{\mathcal N}({\rm BLA})/dz$ for the lowest S/N data in
our sample. To reduce the uncertainty from the broadening of the \hi\ absorbers found in 
low S/N spectra, we restricted our sample to absorbers with $\sigma_b/b, \sigma_N/N \le 0.3$. 
To estimate $ \Omega({\rm BLA})$, we consider the following $b$-value intervals:  
$[40,150]$, $[40,65]$, and $[65,150]$ \km. In the $[40,65]$ \km\ interval, \citet{richter06a} 
found that  about a $1/3$ of the BLAs traces cool photoionized gas ($T< 2\times 10^4$ K), 
i.e. $b_{\rm nt} \gg b_{\rm th}$ for these systems. If all the BLAs are assumed to be
thermally broadened, we would overestimate  $ \Omega({\rm BLA}) $. We therefore randomly pick 2/3 of 
the BLAs in the $[40,65]$ \km\  $b$-range  to estimate  $f_{\rm H}$.
(We note that the baryon content of the photoionized BLAs with $T\la 10^4$ K is less than 1--2\% according
to Richter et al. 2006a.) In the $[65,150]$ \km\ $b$-range, the number of BLAs tracing cool 
photoionized systems is small (less than $\sim$5\%) and we assume that all the BLAs in
this range of $b$-values are mostly thermally broadened. 
With the conditions listed above, there are 12 systems in the 
$[65,150]$ \km\ interval and 31 in the  $[40,65]$ \km\ $b$-range. 
In Table~\ref{t5}, we summarize our estimates of $ \Omega({\rm BLA})/\Omega_{\rm b} $ 
using Eqs.~\ref{e-obla}, \ref{eq-fcol}, and \ref{eq-fcp} where $T$ was derived
assuming  $b_{\rm th} \approx 0.9 b_{\rm obs}$ following the simulation of
\citet{richter06a}. The largest difference in the estimates of the 
baryon budget from CIE and Richter et al.'s model is for low $b$-values, 
where photoionization plays a more important role. 
We note that that if we restricted our sample to systems with 
$\sigma_b/b, \sigma_N/N \le 0.2$, $ \Omega({\rm BLA})/\Omega_{\rm b}$ would decrease
by a factor $\sim$1.2, but would increase by a factor $\sim$1.2--2.4 if $\sigma_b/b, \sigma_N/N \le 0.4$.

As for the estimate of the baryon budget of the NLAs, with our current knowledge
it appears a very difficult task to estimate an error on the baryon budget in
BLAs. It is model dependent. \citet{richter06a} provided estimates of the amount
of the BLAs that probe cool photoionized gas and the amount of thermal 
broadening in BLAs. These estimates may change with refined cosmological 
models. If we assume that the broadening for {\em all} the BLAs is purely  thermal 
($b_{\rm th} = b_{\rm obs}$), the estimates of $ \Omega({\rm BLA})/\Omega_{\rm b} $ in the
$[40,150]$ \km\ $b$-range would increase by a factor 1.6, i.e. in the CIE model,
 $ \Omega({\rm BLA})/\Omega_{\rm b} \simeq 0.13$, and in the collisional ionization plus
photoionization model of \citet{richter06a},  $ \Omega({\rm BLA})/\Omega_{\rm b} \simeq 0.32$.
The broadening could be due to several unresolved components in some cases 
or the non-thermal broadening could be important. This is partly accounted for
since we remove a 1/3 of the BLAs in the  $[40,65]$ \km\ $b$-range. 
Since we discarded systems with large errors in $b$ and $N$, for
the current data set, a single Gaussian fit provides a good representation of
the observed profiles. However, as we discussed in \S\ref{sec-bn}, the formal 
errors given by the fit may not be sufficient especially for the weaker and broader
systems. \citet{richter06} and the other papers mentioned in Table~\ref{t1}
present several examples of BLAs that are well fitted with a single Gaussian with sometimes 
the presence of several \hi\ Lyman series transitions that further constrain the fit parameters 
(see also the discussion in \S\ref{sec-descbevol}). To further explore the quality 
of the profile fitting,  the Cosmic Origin Spectrograph (COS) could really improve the situation 
by increasing the number of sightlines and increasing the S/N in the spectra. 
Additional theoretical simulations of the low-$z$ IGM with higher resolution 
and refined physics combined with the analysis of simulated spectra should
improve our understanding of the BLAs.

\begin{deluxetable}{ccc}
\tabcolsep=3pt
\tablecolumns{4}
\tablewidth{0pt} 
\tabletypesize{\scriptsize}
\tablecaption{Estimates of the Baryon Density in BLAs \label{t5}} 
\tablehead{\colhead{$b$ [\km]}  &  \colhead{$\frac{\Omega_{\rm CIE}({\rm BLA)}^a}{\Omega_{\rm b}}$}   &\colhead{$\frac{\Omega_{\rm C+P}({\rm BLA)}^b}{\Omega_{\rm b}}$} }
\startdata
$[40,65]$    	&  0.03   &   0.11	\\
$[65,150]$   	&  0.05   &   0.09	\\
$[40,150]$   	&  0.08   &   0.20	\\     
\enddata	
\tablecomments{Only data with $\sigma_b/b, \sigma_N/N \le 0.3$ and $\log N_{\rm H I} \ge 13.2$ 
are included in the various samples. In the $[40,65]$ \km\ interval, only 2/3 of the BLAs are 
considered. We derive the temperature of the absorbers assuming
$b_{\rm th} = 0.9 b_{\rm obs}$. The baryon density in the interval $[40,150]$ \km\ is 
the sum of $\Omega($BLA) in the two other intervals. 
$a$: Obtained from Eqs.~\ref{e-obla} and \ref{eq-fcol}; 
baryonic density for the BLAs  assuming that they are collisionally ionized and CIE applies.
$b$: Obtained from Eqs.~\ref{e-obla} and \ref{eq-fcp}; baryonic density for 
BLAs  assuming that they are collisionally ionized and photoionized using the results from the 
hydrodynamical simulations of \citet{richter06a} (see \S\ref{s-bla} for more details). 
$\Omega_{\rm b} = 0.044$ \citep{spergel03}. 
}
\end{deluxetable}
 
In summary, the CIE estimate provides a firm and conservative limit for the baryon budget in 
collisionally ionized BLAs, $\Omega({\rm BLA)}/\Omega_{\rm b} \ga 0.1$ for systems with $40 < b \la 150$ 
\km,  $\log N_{\rm H I} \ge 13.2$, $\sigma_b/b, \sigma_N/N \le 0.3$.
The BLA simulations show that the ionization fraction in BLAs is governed
by collisional ionization and photoionization. Therefore, $\Omega({\rm BLA)}/\Omega_{\rm b} \ga 0.2$
(with the same constraints that those listed above) should reflect better the baryon
budget of the BLAs in the low-$z$ IGM. This is likely a lower estimate 
as well since the existing observations do not have the S/N to detect the broader systems  
(where most of the baryons are believed to exist) and the numerous weaker BLAs (see Fig.~\ref{bcomp1}),
and are not complete at $b\ga 80$ \km.

\subsection{Baryon Budget of the Low Redshift Universe}
At redshift 2--4, observations of the Ly$\alpha$ forest 
show that about 80--90\% of the total baryon budget is found in the 
cool photoionized phase of IGM \citep{rauch97,weinberg97,kim01}. 
In the low redshift universe, \citet{fukugita98} show
that  9--10 \% of the total baryons are found in galaxies (stars, neutral gas,
molecular gas) and $\sim$6 \% of the baryons are associated
with the hot plasma in clusters of galaxies. 

Our analysis of the low redshift Ly$\alpha$ forest shows the presence 
of NLAs and BLAs. The NLAs trace the cool photoionized IGM, while 
the BLAs are likely to probe the cool photoionized IGM and the hot highly-ionized gas in the WHIM. 
There are still many uncertainties that are not controlled in the determination of the 
baryon content in the low redshift Ly$\alpha$ forest. But at least 20\% of 
the baryons are found in the denser parts of the photoionized phase 
(NLAs, $b\le 40$ \km, $\log N_{\rm HI} \ge 13.2$, and 
$\sigma_b/b, \sigma_N/N \le 0.4$) and at least 
$\sim$10\% are found in the WHIM (BLAs, $40< b \le 150$ \km, 
$\log N_{\rm H I} \ge 13.2$, $\sigma_b/b, \sigma_N/N \le 0.3$), 
corresponding to at least 30\% of the total baryon budget. We believe that these limits are very 
conservative since they do not take into account the low column density photoionized
absorbers and the broadest BLAs. Furthermore recent simulations of the 
BLAs in low redshift universe show that the BLAs may trace both photoionized and collisionally
ionized gas. Using the simulations of \citet{richter06a}, we find that the BLAs include
at least 20\% of the baryons. We have also seen that $\Omega$ in the cool photoionized IGM 
is dominated by the low column density \hi\ absorbers (i.e. absorbing gas with overdensity 
$\rho/\bar{\rho} \la 2$). If the weak systems follow the same differential column density distribution function
as the stronger absorbers, the cool photoionized gas traced by the NLAs with 
$\log N_{\rm H I} \ge 12.4$ can contain  $\sim$30\% of the total baryons. 
Combining these NLA and BLA budgets, the low redshift IGM contains
at least $\sim$50\% of the baryons. This is much larger than 
the known amount of baryons in galaxies and clusters of galaxies in the low redshift 
universe. High S/N data will allow us to search for the 
weakest Ly$\alpha$ absorbers ($\la 10^{12.5}$ cm$^{-2}$) and estimate their column density distribution, 
search for the very broad absorbers and the weak BLAs, and provide
a better understanding of their physical nature. 
It is, however, already apparent from this study that the IGM traced 
by the narrow and broad Ly$\alpha$ absorption lines in the low redshift 
universe is a major reservoir of baryons.

\section{Summary}\label{sum}

We analyze the physical properties of the low-$z$ IGM traced by \hi\ absorbers from a sample of 
7 QSOs observed with the E140M mode of STIS ($R \sim 44000$) and with \fuse\ ($R \sim 15000$). 
These seven lines of sight were fully analyzed
and presented in recent and future papers (see references in Table~\ref{t1} and
the Appendix of this paper). 
Our sample has a total unblocked redshift path of 2.064 corresponding to a 
total absorption distance of 2.404 and is complete for $\log N_{\rm H I} \ga 13.2$ provided $b \la 80$ \km. 
With high spectral resolution the column density ($N_{\rm H I}$) and the Doppler parameter ($b$) 
were independently estimated, providing the opportunity to directly study the 
distribution and evolution of $b$.  The spectral resolution of the STIS E140M \hi\ observations in our sample is similar to 
that of the high redshift ($z\ga 1.5$) observations obtained with Keck/HIRES and VLT/UVES. This
similarity allows a relatively simple study of the evolution of the $b$-parameter 
of the neutral hydrogen gas in the intergalactic medium. 
Because the nominal lower  temperature of the WHIM is $T\sim 10^5$ K, we consider two different 
populations of \hi\ absorbers throughout this work: narrow Ly$\alpha$ absorbers (NLAs) have $b\le40$ \km\ and the 
broad Ly$\alpha$ absorbers (BLAs) have  $b>40$ \km.
Our main findings applicable at redshifts $z\la 0.4$ are summarized as follows: 

1. The \hi\ Doppler width distribution  has a high velocity tail that develops at $b\approx 40$--60 \km. 
The $b$--$N_{\rm H I}$ distribution reveals for the NLAs an increase of $b$ with increasing $N_{\rm HI}$ 
with a very large scatter.  We find that most of the BLAs are found in the \hi\ column density 
range [13.2,14.0] dex and the broader systems are found for $13.1 \la \log N_{\rm H I} \la 13.5$. 
Recent cosmological simulations of BLAs show a good agreement with the observations, 
but clearly indicate that broader absorbers and many weak BLAs ($\log N_{\rm H I} \la 13.1$)
remain to be discovered. 

2. We find  $d{\mathcal N}({\rm BLA})/dz = 30 \pm 4 $ 
for absorbers with $40 < b \la 150$ \km, $\log N_{\rm HI} \ge 13.2$, and 
$\sigma_b/b, \sigma_N/N \le 0.4$.   The  narrow \hi\ absorbers are more 
frequent, with $d{\mathcal N}({\rm NLA})/dz = 66 \pm 6 $ 
for absorbers with $ b \le 40$ \km, $\log N_{\rm HI} \ge 13.2$, and 
$\sigma_b/b, \sigma_N/N \le 0.4$. Very narrow absorbers ($b\la 15$ \km) 
with  $\log N_{\rm HI} \ge 13.2$ are scarce. 
The number of weak NLAs and BLAs  ($\log N_{\rm H I} \le 14$)
is far larger than the number of strong NLAs and BLAs, respectively.  

3. For the narrow absorbers with $13.2 \le \log N_{\rm H I} \le 16.5 $ and
$\sigma_b/b, \sigma_N/N \le 0.4$, we find that 
the column density distribution, $f(N_{\rm H I}) \propto N^{-\beta}_{\rm H I}$,
can be fitted with $\beta = 1.76 \pm 0.06$.  For the entire sample (i.e. including NLAs and BLAs), 
the slope changes to 1.84  for the same column density range. 
We confirm an increase of $\beta$ with decreasing $z$
when our results are compared to higher redshift analyses. 
There is some weak evidence for a break at $\log N_{\rm H I } \sim 14.4$, 
but the location of this break is uncertain because of the small number statistics for the 
higher column density lines.

4. We argue that various samples probing different redshift ranges can be directly 
compared when the conditions $0 < b \le 100$ \km, $13.2 \le \log N_{\rm HI} \le 14.0$, and 
$\sigma_b/b, \sigma_N/N \le 0.4$ are set. 
The distribution of $b$ for the broad absorbers has a distinctly more prominent high velocity 
tail at low $z$ than at high $z$ ($1.5 \la z \la 3.6$). The median and mean $b$-values are systematically 
larger by $\sim$15--30\% at  $z \la 0.5$ than at $1.5 \la z\la 3.6$. 
The ratio of the number density of BLAs to NLAs at low $z$ is larger than at high $z$
by a factor $\sim$3. This suggests that 
a larger fraction of the low-$z$ universe is hotter than at $1.5 \la z\la 3.6$ and/or 
the low-$z$ universe is more kinematically disturbed than the high-$z$ universe. 

5. The NLAs trace the cool photoionized IGM at $T\la 10^4$ K, and we find that 
$ \Omega({\rm NLA})/\Omega_{\rm b} \simeq 0.2\, (N_{\rm min}/[10^{13.2} \, {\rm cm}^{-2}])^{-0.18}$
(when $f(N_{\rm H I}) \propto N^{-1.76}_{\rm H I}$ and assuming $N_{\rm H I}\propto T^{-0.42}$), 
where $N_{\rm min}$ is the lowest \hi\ column density in the sample considered 
to estimate  $ \Omega({\rm NLA})$. The contribution to the baryon budget of the denser parts of 
the photoionized Ly$\alpha$ forest with $\log N_{\rm H I} \ga 13.2$ is $\sim$20\%. 
If the weakest \hi\ absorber with $\log N_{\rm H I} = 12.4$ follows the same $f(N_{\rm H I})$,  
the cool photoionized gas with  $\log N_{\rm H I} \ga 12.4$ contains  about 30\% of the total baryons.

6. The BLAs trace mostly the highly-ionized gas in the WHIM at $T\simeq 10^5$--$10^6$ K 
if the width of the BLAs is dominated by thermal motions. The assumption of 
collisional ionization equilibrium provides a firm lower limit to the amount of the baryons 
in the WHIM traced by the BLAs ($40 < b \la 150$ \km,  $\log N_{\rm H I} \ge 13.2$, 
$\sigma_b/b, \sigma_N/N \le 0.3$): $ \Omega({\rm BLA})/\Omega_{\rm b} \ga 10$\%. 
If a hydrodynamical simulation including the effects of collisional ionization and 
photoionization is used to estimate the ionization correction
for the BLAs, we find that at least  20\% of the baryons are in the BLAs.
 
7. Our most conservative estimate of the baryon budget in absorbers with $\log N_{\rm H I} \ga 13.2$
shows that the Ly$\alpha$ forest has at least about 30\% of the baryons (NLAs+BLAs), 
far larger than the baryon budget in galaxies. We suggest that the low redshift IGM with 
$T\la 10^6$ K traced by the NLAs with $\log N_{\rm H I} \ga 12.4$ and the BLAs 
with $\log N_{\rm H I} \ga 13.2$ could contain at least 50\% of the baryons (this 
is a lower limit since the broader BLAs where most the baryons reside
remain to be discovered).  The estimates of the amount of baryons still rely, however, 
on critical assumptions (e.g. pure thermal broadening versus other broadening mechanisms, 
temperature of the IGM, behavior of the weak narrow systems) 
that produce uncertainties that are currently not well controlled. 
Observationally, if COS is deployed, 
our understanding of the BLAs, in particular, will improve given the high S/N
spectra that COS should obtain (but with approximately  2 times lower resolution 
than for the STIS observations presented here). 
The present paper and future observations of the low-$z$ IGM 
should motivate more precise cosmological simulations of the low-
and  high-$z$ IGM that are needed for our understanding of the intrinsic 
properties of the BLAs and NLAs.

\acknowledgments

We thank the {\em HST}\ and {\em FUSE}\ mission operation teams
for their continuous efforts to provide excellent UV and FUV 
spectroscopic data to the astronomical community. 
Support for this research was provided by NASA through grant 
HST-AR-10682.01-A. N.L. appreciates support from the University
of Notre Dame for part of this work. 
P.R. acknowledges financial support by the German
\emph{Deut\-sche For\-schungs\-ge\-mein\-schaft}, DFG,
through Emmy-Noether grant Ri 1124/3-1.
TMT appreciates support from NASA grants NNG04GG73G and
HST-GO-9184.08-A. B.P.W. acknowledges grants HST-GO-00754.01-A and NASA NNG04GD856.
This research has made use of the NASA
Astrophysics Data System Abstract Service and the Centre de Donn\'ees de Strasbourg (CDS).

\appendix 

\begin{figure}[b]
\epsscale{0.8} 
\plotone{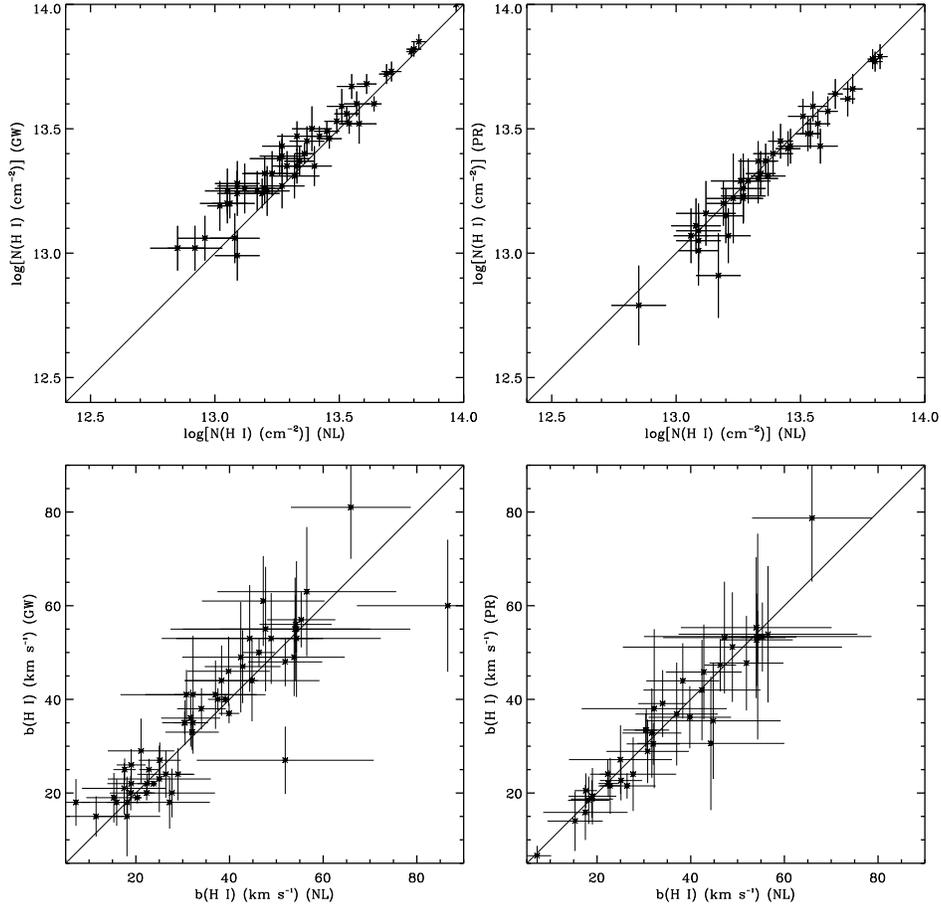}
\caption{Comparison of \hi\ column densities and Doppler parameters for the weak 
absorbers ($\log N_{\rm H I} \le 14$).  The left-hand side
compares the fit results from Williger et al. (2006, GW) and N. Lehner (NL). The
right-hand side compares the independent measurements produced by P. Richter (PR) and N. Lehner (NL). 
Identical results would lie along the solid line.
\label{compmeas}}
\end{figure}

We have revisited the \hi\ measurements presented by \citet{williger06}
toward PKS\,0405--123. We were first motivated to produce a new analysis of the PKS\,0405--123 
{\em FUSE}\ and STIS/E140M spectra because
\citet{williger06} noted that several of their BLAs may be noise features. For our work, we 
needed to clearly differentiate a real detection from a noise feature. While reanalyzing
the BLAs, we noted that some narrow \hi\ lines were either not present 
in the spectra at the 3$\sigma$ level (although Williger et al. claimed
to report only 4$\sigma$ features) or were misidentified. 
We have not identified the ultimate origin of the
problem, but we suspect that the difference with Williger et al. can
ultimately be traced to a difference in the manner in which the
continuum placement was done. Williger et al. fitted a single
global continuum, while we determine continua separately
placed within $\pm$1000--3000 \km\ of each absorption line.
We therefore produce here a new line list and new measurements
for \hi\ toward PKS\,0405-0123 in Table A1.
This reanalysis also provides an overall coherent 
data sample, since the other sightlines were analyzed following the same methodology. 
In Table~A1, we report our measurements for 
\hi\ toward PKS\,0405--123. These \hi\ absorbers are detected at least at the 3$\sigma$ level. 
We list in this table the observed wavelengths and the rest-frame equivalent widths of the Ly$\alpha$
transition. The listed column densities and Doppler parameters were measured 
in the rest-frame by N. Lehner (NL) following the method described in \citet{lehner06}. 
The weak absorbers ($\log N_{\rm H I} < 14$) were also independently measured 
by P. Richter (PR) using the method described in \citet{richter04}. In particular, 
different profile fitting software was used by NL and PR. 
In Fig.~\ref{compmeas}, we compare the column densities and $b$-values (for the
weak absorbers) presented by \citet{williger06} to those listed in Table~A1 
(left panels). On the right-hand side, we present the measurements obtained
by PR against those listed in Table~A1. For both $b$ and $N_{\rm H I}$, 
there is an excellent agreement between  PR and NL measurements
at the 1$\sigma$ level. In contrast, Williger et al. (GW) found generally larger
column densities and Doppler parameters. The main difference between 
our analysis (NL and PR) and GW's analysis is that we systematically and 
{\em independently} chose the continua within $\pm 1000$--3000 \km\ of the \hi\ absorption line, 
while GW fitted the continuum using a semi-automatic procedure (AUTOVP) combined with custom 
fitting in the vicinity of the QSO emission lines. Since these weak
absorbers are generally detected solely in Ly$\alpha$, the continuum 
placement is critical.  Our method for the continuum 
placement was followed for all the other sightlines presented in this work. 
For the absorbers where Ly$\alpha$ and Ly$\beta$ are detected, our measurements 
and those made by GW agree within 1$\sigma$.

\begin{figure}[tbp]
\epsscale{0.5} 
\plotone{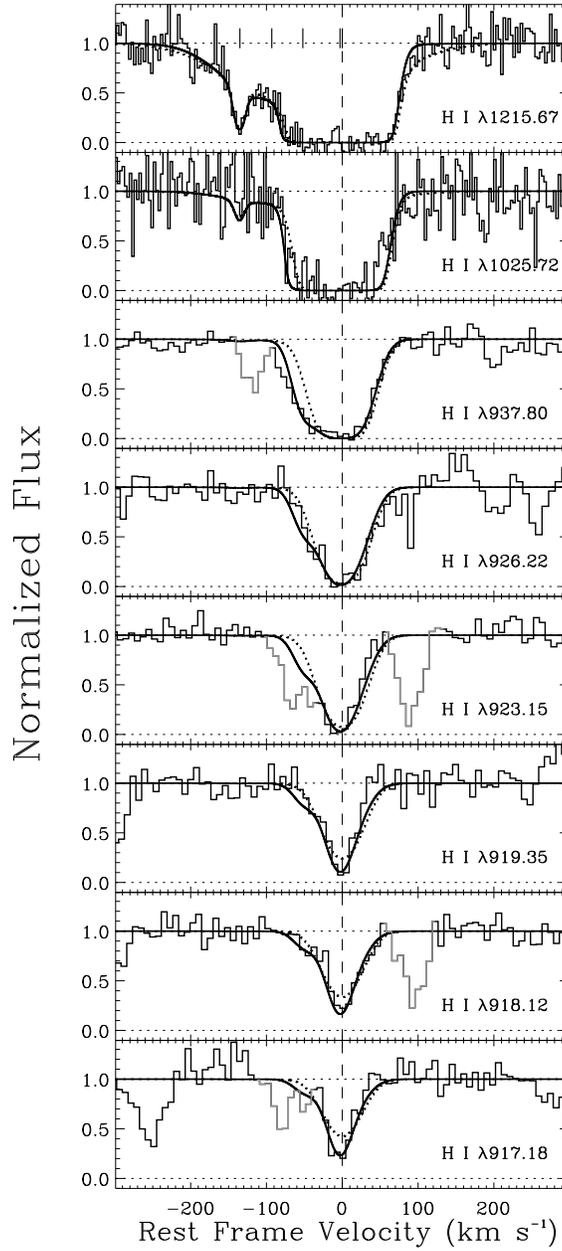}
\caption{Profile fitting to the system at $z= 0.167$. The solid thick 
line shows the fit adopted in Table~A1. The dotted thick line shows the fit adopted by \citet{williger06}. 
The light parts of the spectra denote blends with other features. The tick marks in the \hi\ $\lambda$1215
panel are the centroids of the 5 components found in our profile fitting. From left to right, the 
tick marks correspond to $z = 0.16661, 0.16678, 0.16694, 0.16712, 0.16714$. Discrepancies of the fits to
the data are discussed in the text and footnote~6 of Table~\ref{ta1}.
\label{fitpks}}
\end{figure}

Absorbers with $\log N_{\rm H I} > 14$ are detected in several 
Lyman series lines. For the absorbers that were fitted 
with a single component ($z=0.03000,0.35099,0.40571,0.40886$), GW and our measurements are in 
agreement within 1$\sigma$. GW fitted the systems listed in Table~A1 
at $z=(0.09657,0.09659)$ and (0.18262,0.18287) with a single component.
We found that a 2-component fit provides
a statistically better fit. The total column densities for those absorbers
are in agreement within 1$\sigma$ with those measured by GW. For the partial 
Lyman limit system at $z=0.167$, we found that 5 components provide
a better fit to this complicated absorber (see Fig.~\ref{fitpks}
and note (6) in Table~A1). We note that our fit is not yet as good
as it should be despite our best effort (see Fig.~\ref{fitpks}). In particular, 
the broad component at $z=0.16678$ may actually be narrower since 
metals-ions such as \ciii\ are detected at this redshift (this directly
affects the measurements of the absorber at $z=0.16661$ where \ciii\ is also
detected).  Finally, we  fitted the system at $z=0.36808$
with 4 components. This is not statistically different from the 2 component
fit presented by GW, but the 4-component fit appears to be a better model
when the kinematics of the metal-ions are taken into account (see note (12) in
Table~A1 for more details). 

In Table~A2, we report the list of features that were identified as Ly$\alpha$ and claimed to be detected
at least at the 4$\sigma$ level by GW. These features are actually either non-detections at the 3$\sigma$
level (according to our measurements) or were misidentified.  The 181 Ly$\alpha$ absorbers toward PKS\,0405--123 
reported by GW now reduce to 74 absorbers (and in a few cases, those are multi-component
absorbers). 
\clearpage

{\LongTables
\begin{deluxetable}{cccccc}
\tabcolsep=3pt
\tablecolumns{4}
\tablewidth{0pt} 
\tabletypesize{\scriptsize}
\tablecaption{IGM \hi\ revised measurements toward PKS\,0405--123$^a$  \label{ta1}} 
\tablehead{\colhead{$\lambda$} &\colhead{$z$} & \colhead{$W_{\rm Ly \alpha}$}  & \colhead{$\log N_{\rm H I}$}  & \colhead{$ b_{\rm H I}$} & \colhead{Note}   \\
		\colhead{(\AA)} &	\colhead{} & \colhead{(m\AA)} & \colhead{(dex)} &\colhead{(\km)} & \colhead{}	}
\startdata
1230.1364  & 0.01190  &  $  53.5 \pm   11.5  $  & $  13.09 \pm 0.09    $     &   $   7.2  \pm 3.0    $ &											    \\ 
1233.8078  & 0.01492  &  $  35.1 \pm   10.9  $  & $  13.09 \pm 0.11    $     &   $   11.5 \pm 5.9   $ & 											    \\ 
1236.1053  & 0.01681  &  $  45.3 \pm   14.1  $  & $  13.08 \pm 0.10    $     &   $   25.0 \pm 11.0   $ &											    \\ 
1240.3360  & 0.02029  &  $  48.7 \pm   14.1  $  & $  13.12 \pm 0.12    $     &   $   32.2 \pm 15.5   $ &											    \\ 
1251.9578  & 0.02985  &  $  65.4 \pm   9.6   $  & $  13.36 \pm 0.13    $     &   $   12.4	:   $ & 								    \\ 
1252.1401  & 0.03000  &  $  255.4\pm  16.6   $  & $  14.37 \pm 0.04    $     &   $   20.3    \pm 0.9	$ &											       \\ 
1254.5228  & 0.03196  &  $  90.5 \pm  19.2   $  & $  13.33 \pm 0.08    $     &   $   54.0    \pm 16.1:  $ &											       \\ 
1272.3689  & 0.04664  &  $  51.4 \pm   12.1  $  & $  13.09 \pm 0.09    $     &   $   27.7    \pm9.2    $ &											       \\
1287.3459  & 0.05896  &  $  41.0 \pm   10.2: $  & $  13.34 \pm 0.10    $     &   $   76.5   \pm 26.8	     $ &  1		       \\
1287.6984  & 0.05925  &  $  48.3 \pm	9.3  $  & $  12.76 \pm 0.18    $     &   $   10.8	 :     $ & 1		       \\ 
1303.4171  & 0.07218  &  $  49.5 \pm   10.8  $  & $  13.09 \pm 0.08    $     &   $   44.8    \pm14.4   $ &						       \\ 
1303.8182  & 0.07251  &  $  44.6 \pm   9.6   $  & $  13.06 \pm 0.07    $     &   $   19.0    \pm5.1    $ &											       \\ 
1306.2252  & 0.07449  &  $  31.8 \pm   8.7   $  & $  12.85 \pm 0.11    $     &   $   17.5    \pm9.0    $ &											       \\ 
1307.1249  & 0.07523  &  $  42.1 \pm   11.6  $  & $  13.05 \pm 0.11    $     &   $   47.7    \pm20.3   $ &											       \\ 
1314.6134  & 0.08139  &  $  236.0\pm   18.7  $  & $  13.79 \pm 0.02    $     &   $   54.3    \pm3.5    $ &											       \\ 
1327.3172  & 0.09184  &  $  472.2\pm 24.4 $	& $  14.60\pm  0.02    $     &   $   39.1 \pm0.9    $ &   2							    \\ 
1333.0673  & 0.09657  &  $  487.2\pm 27.9 $	& $  14.57\pm  0.04    $     &   $   30.3 \pm2.3    $ &  3	  \\
1333.0916  & 0.09659  &  \nodata		& $  13.90 \pm 0.18    $     &   $   69.8 \pm19.8   $ &  3	  \\ 
1333.7845  & 0.09716  &  $   50.9\pm   6.7  $	& $  13.02\pm  0.05:   $     &   $   29.0 \pm3.2    $ & 					    \\ 
1340.8597  & 0.10298  &  $   96.3 \pm  19.2 $  &  $  13.40 \pm 0.07    $     &   $   86.6 \pm 19.4	  $ &	  \\ 
1374.9470  & 0.13102  &  $   108.9\pm  15.1 $	& $  13.46\pm  0.05	$    &   $   51.9 \pm7.9    $ & 										    \\ 
1376.5395  & 0.13233  &  $   142.5\pm13.5 $	& $  13.64\pm  0.03	$    &   $   22.3 \pm1.7    $ & 								    \\ 
1377.4270  & 0.13306  &  $   72.9  \pm 11.5 $	& $  13.29\pm  0.06	$    &   $   31.7 \pm6.3    $ & 										      \\ 
1378.2902  & 0.13377  &  $   90.3  \pm 16.2 $	& $  13.34 \pm 0.06	  $  &   $   42.8 \pm  8.1    $ &											      \\ 
1381.5604  & 0.13646  &  $   76.8  \pm 13.2 $	& $  13.34 \pm 0.06	  $  &   $   53.8 \pm  10.8   $ & 4								      \\ 
1399.6739  & 0.15136  &  $   49.2  \pm 10.4 $	& $  13.20 \pm 0.07	  $  &   $   37.0 \pm  8.8    $ &											      \\ 
1400.7436  & 0.15224  &  $   114.8 \pm 13.2 $	& $  13.54 \pm 0.04	  $  &   $   22.4 \pm  2.1    $ &										      \\ 
1401.7162  & 0.15304  &  $   223.3 \pm 16.8 $	& $  13.80 \pm 0.03	  $  &   $   46.3 \pm  3.4    $ &											      \\ 
1411.6482  & 0.16121  &  $   157.5:\pm 16.1 $	& $  13.71 \pm 0.04	  $  &   $   54.1 \pm  7.7    $ &  5									     \\ 
1412.0007  & 0.16150  &  $   120.7:\pm 13.1 $	& $  13.27 \pm 0.09	  $  &   $   18.2 \pm  4.4    $ &  5									     \\ 
1413.8485  & 0.16302  &  $   110.7 \pm 16.5 $	& $  13.42 \pm 0.05	  $  &   $   34.0 \pm  5.2    $ &											      \\ 
1418.2128  & 0.16661  &  \nodata		& $  13.29 \pm 0.08:	  $  &   $   7.9  \pm 2.1:    $ & 6 										      \\ 
1418.4194  & 0.16678  &  \nodata		& $  13.91 \pm 0.04:	    $&   $   74.9 \pm 7.2 :   $  & 6	  \\ 
1418.6139  & 0.16694  &  \nodata		& $  15.33 \pm 0.20	   $ &   $   13.5 \pm 3.7     $ &   6		\\ 
1418.8327  & 0.16712  &  \nodata		& $  16.30 \pm 0.09	  $  &   $   12.5 \pm 5.6     $ & 6				    \\ 
1418.8571  & 0.16714  &  $   844.0 \pm 41.6 $	& $  16.27 \pm 0.13	  $  &   $   29.9 \pm 1.2     $ & 6				    \\ 
1432.9832  & 0.17876  &  $   155.2 \pm 19.6 $	& $  13.61 \pm 0.04	  $  &   $   55.3 \pm  7.3    $ &  4								      \\ 
1437.6757  & 0.18262  &  $   684.3 \pm 31.2 $	& $  14.83 \pm 0.03	  $  &   $   32.6 \pm  1.4    $ & 7				      \\
1437.9796  & 0.18287  &  \nodata		& $  14.17 \pm 0.04	  $  &   $   25.4 \pm  2.3    $ & 7				      \\ 
1447.6928  & 0.19086  &  $  51.6 \pm 13.3    $  & $  13.17 \pm 0.09	  $  &   $   44.3  \pm 15.7   $ &			  \\ 
1471.8239  & 0.21071  &  $   69.8  \pm 15.4 $	& $  13.27 \pm 0.08	  $  &   $   30.8 \pm  8.8    $ &											      \\ 
1472.4073  & 0.21119  &  $   39.1  \pm 10.1 $	& $  13.09 \pm 0.08	  $  &   $   21.1 \pm  7.1    $ &										    \\ 
1478.2426  & 0.21599  &  $   85.6  \pm 11.9 $	& $  13.36 \pm 0.04	  $  &   $   17.6 \pm  2.4    $ &									    \\ 
1493.1224  & 0.22823  &  $   51.3  \pm 11.5 $	& $  13.19 \pm 0.07	  $  &   $   18.9 \pm  4.5    $ &											      \\ 
1508.1237  & 0.24057  &  $   68.4  \pm 14.8 $	& $  13.27 \pm 0.09	  $  &   $   56.5 \pm  19.1   $ &									      \\ 
1513.6673  & 0.24513  &  $   50.7  \pm 12.8 $	& $  13.23 \pm 0.11	  $  &   $   54.3 \pm  24.3   $ & 8							      \\ 
1514.1656  & 0.24554  &  $   154.4 \pm 15.3 $	& $  13.69 \pm 0.03	  $  &   $   22.8 \pm  2.1    $ & 8							      \\ 
1521.4717  & 0.25155  &  $   52.9  \pm 12.3 $	& $  13.20 \pm 0.09	  $  &   $   27.2 \pm  8.6    $ &											      \\ 
1523.5992  & 0.25330  &  $   38.7  \pm 11.8 $	& $  13.05 \pm 0.09	  $  &   $   18.1 \pm  7.1    $ &											      \\ 
1530.0544  & 0.25861  &  $   84.4  \pm 15.5 $	& $  13.37 \pm 0.07	  $  &   $   39.8 \pm  8.8    $ &											      \\ 
1532.2791  & 0.26044  &  $   107.1 \pm 15.3 $	& $  13.45 \pm 0.05	  $  &   $   32.0 \pm  5.7    $ &											      \\ 
1540.9347  & 0.26756  &  $   36.7  \pm 10.9 $	& $  12.96 \pm 0.11	  $  &   $   15.9 \pm  7.7    $ &								      \\ 
1566.2449  & 0.28838  &  $   62.2  \pm 17.1 $	& $  13.32 \pm 0.10	  $  &   $   51.9 \pm  18.9   $ &								      \\ 
1574.5723  & 0.29523  &  $   81.6  \pm 19.8 $	& $  13.33 \pm 0.08	  $  &   $    47.2\pm	13.1  $ &											      \\ 
1577.5263  & 0.29766  &  $   236.5 \pm 20.3 $	& $  13.97 \pm 0.03	  $  &   $    32.2\pm	2.4   $ &  9										      \\ 
1579.2039  & 0.29904  &  $   75.6  \pm 21.5 $	& $  13.26 \pm 0.12	  $  &   $    48.9\pm	23.4  $ &											      \\ 
1604.7208  & 0.32003  &  $   105.2 \pm 18.2 $	& $  13.49 \pm 0.05	  $  &   $    19.0\pm	3.1   $ &									      \\ 
1610.7628  & 0.32500  &  $   115.6 \pm 22.4 $	& $  13.55 \pm 0.06	  $  &   $    65.9\pm	12.8  $ &											      \\ 
1613.5831  & 0.32732  &  $   79.5  \pm 19.1 $	& $  13.30     :	  $  &   $    30.9   :        $ &						      \\ 
1621.7281  & 0.33402  &  $   194.3 \pm 20.4 $	& $  13.82 \pm 0.03	  $  &   $    30.4\pm	2.4   $ &  10										      \\ 
1631.2590  & 0.34186  &  $   126.5 \pm 19.6 $	& $  13.51 \pm 0.06	  $  &   $   38.3\pm	7.8   $ &  11							 \\ 
1631.8425  & 0.34234  &  $   85.5  \pm 18.1 $	& $  13.39 \pm 0.08	  $  &   $   42.4\pm   12.5   $ &  11							 \\ 
1642.3581  & 0.35099  &  $   369.8 \pm 30.4 $	& $  14.25 \pm 0.03	  $  &   $   37.5\pm   2.1    $ &											    \\ 
1642.9902  & 0.35151  &  $   118.9 \pm 19.2 $	& $  13.53 \pm 0.05	  $  &   $   25.1\pm   4.5    $ &											    \\ 
1643.7318  & 0.35212  &  $   121.5 \pm 21.5 $	& $  13.57 \pm 0.05	  $  &   $   30.5\pm   4.9    $ &											    \\ 
1654.2716  & 0.36079  &  \nodata		& $  14.66 \pm 0.14	  $  &   $   37.7 \pm	4.0   $ &  12		\\ 
1654.2838  & 0.36080  &  $   767.0 \pm 48.9 $	& $  15.10 \pm 0.06	  $  &   $   18.2 \pm	1.5   $ &  12		     \\ 
1654.7093  & 0.36115  &  \nodata		& $  13.63 \pm 0.17	  $  &   $   22.7 \pm	7.6   $ &  12		\\ 
1655.1347  & 0.36150  &  \nodata		& $  13.71 \pm 0.10	  $  &   $   44.3 \pm  10.1   $ &  12		\\ 
1657.3847  & 0.36335  &  $   103.3 \pm 26.9 $	& $ (<)13.55  \pm 0.09	  $  &   $   26.4 \pm	4.5   $ &  13		     \\   	 
1685.7938  & 0.38672  &  $   37.5 \pm 10.9  $	& $  13.21 \pm 0.09	  $  &   $   15.3\pm   5.9    $ &											     \\ 
1708.8795  & 0.40571  &  $   460.5 \pm 37.0  $  & $  14.98 \pm 0.02	  $  &   $   33.0 \pm  1.5    $&   14								       \\ 
1712.7088  & 0.40886  &  $   341.6 \pm 37.8 $	& $  14.38 \pm 0.03	  $  &   $   39.9 \pm	2.2   $&   15										       \\ 
1713.5598  & 0.40956  &  $   95.6  \pm 26.2 $	& $  13.58 \pm 0.07	  $  &   $   26.4 \pm	6.1   $& 
\enddata		
\tablecomments{
$a$: The wavelength, $\lambda$, is the observed redshifted wavelength of Ly$\alpha$. The equivalent width, $W_{\rm Ly \alpha}$,
is the rest-frame equivalent width of Ly$\alpha$. A colon means that the result is uncertain.  
\\
(1) The Ly$\alpha$ for the systems at $z=0.05925$ and 0.05896 were fitted simultaneously.
\\
(2) We fitted simultaneously detected \hi\ $\lambda\lambda$1215, 1025, 972, 937 (\hi\ $\lambda$949 is blended). 
\\
(3) We  fitted simultaneously detected \hi\ $\lambda\lambda$1215, 1025, 972, 937 (\hi\ $\lambda\lambda$972, 949 are blended). 
We found that a 2-component fit improves the modeling of Ly$\alpha$, yielding a narrow and broad component. 
If the BLA is mostly thermally broadened, it yields a temperature of $T \sim 3 \times 10^5$ K.  \ovi\ 
is found in this absorber \citep{prochaska04}.  
The temperature found from the \hi\ broadening is consistent with \ovi\ being mainly collisionally ionized. 
\\
(4) This profile appears asymmetric, but within the S/N a single component fit is suitable. This system
is, however, not accounted in the BLAs in this paper because of the asymmetry. 
\\
(5) The  Ly$\alpha$ for the systems at $z=0.16121$ and 0.16150 were fitted simultaneously.
\\
(6) The systems at $z=0.16661$, 0.16678, 0.16694, 0.16712, and 0.16714 were fitted simultaneously. 
The equivalent width reported at $z=0.16714$ is the total equivalent width for these absorbers.
The systems at  $z=0.1671$ are detected from Ly$\alpha$ down to the Lyman Limit.  
\citet{prochaska04} derived the total column density of the system at $z\simeq 0.1671$ 
using the flux decrement at the Lyman limit. \citet{williger06} 
fixed the column density found by \citet{prochaska04} for the systems at $z=0.16714$  and used 
\hi\ $\lambda\lambda$1215, 1025 to derive $b$. They fitted a very broad absorber at $z=0.16692$ 
and a very narrow absorber at $z=0.16628$. This fit appears satisfactory for 
\hi\ $\lambda\lambda$1215, 1025, but not for the other transitions (see Fig.~\ref{fitpks}). 
In particular, the core of the higher \hi\ Lyman series is not reproduced. To reproduce
the core of the line, we find that a 2-component fit is necessary: a very narrow component 
with $b=12$ \km\ and a broader (but still relatively narrow) component with  $b=30$ \km. Furthermore
the \cii\ and \ciii\ kinematics (see  Figs.~9 and 10 in Prochaska et al. 2004)
indicate 4 distinct components at  $-136,-93,-40,0$ \km. We, therefore, fitted 
5 components set initially at $-136,-93,-40,-5,5$ \km. The velocity of the component at 
$-40$ \km\ was not allowed to vary; all the other parameters for the various components were 
allowed to vary. For the fit, we used 
simultaneously detected \hi\ $\lambda\lambda$1215, 1025, 937, 930, 926, 923, 919, 918, 917 lines. 
\hi\ $\lambda\lambda$972 and 949 are contaminated, while \hi\ $\lambda$920 suffers
from fixed-pattern noise. We did not use higher Lyman series transitions because the continuum 
placement becomes uncertain due to the flux decrement. The resulting fit is shown in Fig. ~\ref{fitpks}.
We  note that \citet{prochaska04} used a curve-of-growth analysis with two components separated
by 40 \km\ based on the \cii\ and \siii\ profiles and found 16.35 and 15.5 dex, which is 
roughly consistent with our results for the systems $z= 0.16694, 0.16712, 0.16714$. While our fit appears
more satisfactory than the one presented by \citet{williger06} who did not take into
account the velocity structure of the metal-ion profiles, the solution may actually be
more complicated. The errors do not reflect that there may be more than
5 components. In particular, 
the broad component at $z=0.16678$ may actually be narrower since 
metal-ions such as \ciii\ are detected at this redshift; this directly
affects the measurements of the absorber at $z=0.16661$ where \ciii\ is also
detected. Therefore we mark the measurements at $z=0.16661$ and 0.16678 as uncertain.
\\
(7) The systems at $z = 0.18261$ and 0.18287 were fitted simultaneously. 
We used detected \hi\ $\lambda\lambda$1215, 972, 949 (\hi\ $\lambda\lambda$1025, 937 are contaminated). 
We note that \citet{williger06} fitted a single component at $z = 0.18271$: a 2-component fit improves 
the fit. The equivalent width is the total equivalent width including both systems. 
\\
(8) The  systems at $z=0.24513$ and 0.24554 were fitted simultaneously. 
We used detected \hi\ $\lambda\lambda$1215,1025 in the profile fitting. 
\\
(9) We  fitted simultaneously detected \hi\ $\lambda\lambda$1215, 1025. We note
that \hi\ $\lambda$972 is also detected and possibly contaminated by a weak Ly$\alpha$ system. 
\\
(10) We  fitted simultaneously detected \hi\ $\lambda\lambda$1215, 1025.
\\
(11) The  Ly$\alpha$ for the systems at $z=0.34186$ and 0.34234 were fitted simultaneously.
\\
(12) The equivalent width is the total equivalent width including the systems 
at $z=0.36079$, 0.36080, 0.36115, 0.36150.  The system at $z=0.36080$ is detected in 
\hi\ $\lambda\lambda$1215, 1025, 972, 949, 937, 930, 923 (\hi\ $\lambda$926 is contaminated). 
We undertook to fit these lines with 2, 3, and 4 components. Statistically, these fits are
comparably good because Ly$\alpha$ is so noisy. The 3- and 4-component fits appear, however, to be better models
than the 2-component fit, especially when one takes into account that
\ovi\ $\lambda\lambda$1031, 1037 appear to be present in the weakest component.
The 3-component fit does not fit the blue edge of the Ly$\alpha$ profile as well
as the 4-component fit, but otherwise the 3- or 4-component fits are
quite similar. We note that our 2-component fit yields similar results to 
those found by the 2-component fit of  \citet{williger06}.
\\
(13) Partially blended with interstellar \ci*\ $\lambda$1657.379.
\\
(14) We  fitted simultaneously detected \hi\ $\lambda\lambda$1215, 1025, 972, 949, 937, 930, 926, 923.
\\
(15) We  fitted simultaneously detected \hi\ $\lambda\lambda$1215, 1025, 972, 949.
}
\end{deluxetable}
}

{\LongTables
\begin{deluxetable}{ccccc}
\tabcolsep=3pt
\tablecolumns{5}
\tablewidth{0pt} 
\tabletypesize{\scriptsize}
\tablecaption{Non-detection and confusion with other ISM/IGM lines listed as  Ly$\alpha$ detection in \citet{williger06} toward PKS\,0405--123$^a$  \label{ta2}} 
\tablehead{\colhead{$\lambda$} &\colhead{$z$} & \colhead{$W_{\rm H I}$}  & \colhead{$\log N_{\rm H I}$} & \colhead{Note}   \\
		\colhead{(\AA)} &	\colhead{} & \colhead{(m\AA)} & \colhead{(dex)} & \colhead{}}
\startdata
1218.8185 & 0.00259  &  $  <137.9	 $  & $  <13.40 	    $ &   \\ 
1220.3381 & 0.00384  &  $  <52.9	 $  & $  <13.00 	    $ &   \\ 
1226.8541 & 0.00920  &  $  <43.4	 $  & $  <12.90 	    $ &   \\ 
1232.5921 & 0.01392  &  $  <42.1	 $  & $  <12.90  	    $ &   \\ 
1234.2820 & 0.01531  &  $  <44.3	 $  & $  <12.91 	    $ &   \\ 
1237.1509 & 0.01767  &  $  <35.9	 $  & $  <12.82 	    $ &   \\ 
1238.7434 & 0.01898  &  $  <39.2	 $  & $  <12.86 	    $ &   \\ 
1240.5912 & 0.02050  &  $   <35.0	 $  & $  <12.81 	    $ &   \\ 
1246.0617 & 0.02500  &  $  <39.9	 $  & $  <12.87 	    $ &   \\ 
1293.0596 & 0.06366  &  $  <37.2	 $  & $  <12.82 	    $ &   \\ 
1295.3086 & 0.06551  &  $  <32.7	 $  & $  <12.78 	    $ &   \\ 
1297.2415 & 0.06710  &  $  <16.4	 $  & $  <12.48 	    $ &   \\ 
1298.5180 & 0.06815  &  $  <32.6	 $  & $  <12.78 	    $ &   \\ 
1299.5634 & 0.06901  &  $  <32.6	 $  & $  <12.78 	    $ &   \\ 
1303.1132 & 0.07193  &  $  <31.9	 $  & $  <12.77 	    $ &   \\ 
1305.5080 & 0.07390  &  $  <36.5	 $  & $  <12.83 	    $ &   \\ 
1306.5413 & 0.07475  &  $  <34.7	 $  & $  <12.81 	    $ &   \\ 
1315.6954 & 0.08228  &  $  <31.7	 $  & $  <12.77 	    $ &   \\ 
1318.8195 & 0.08485  &  $   <31.9	 $  & $  <12.77 	    $ &   \\ 
1319.7921 & 0.08565  &  $   <32.1	 $  & $  <12.77 	    $ &   \\ 
1329.2014 & 0.09339  &     \nodata	    &  \nodata  	      & 1 \\ 
1338.9997 & 0.10145  &  $   <32.4	 $  & $  <12.77 	    $ &   \\ 
1346.9259 & 0.10797  &  $   <32.8	 $  & $  <12.78 	    $ &   \\ 
1358.1951 & 0.11724  &  $   <32.5	 $  & $  <12.78 	    $ &   \\ 
1359.6053 & 0.11840  &  $   <33.5	 $  & $  <12.79 	    $ &   \\ 
1360.6994 & 0.11930  &  $   <31.8	 $  & $  <12.77 	    $ &   \\ 
1361.0399 & 0.11958  &  $   <32.4	 $  & $  <12.78 	    $ &   \\ 
1384.8912 & 0.13920  &  $   <32.0	 $  & $  <12.77 	    $ &   \\ 
1422.4069 & 0.17006  &  $   <31.3	 $  & $  <12.76 	    $ &   \\ 
1422.7230 & 0.17032  &  $   <31.9	 $  & $  <12.77 	    $ &   \\ 
1422.9660 & 0.17052  &  $   <32.0	 $  & $  <12.77 	    $ &   \\ 
1424.1210 & 0.17147  &  $   <31.9	 $  & $  <12.77 	    $ &   \\ 
1424.7166 & 0.17196  &  $   <31.9	 $  & $  <12.77 	    $ &   \\ 
1425.0570 & 0.17224  &  $   <31.8	 $  & $  <12.77 	    $ &   \\ 
1427.1357 & 0.17395  &  $   <32.2	 $  & $  <12.77 	    $ &   \\ 
1435.7428 & 0.18103  &  $   <31.9	 $  & $  <12.77 	    $ &   \\ 
1439.6572 & 0.18425  &  $   <37.7	 $  & $  <12.84 	    $ &   \\ 
1440.2530 & 0.18474  &  $   <37.2	 $  & $  <12.82 	    $ &   \\ 
1440.6663 & 0.18508  &  $   <37.4	 $  & $  <12.83 	    $ &   \\ 
1449.0909 & 0.19201  &  $   <39.5	 $  & $  <12.86 	    $ &   \\
1449.7472 & 0.19255  &  \nodata 	    &  \nodata  	      & 2  \\
1452.2637 & 0.19462  &  \nodata 	    &  \nodata  	      & 2  \\ 
1453.4673 & 0.19561  &  $   <39.2	 $  & $  <12.86 	    $ &   \\ 
1454.8895 & 0.19678  &  $   <36.6	 $  & $  <12.83 	    $ &   \\ 
1463.6910 & 0.20402  &  $   <35.5	 $  & $  <12.82 	    $ &   \\ 
1464.4933 & 0.20468  &  $   <35.0	 $  & $  <12.81 	    $ &   \\ 
1465.3808 & 0.20541  &  $   <36.6	 $  & $  <12.83 	    $ &   \\ 
1467.3259 & 0.20701  &  $   <32.6	 $  & $  <12.77 	    $ &   \\ 
1467.8486 & 0.20744  &  $   <33.4	 $  & $  <12.79 	    $ &   \\ 
1474.5713 & 0.21297  &  $   <42.4	 $  & $  <12.89 	    $ &   \\ 
1474.9602 & 0.21329  &  $   <42.3	 $  & $  <12.89 	    $ &   \\ 
1476.2732 & 0.21437  &  $   <42.6	 $  & $  <12.89 	    $ &   \\ 
1477.0876 & 0.21504  &  $   <34.6	 $  & $  <12.80 	    $ &   \\ 
1478.8383 & 0.21648  &  $   <37.5	 $  & $  <12.84 	    $ &   \\ 
1479.5311 & 0.21705  &  $   <36.2	 $  & $  <12.82 	    $ &   \\ 
1480.0175 & 0.21745  &  $   <35.7	 $  & $  <12.82 	    $ &   \\ 
1480.6374 & 0.21796  &  $   <34.4	 $  & $  <12.80 	    $ &   \\ 
1481.2574 & 0.21847  &  $   <35.6	 $  & $  <12.82 	    $ &   \\ 
1482.9472 & 0.21986  &  $   <35.8	 $  & $  <12.82 	    $ &   \\ 
1484.4182 & 0.22107  &  $   <38.7	 $  & $  <12.85 	    $ &   \\ 
1487.4573 & 0.22357  &  $   <39.6	 $  & $  <12.86 	    $ &   \\ 
1488.6122 & 0.22452  &  $   <39.4	 $  & $  <12.86 	    $ &   \\ 
1489.4268 & 0.22519  &  $   <39.6	 $  & $  <12.86 	    $ &   \\ 
1490.0102 & 0.22567  &  $   <39.3	 $  & $  <12.86 	    $ &   \\ 
1493.5965 & 0.22862  &  $   <38.7	 $  & $  <12.85 	    $ &   \\ 
1494.5447 & 0.22940  &  $   <38.7	 $  & $  <12.85 	    $ &   \\ 
1495.1282 & 0.22988  &  $   <39.1	 $  & $  <12.86 	    $ &   \\ 
1499.8450 & 0.23376  &  $   <39.6	 $  & $  <12.86 	    $ &   \\ 
1501.5835 & 0.23519  &  $   <39.2	 $  & $  <12.86 	    $ &   \\ 
1506.5190 & 0.23925  &  $   <40.1	 $  & $  <12.87 	    $ &   \\ 
1507.1147 & 0.23974  &  $   <40.9	 $  & $  <12.88 	    $ &   \\ 
1510.3605 & 0.24241  &  $   <39.6	 $  & $  <12.86 	    $ &   \\ 
1510.7740 & 0.24275  &  $   <39.4	 $  & $  <12.86 	    $ &   \\ 
1517.8612 & 0.24858  &  $   <52.5	 $  & $  <12.99 	    $ &   \\ 
1538.0050 & 0.26515  &  $   <43.7	 $  & $  <12.91 	    $ &   \\ 
1541.2264 & 0.26780  &  $   <45.6	 $  & $  <12.92 	    $ &   \\ 
1543.0499 & 0.26930  &  \nodata 	    &  \nodata  	      & 3  \\ 
1544.3142 & 0.27034  &  $   <45.1	 $  & $  <12.92 	    $ &   \\ 
1544.9585 & 0.27087  &  $   <45.1	 $  & $  <12.92 	    $ &   \\ 
1545.5785 & 0.27138  &  $   <42.7	 $  & $  <12.90 	    $ &   \\ 
1561.5525 & 0.28452  &  $   <47.0	 $  & $  <12.94 	    $ &   \\ 
1568.7249 & 0.29042  &  $   <49.7	 $  & $   <12.96	    $ &   \\ 
1569.4422 & 0.29101  &  $   <49.1	 $  & $   <12.96	    $ &   \\ 
1573.3566 & 0.29423  &  $   <42.5	 $  & $   <12.89	    $ &   \\ 
1580.3102 & 0.29995  &  $   <57.4	 $  & $   <13.02	    $ &   \\ 
1586.4007 & 0.30496  &  $   <57.4	 $  & $   <13.02	    $ &   \\ 
1588.5768 & 0.30675  &  $   <47.5	 $  & $   <12.94	    $ &   \\ 
1589.1116 & 0.30719  &  $   <45.5	 $  & $   <12.92	    $ &   \\ 
1592.3453 & 0.30985  &  $   <48.9	 $  & $   <12.95	    $ &   \\ 
1594.6187 & 0.31172  &  $   <48.9	 $  & $   <12.95	    $ &   \\ 
1595.1779 & 0.31218  &  $   <50.1	 $  & $   <12.96	    $ &   \\ 
1596.1139 & 0.31295  &  \nodata 	    &  \nodata  	      & 4    \\ 
1596.5880 & 0.31334  &  \nodata 	    &  \nodata  	      & 4    \\ 
1597.1716 & 0.31382  &  \nodata 	    &  \nodata  	      & 4    \\ 
1606.0945 & 0.32116  &  $   <48.5	 $  & $   <12.95	    $ &   \\ 
1614.9324 & 0.32843  &  $   <62.0	 $  & $   <13.06	    $ &   \\ 
1615.6011 & 0.32898  &  $   <59.9	 $  & $   <13.04	    $ &   \\ 
1623.9163 & 0.33582  &  $   <44.5	 $  & $   <12.91	    $ &   \\ 
1624.9982 & 0.33671  &  $   <45.7	 $  & $   <12.92	    $ &   \\ 
1625.6182 & 0.33722  &  $   <44.9	 $  & $   <12.92	    $ &   \\ 
1628.5358 & 0.33962  &  $   <46.2	 $  & $   <12.93	    $ &   \\ 
1636.8388 & 0.34645  &  $   <52.5	 $  & $   <12.99	    $ &   \\ 
1639.2459 & 0.34843  &  $   <73.5	 $  & $   <13.13	    $ &   \\ 
1668.3856 & 0.37240  &  $   <61.3	 $  & $   <13.05	    $ &   \\ 
1673.1509 & 0.37632  &  $   <83.0	 $  & $   <13.18	    $ &   \\ 
1678.6093 & 0.38081  &  $   <51.8	 $  & $   <12.99	    $ &   \\ 
1684.1528 & 0.38537  &  $   <49.3	 $  & $   <12.96	    $ &   \\ 
1685.4779 & 0.38646  &  $   <50.4	 $  & $   <12.97	    $ &   \\ 
1696.8565 & 0.39582  &  $   <64.2	 $  & $   <13.07	    $ &   \\ 
1700.7588 & 0.39903  &  $   <62.2	 $  & $   <13.06	    $ &   \\ 
1702.0596 & 0.40010  &  $   <62.0	 $  & $   <13.06	    $ &   \\
\enddata		
\tablecomments{$a$: The upper limits are 3$\sigma$ and were estimated over a
velocity range $[-40,40]$ \km. \\
(1)  Misidentification: interstellar \ci*\ $\lambda$1329. \\
(2)  The spectrum is erratic near this wavelength.  \\
(3)  Misidentification: \ovi\ $\lambda$1031 at $z = 0.4951$. \\
(4)  There is undulation in the spectrum, making any identification with a BLA very uncertain.
    }
\end{deluxetable}
}

\begin{thebibliography}{}
\bibitem[Aracil et al.(2006a)]{aracil06} 
	Aracil, B., Tripp, T.~M., Bowen, D.~V., Prochaska, J.~X., Chen, H.-W., 
	\& Frye, B.~L.\ 2006a, \mnras, 367, 139 
\bibitem[Aracil et al.(2006b)]{aracil06a} 
	Aracil, B., Tripp, T.~M., Bowen, D.~V., Prochaska, J.~X., Chen, H.-W., \& Frye, B.~L.\ 2006b, \mnras, 372, 959 
\bibitem[Bowen et al.(2002)]{bowen02} 
	Bowen, D.~V., Pettini, M., \& Blades, J.~C.\ 2002, \apj, 580, 169 
\bibitem[Burles et al.(2001)]{burles01} 
	 Burles, S., Nollett, K.~M., \& Turner, M.~S.\ 2001, ApJ, 552, L1
\bibitem[Bahcall et al.(1991)]{bahcall91} 
	Bahcall, J.~N., Jannuzi, B.~T., Schneider, D.~P., Hartig, G.~F., Bohlin, R., \& 
	Junkkarinen, V.\ 1991, \apjl, 377, L5 
\bibitem[Cen \& Fang(2006)]{cen06a} 
	Cen, R., \& Fang, T.\ 2006, \apj, 650, 573 
\bibitem[Cen \& Ostriker(1999)]{cen99} 
	Cen, R., \& Ostriker, J.~P.\ 1999, ApJ 514, 1
\bibitem[Cen \& Ostriker(2006)]{cen06b} 
	Cen, R., \& Ostriker, J.~P.\ 2006, \apj, 650, 560 
\bibitem[Cristiani \& D'Odorico(2000)]{cristiani00} 
	Cristiani, S., \& D'Odorico, V.\ 2000, \aj, 120, 1648 
\bibitem[Danforth \& Shull(2005)]{danforth05} 
	Danforth, C.~W., \& Shull, J.~M.\ 2005, \apj, 624, 555 
\bibitem[Dav{\' e} et al.(1999)]{dave99} 
	Dav{\' e}, R., Hernquist, L., Katz, N., \& Weinberg, D.~H.\ 1999, ApJ, 511, 521 
\bibitem[Dav{\'e} et al.(2001)]{dave01a} 
	Dav{\'e}, R., et al.\ 2001, \apj, 552, 473 
\bibitem[Dav{\' e} \& Tripp(2001)]{dave01} 
	Dav{\' e}, R., \& Tripp, T.~M.\ 2001, \apj, 553, 528 
\bibitem[D'Odorico \& Petitjean(2001)]{dodo01} 
	D'Odorico, V., \& Petitjean, P.\ 2001, \aap, 370, 729 
\bibitem[Fang \& Bryan(2001)]{fang01} 
	Fang, T., \& Bryan, G.~L.\ 2001, \apjl, 561, L31 
\bibitem[Fang et al.(2006)]{fang06} 
	Fang, T., Mckee, C.~F., Canizares, C.~R., \& Wolfire, M.\ 2006, \apj, 644, 174 
\bibitem[Fukugita et al.(1998)]{fukugita98} 
	Fukugita, M., Hogan, C.~J., \& Peebles, P.~J.~E.\ 1998, \apj, 503, 518
\bibitem[Ganguly et al.(2003)]{ganguly03} 
	Ganguly, R., Masiero, J., Charlton, J.~C., \& Sembach, K.~R.\ 2003, \apj, 598, 922 
\bibitem[Haardt \& Madau(1996)]{haard96}
	Haardt, F., \& Madau, P.\ 1996, \apj, 461, 20 
\bibitem[Hu et al.(1995)]{hu95} 
	Hu, E.~M., Kim, T., Cowie, L.~L., Songaila, A., \& Rauch, M.\ 1995, \aj, 110, 1526 
\bibitem[Impey et al.(1999)]{impey99} 
	Impey, C.~D., Petry, C.~E., \& Flint, K.~P.\ 1999, \apj, 524, 536
\bibitem[Janknecht et al.(2006)]{jank06} 
	Janknecht, E., 	Reimers, D., Lopez, S., \& Tytler, D.\ 2006, \aap, in press  [astro-ph/0608342]
\bibitem[Kaastra et al.(2006)]{kaastra06} 
	Kaastra, J.~S., Werner, N., den Herder, J.~W.~A., Paerels, F.~B.~S., de Plaa, J., Rasmussen, A.~P., 
	\& de Vries, C.~P.\ 2006, \apj, submitted  [astro-ph/0604519]
\bibitem[Kim et al.(2002a)]{kim02} 
	Kim, T.-S., Carswell, R.~F., Cristiani, S., D'Odorico, S., \& Giallongo, E.\ 
	2002a, \mnras, 335, 555 
\bibitem[Kim et al.(2001)]{kim01} 
	Kim, T.-S., Cristiani, S., \& D'Odorico, S.\ 2001, \aap, 373, 757 
\bibitem[Kim et al.(2002b)]{kim02a} 
	Kim, T.-S., Cristiani, S., \& D'Odorico, S.\ 2002b, \aap, 383, 747 
\bibitem[Kim et al.(1997)]{kim97} 
	Kim, T.-S., Hu, E.~M., Cowie, L.~L., \& Songaila, A.\ 1997, \aj, 114, 1 
\bibitem[Kirkman \& Tytler(1997)]{kirkman97} 
	Kirkman, D., \& Tytler, D.\ 1997, \apj, 484, 672 
\bibitem[Lehner et al.(2006)]{lehner06} 
	Lehner, N., Savage, B.~D., Wakker, B.~P., Sembach, K.~R., \& Tripp, T.~M.\ 2006, ApJS, 164, 1
\bibitem[Lu et al.(1996)]{lu96} 
	Lu, L., Sargent, W.~L.~W., Womble, D.~S., \& Takada-Hidai, M.\ 1996, \apj, 472, 509 
\bibitem[Morris et al.(1991)]{morris91} 
	Morris, S.~L., Weymann, R.~J., Savage, B.~D., \& Gilliland, R.~L.\ 1991, \apjl, 377, L21 
\bibitem[Nicastro et al.(2005)]{nicastro05} 
	Nicastro, F., et al.\ 2005, \nat, 433, 495 
\bibitem[Oegerle et al.(2000)]{oegerle00} 
	Oegerle, W.~R., et al.\ 2000, \apjl, 538, L23 
\bibitem[O'Meara et al.(2001)]{omeara01} 
	O'Meara, J.~M., Tytler, D., Kirkman, D., Suzuki, N., Prochaska, J.~X., Lubin, D., \& 
	Wolfe, A.~M.\ 2001, \apj, 552, 718 
\bibitem[Penton et al.(2000)]{penton00} 
	Penton, S.~V., Shull, J.~M., \& Stocke, J.~T.\ 2000, \apj, 544, 150 
\bibitem[Penton et al.(2004)]{penton04} 
	Penton, S.~V., Stocke, J.~T., \& Shull, J.~M.\ 2004, \apjs, 152, 29 
\bibitem[Petitjean et al.(1993)]{petitjean93} 
	Petitjean, P., Webb, J.~K., Rauch, M., Carswell, R.~F., \& Lanzetta, K.\ 1993, \mnras, 262, 499 
\bibitem[Prochaska et al.(2004)]{prochaska04} 
	Prochaska, J.~X., Chen, H.-W., Howk, J.~C., Weiner, B.~J., \& Mulchaey, 
	J.\ 2004, \apj, 617, 718 
\bibitem[Rasmussen et al.(2006)]{rasmussen06} 
	Rasmussen, A.~P., Kahn, S.~M., Paerels, F., Willem den Herder, J., Kaastra, J., \& de Vries, 
	C.\ 2006, \apj, submitted [astro-ph/0604515]
\bibitem[Rauch et al.(1997)]{rauch97} 
	Rauch, M., et al.\ 1997, \apj, 489, 7 
\bibitem[Richter et al.(2006a)]{richter06a} 
	Richter, P., Fang, T., \& Bryan, G.~L.\ 2006a, \aap, 451, 767 
\bibitem[Richter et al.(2006b)]{richter06} 
	Richter, P., Savage, B.~D., Sembach, K.~R.,  \& Tripp, T.~M. \ 2006b, \aap, 445, 827 	
\bibitem[Richter et al.(2004)]{richter04} 
	Richter, P., Savage, B.~D., Tripp, T.~M., \& Sembach, K.~R.\ 2004, \apjs, 153, 165 
\bibitem[Savage et al.(2005)]{savage05} 
	Savage, B.~D., Lehner, N., Wakker, B.~P., Sembach, 
	K.~R., \& Tripp, T.~M.\ 2005, \apj, 626, 776 
\bibitem[Savage et al.(2002)]{savage02} 
	Savage, B.~D., Sembach, K.~R., Tripp, T.~M., \& Richter, P.\ 2002, \apj, 564, 631 
\bibitem[Schaye(2001)]{schaye01} 
	Schaye, J.\ 2001, \apj, 559, 507 
\bibitem[Schaye et al.(1999)]{schaye99} 
	Schaye, J., Theuns, T., Leonard, A., \& Efstathiou, G.\ 1999, \mnras, 310, 57 
\bibitem[Sembach et al.(2004)]{sembach04} 
	Sembach, K.~R., Tripp, T.~M., Savage, B.~D., \& Richter, P.\ 2004, \apjs, 155, 351
\bibitem[Spergel et al.(2003)]{spergel03} 
	Spergel, D.~N., et al.\ 2003, \apjs, 148, 175
\bibitem[Shull et al.(2000)]{shull00} 
	Shull, J.-M., et al.\ 2000, \apjl, 538, L13
\bibitem[Sutherland \& Dopita(1993)]{sutherland93} 
	Sutherland, R.~S.~\& Dopita, M.~A.\ 1993, \apjs, 88, 253 
\bibitem[Theuns et al.(1998)]{theuns98} 
	Theuns, T., Leonard, A., \& Efstathiou, G.\ 1998, \mnras, 297, L49 
\bibitem[Tripp et al.(2001)]{tripp01} 
	Tripp, T.~M., Giroux, 	M.~L., Stocke, J.~T., Tumlinson, J., \& Oegerle, W.~R.\ 
	2001, \apj, 563, 724 
\bibitem[Tripp et al.(1998)]{tripp98} 
	Tripp, T.~M., Lu, L., \& Savage, B.~D.\ 1998, \apj, 508, 200 
\bibitem[Tripp et al.(2006)]{tripp04} 
	Tripp, T.~M., Bowen, D.~V., Sembach, K.~R., Jenkins, E.~B., Savage, B.~D., 
	\& Richter, P.\ 2006, Astronomical Society of the Pacific Conference Series, 348, 341 
\bibitem[Tripp \& Savage(2000)]{tripp00a} 
	Tripp, T.~M.~\& Savage, B.~D.\ 2000, \apj, 542, 42 
\bibitem[Tripp, Savage, \& Jenkins(2000)]{tripp00b} 
	Tripp, T.~M., Savage, B.~D., \& Jenkins, E.~B.\ 2000, \apjl, 534, L1 
\bibitem[Tytler(1987)]{tytler87} 
	Tytler, D.\ 1987, \apj, 321, 49 
\bibitem[Weinberg et al.(1997)]{weinberg97} 
	Weinberg, D.~H., Miralda-Escude, J., Hernquist, L., \& Katz, N.\ 1997, \apj, 490, 564 
\bibitem[Weymann et al.(1998)]{weymann98} 
	Weymann, R.~J., et al.\ 1998, \apj, 506, 1 
\bibitem[Williger et al.(2006)]{williger06} 
	Williger, G.~M., Heap, S.~R., Weymann, R.~J., Dav{\'e}, R., Ellingson, E., Carswell, R.~F., Tripp, 
	T.~M., \& Jenkins, E.~B.\ 2006, \apj, 636, 631 
\bibitem[Yuan et al.(2002)]{yuan02} 
	Yuan, Q., Green, R.~F., Brotherton, M., Tripp, T.~M., Kaiser, M.~E., \& Kriss, G.~A.
	\ 2002, \apj, 575, 687 
\end{thebibliography}
\end{document}